\documentclass{aa}  
\usepackage{graphicx}
\usepackage{txfonts}
\usepackage{hyperref}
 \usepackage{lscape}
\begin{document} 

   \title{Radio continuum emission in the northern Galactic plane:
Sources and spectral indices from the THOR survey\thanks{The full continuum catalog, full table of Table~\ref{table_hiifits} and all the fits files of the continuum data are available at the project website \url{http://www2.mpia-hd.mpg.de/thor/DATA/www/}}}


   \author{Y. Wang \inst{1}
          \and
          S. Bihr\inst{1}
          \and
          M. Rugel\inst{1}
          \and
          H. Beuther\inst{1}
          \and
          K. G. Johnston\inst{2}
          \and
          J. Ott\inst{3}
          \and
          J. D. Soler\inst{1}
          \and
          A. Brunthaler\inst{4}
          \and
          L. D. Anderson\inst{5,6,7}
          \and
          J. S. Urquhart\inst{8}
          \and
          R. S. Klessen\inst{9,10}
          \and
          H. Linz\inst{1}
           \and
          N. M. McClure-Griffiths\inst{11}
          \and
          S. C. O. Glover\inst{9}
          \and
          K. M. Menten\inst{4}
          \and
          F. Bigiel\inst{9}
          \and
          M. Hoare\inst{2}
          \and
          S. N. Longmore\inst{12}
          }

   \institute{Max-Planck Institute for Astronomy, K\"onigstuhl 17, 69117 Heidelberg, Germany\\
              \email{wang@mpia.de}
            \and
             School of Physics and Astronomy, University of Leeds, Leeds LS2 9JT, UK
             \and
             National Radio Astronomy Observatory, PO Box O, 1003 Lopezville Road, Socorro, NM 87801, USA
             \and
             Max-Planck-Institut f\"ur Radioastronomie, Auf dem H\"ugel 69, 53121 Bonn, Germany
             \and
             Department of Physics and Astronomy, West Virginia University, Morgantown, WV 26506, USA
             \and
             Adjunct Astronomer at the Green Bank Observatory, P.O. Box 2, Green Bank WV 24944, USA
             \and
             Center for Gravitational Waves and Cosmology, West Virginia University, Chestnut Ridge Research Building, Morgantown, WV 26505, USA
             \and
             School of Physical Sciences, University of Kent, Ingram Building, Canterbury, Kent CT2 7NH, UK
             \and
             Universit\"at Heidelberg, Zentrum f\"ur Astronomie, Institut f\"ur Theoretische Astrophysik, Albert-Ueberle-Str. 2, 69120 Heidelberg, Germany
             \and
             Universit\"at Heidelberg, Interdisziplin\"ares Zentrum fur Wissenschaftliches Rechnen, INF 205, 69120, Heidelberg, Germany
             \and
             Research School of Astronomy and Astrophysics, The Australian National University, Canberra, ACT, Australia
             \and
   	     Astrophysics Research Institute, Liverpool John Moores University, IC2, Liverpool Science Park, 146 Brownlow Hill, Liverpool L3 5RF, UK
             }

   \date{Received xxxx; accepted xxxx}

 
  \abstract
   {Radio continuum surveys of the Galactic plane can find and characterize \ion{H}{ii} regions, supernova remnants (SNRs), planetary nebulae (PNe), and extragalactic sources. A number of surveys at high angular resolution ($\leq$~25$^{\prime\prime}$) at different wavelengths exist to study the interstellar medium (ISM), but no comparable high-resolution and high-sensitivity survey exists at long radio wavelengths around 21~cm.}
   {Our goal is to investigate the 21~cm radio continuum emission in the northern Galactic plane at $<$25\arcsec\ resolution. }
   {We observed a large fraction of the Galactic plane in the first quadrant of the Milky Way ($l=14.0-67.4\degr$ and $|b| \leq 1.25\degr$) with the {\it Karl G. Jansky} Very Large Array (VLA) in the C-configuration covering six continuum spectral windows. These data provide a detailed view on the compact as well as extended radio emission of our Galaxy and thousands of extragalactic background sources.}
   {We used the BLOBCAT software and extracted 10916 sources. After removing spurious source detections caused by the sidelobes of the synthesised beam, we classified 10387 sources as reliable detections.
We smoothed the images to a common resolution of 25$^{\prime\prime}$ and extracted the peak flux 
density of each source in each spectral window (SPW) to determine the spectral indices $\alpha$ (assuming $I(\nu)\propto\nu^\alpha$).
By cross-matching with catalogs of \ion{H}{ii} regions, SNRs, PNe, and pulsars, we found radio counterparts for 840 \ion{H}{ii} regions, 52 SNRs, 164 PNe, and 38 pulsars. We found 79 continuum sources that are associated with X-ray sources. We identified 699 ultra-steep spectral sources ($\alpha < -1.3$) that could be high-redshift galaxies. Around 9000 of the sources we extracted are not classified specifically, but based on their spatial and spectral distribution, a large fraction of them is likely to be extragalactic background sources. More than 7750 sources do not have counterparts in the SIMBAD database, and more than 3760 sources do not have counterparts in the NED database. }
{Studying the long wavelengths cm continuum emission and the associated spectral indices allows us to characaterize a large fraction of Galactic and extragalactic radio sources in the area of the northern inner Milky Way. This database will be extremely useful for future studies of a diverse set of astrophysical objects.}

\maketitle
%
\section{Introduction}
A number of surveys at high angular resolution ($\leq$~20$^{\prime\prime}$) at different wavelengths exist to study the interstellar medium (ISM), 
from infrared  (e.g., UKIDSS\footnote{UKIRT Infrared Deep Sky Survey},  \citealt{lucas2008}; {\it Spitzer}/GLIMPSE\footnote{Galactic Legacy Infrared Midplane Survey Extraordinaire }, \citealt{benjamin2003,churchwell2009}, {\it Spitzer}/MIPSGAL\footnote{A 24 and 70 Micron Survey of the Inner Galactic Disk with MIPS}, \citealt{carey2009} ), to (sub)mm (e.g., ATLASGAL\footnote{APEX Telescope Large Area Survey of the Galaxy} and BGPS\footnote{Bolocam Galactic Plane Survey}, \citealt{schuller2009,rosolowsky2010,aguirre2011,csengeri2014}) and radio (e.g. MAGPIS\footnote{Multi-Array Galactic Plane Imaging Survey}, CORNISH\footnote{the Co-Ordinated Radio `N' Infrared Survey for High-mass star formation}, \citealt{helfand2006,hoare2012}) wavelengths. Previously, the best 21~cm \ion{H}{i} line survey was the HI Very Large Array Galactic Plane Survey (VGPS, \citealt{stil2006}) which has a resolution of 60$^{\prime\prime}$, significantly more coarse than the resolution of the aforementioned surveys. This was one of the motivations for initiating ``The \ion{H}{i}, OH, Recombination line survey of the Milky Way (THOR) \footnote{\url{http://www.mpia.de/thor/Overview.html}}'' \citep{beuther2016}. Using the {\it Karl G. Jansky} Very Large Array (VLA) in C-configuration, we achieve a spatial resolution of $<25^{\prime\prime}$. The WIDAR correlator at the VLA allows us to observe many spectral lines simultaneously, in particular several molecular OH transitions, a series of H$n\alpha$ radio recombination lines (RRLs, $n=$151 to 186), as well as eight spectral windows (SPWs) to cover the continuum emission between 1 and 2 GHz \citep{bihr2015, beuther2016, bihr2016, rugel2018}. We observed a large fraction of the Galactic plane in the first quadrant of the Milky Way ($l=14.0-67.4^\circ$ and $|b| \leq 1.25^\circ$) in several semesters (from 2012 to 2014). The continuum data from the first half of the survey ($l=14.0-37.9^\circ$ and $l=47.1-51.2^\circ$) have been published by \citet{bihr2016}. In this paper, we combine all the continuum data and present the results from the full survey.

The radio continuum emission from 1 to 2~GHz is dominated by thermal free-free emission from electrons and non-thermal synchrotron emission of the relativistic electrons in magnetic fields \citep[e.g.,][]{wilson2013}. These can be distinguished by the spectral index $\alpha$, assuming $I(\nu) \propto \nu^\alpha$, where $I(\nu)$ is the intensity at frequency $\nu$. The thermal free-free emission shows an almost flat ($\alpha=-0.1$) spectrum if it is optically thin, or positive spectral index if it is optically thick with values varying between --0.1 and 2 \citep[e.g.,][]{keto2003, wilson2013}. \ion{H}{ii} regions and planetary nebulae are often the sources for thermal free-free emission. In contrast to this, synchrotron emission shows a negative spectral index whose value depends, amongst other things, on the particle energy distribution. Towards extragalactic jets powered by an active galactic nucleus (AGN), one often finds the synchrotron emission with a spectral index $\alpha<-0.5$ \citep[e.g.,][]{hey1971,rybicki1979}. Galactic SNRs often show synchrotron emission with a spectral index around $-0.5$ \citep[e.g.,][]{bhatnagar2011, green2014, reynoso2015}. Thus the spectral index can help us to characterize the nature of the continuum sources we detected in the survey. This allows us to determine whether they are Galactic or extragalactic, which is crucial for \ion{H}{i} and OH absorption studies. 

Compact galactic radio sources associated with X-ray emission are usually Pulsar Wind Nebulae \citep[e.g.,][]{brisken2005, miller2005}, or X-ray binaries \citep[XRB or microquasars; ][]{mirabel1998, mirabel1999}. By investigating the X-ray and radio flux ratios, the spectral indices and observations at other (optical/infrared) wavelengths of the Galactic sources in detail, we can also constrain the type of the sources, i.e., low mass XRB, PN, pulsar etc. \citep[e.g.,][]{seaquist1993, maccarone2012, tetarenko2016}. With the high-angular resolution ($<25\arcsec$) of our THOR continuum data, we can not only derive the spectral indices of the sources, but also further study the variation in frequency and space.

The paper is structured as follows: In Sect.~\ref{sect_obs}, we present the observations and data reduction. Sect.~\ref{sect_extract} presents the methods we used to extract sources and determine the spectral indices. Sect.~\ref{sect_cata} describes the continuum catalog, and the distribution of the continuum sources we extracted. The nature of continuum sources are discussed in Sect.~\ref{sect_discuss}. The conclusions and summary are presented in Sect~\ref{sect_con}. The appendix gives additional information of the continuum observations and tables.  
\section{Observations and data reductions}
\label{sect_obs}
We observed a part of the first quadrant of the Galactic plane ($l=14.0-67.4^\circ$ and $\lvert b \rvert \leq 1.25^\circ$) with the {\it Karl G. Jansky} Very Large Array (VLA) in C configuration at L band from 1 to 2~GHz. The detailed observing strategy and data reduction is discussed and described in \citet{bihr2016} and \citet{beuther2016}. With the WIDAR correlator, we cover the \ion{H}{i} 21~cm line, 4 OH lines, 19 H$\alpha$ recombination lines, as well as eight continuum bands, i.e., SPWs. Each continuum SPW has a band width of 128~MHz. Due to strong contamination from radio frequency interference (RFI), two SPWs around 1.2 and 1.6~GHz were not usable and discarded. The remaining six SPWs are centered at 1.06, 1.31, 1.44, 1.69, 1.82 and 1.95~GHz. For the fields at $l=23.1-24.3^\circ$ and $25.6-26.8^\circ$, the SPW around 1.95~GHz is also severely affected by RFI and is therefore flagged \citep[see][]{bihr2016}. 

The calibration and imaging were performed with the CASA\footnote{\url{http://casa.nrao.edu}} software package. We employed the RFlag algorithm which was first introduced to AIPS by E. Greisen in 2011 to flag outliers in each visibility dataset before imaging. For imaging, we chose a pixel size of 2.5$^{\prime\prime}$ and $robust=0$ as a weighting parameter, which results in a synthesized beam size varying from $9^{\prime\prime}$ to $25^{\prime\prime}$ depending on the SPW and the declination of different observing blocks and uv-coverage (see Table~\ref{tab_beams} for details). We also used the multiscale CLEAN in CASA in order to recover the large scale structure better, and chose the scale parameters as 1, 3 and 6 times the resolution element. The cleaning process was set to stop at a threshold of 5~mJy~beam$^{-1}$ or $10^5$ iterations, whichever was reached first. The noise of our data is dominated by the artifacts resulting from residual sidelobes, and varies from $\sim0.3$ to $>1$~mJy~beam$^{-1}$ depending on the frequency and sky position. The thermal noise is $\sim0.1$~mJy ~beam$^{-1}$. We will discuss the noise level of the images in Sect.~\ref{sect_extract}. 

\begin{figure*}
\centering
   \includegraphics[width= 0.95\hsize]{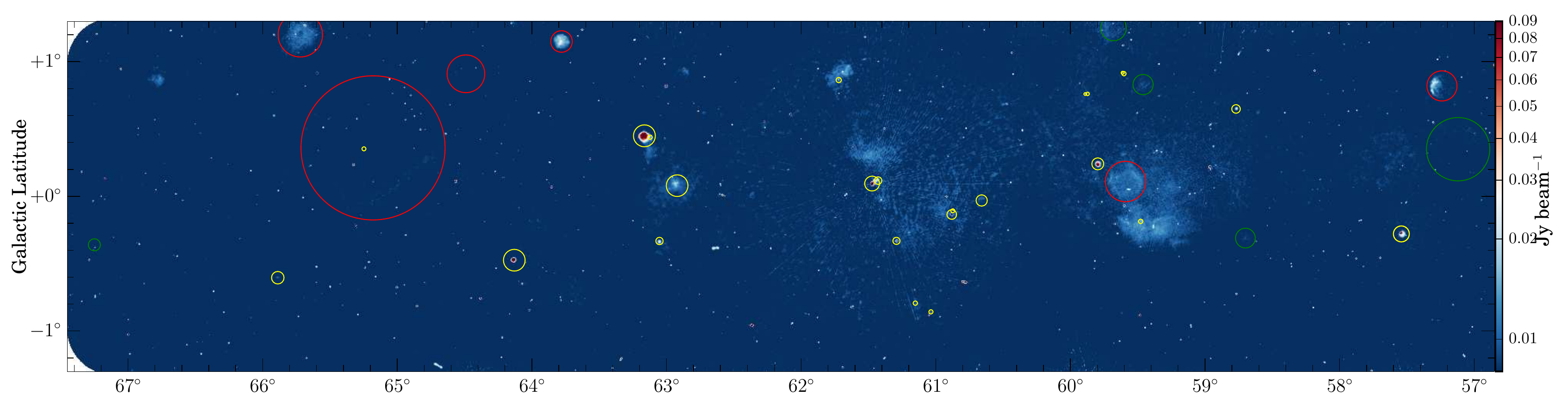}
   \includegraphics[width= 0.95\hsize]{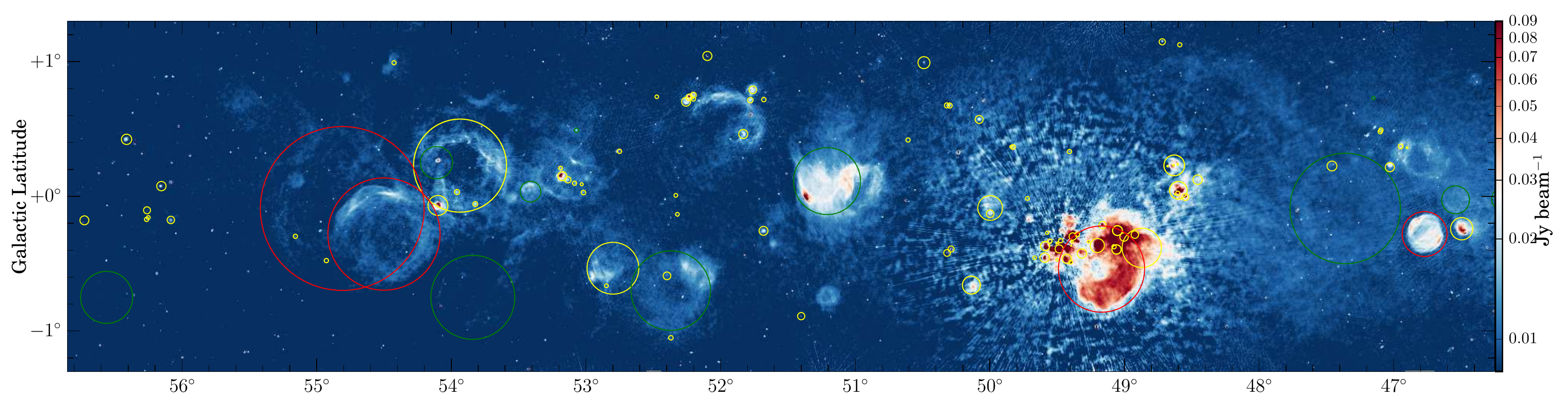}
   \includegraphics[width= 0.95\hsize]{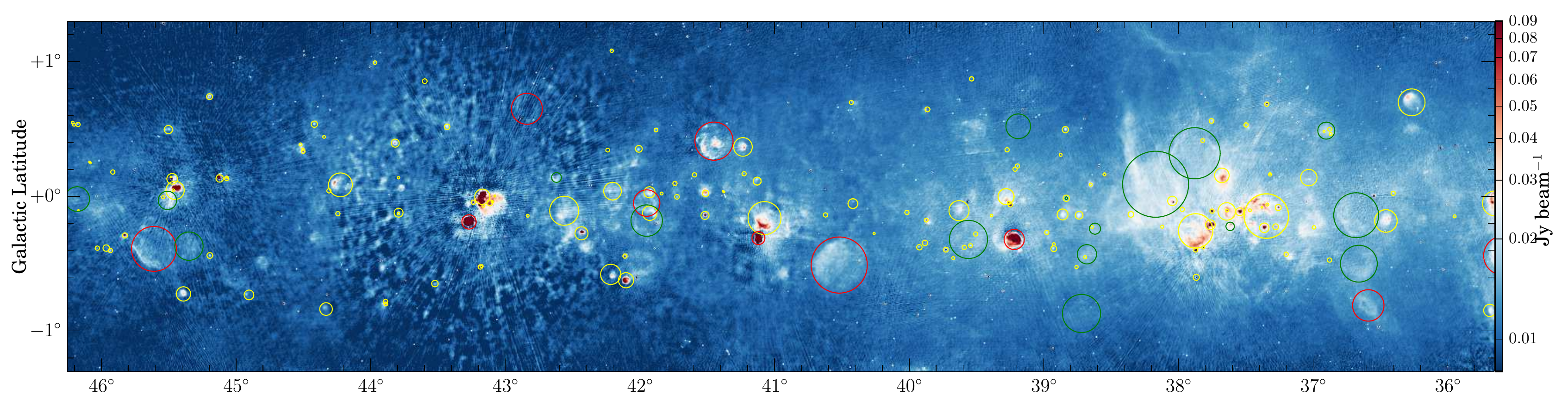}
   \includegraphics[width= 0.95\hsize]{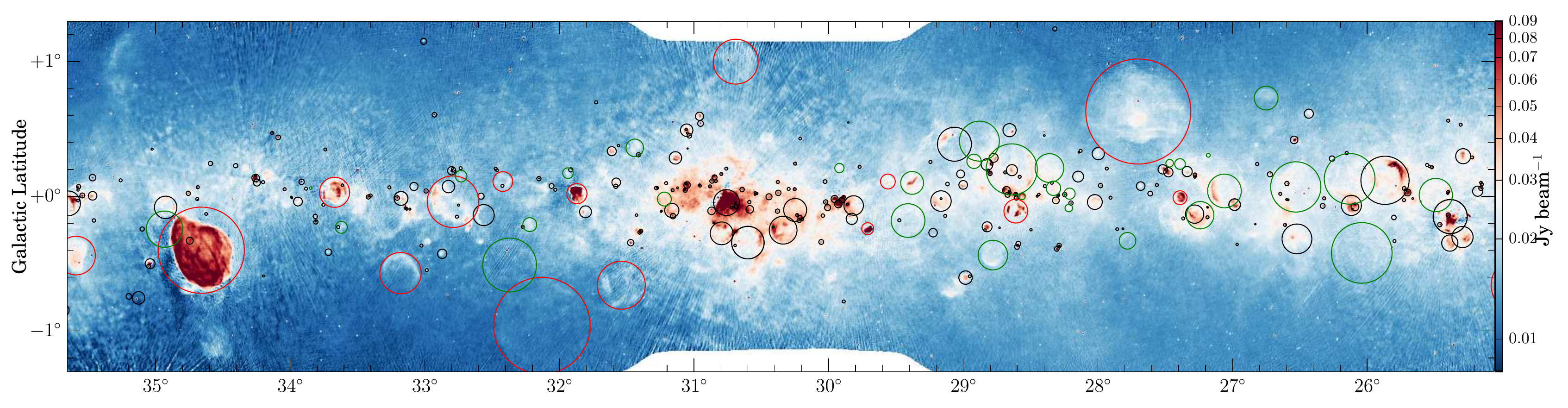}
   \includegraphics[width= 0.95\hsize]{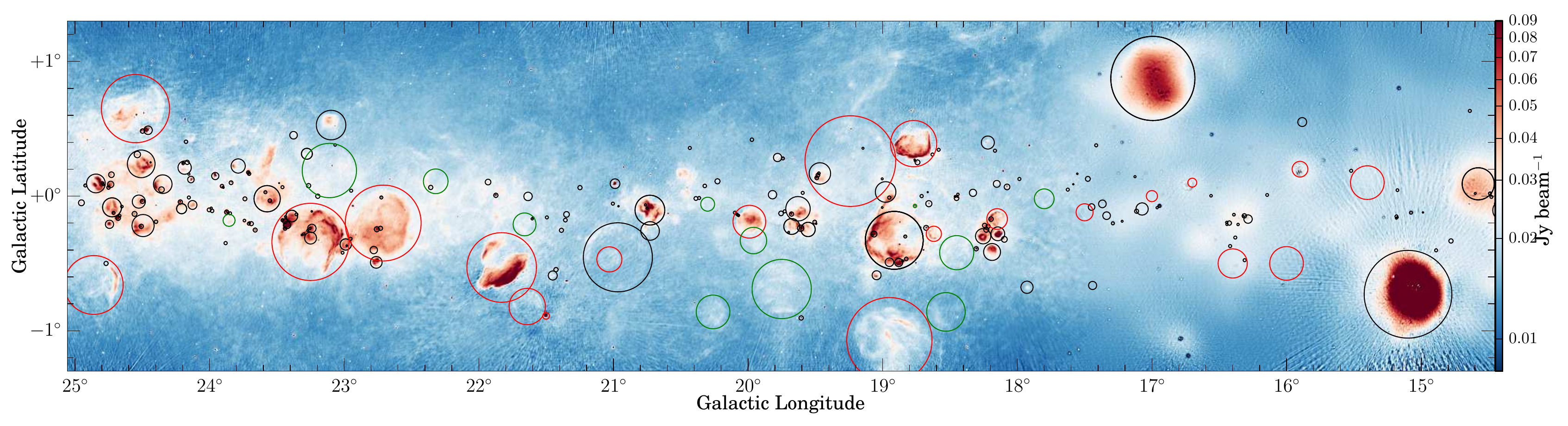}
   \caption{THOR+VGPS 1.4~GHz continuum map of the whole survey. The yellow and black circles mark the {\it WISE} \ion{H}{ii} regions \citep{anderson2017} matched with our continuum sources (see also Sect.~\ref{sect_hii}). The red circles mark the SNRs from \citet{green2014}. The green circles mark the new SNR candidates identified in this combined dataset \citep{anderson2017}. The synthesized beam size is 25\arcsec. For regions $l<17.5\degr$, complementary D-configuration data do not exist. Therefore, in these regions we have only combined the THOR data with the Effelsberg data.}
    \label{fig_wholemap}
\end{figure*}

\paragraph{THOR+VGPS data:} The 1.4~GHz continuum data from the VGPS survey \citep{stil2006} combined VLA D-configuration with single-dish observations from Effelsberg, and have an angular resolution of 60\arcsec. The $14.0\degr<l<17.5\degr$ Galactic longitude range in the VGPS data is just comprised of single-dish observations. We smoothed the THOR continuum data at 1.4~GHz to a resolution of 25\arcsec, and used the task ``feather'' in CASA to combine the THOR data with those from VGPS. While this combined dataset retains the high angular resolution of the THOR observations, it can recover the large scale structure. The combined image of the whole survey at 1.4~GHz in Fig.~\ref{fig_wholemap} shows that large scale \ion{H}{ii} regions and SNRs dominate the extended radio emission in the inner Galactic plane ($l<55\degr$), while compact sources are distributed across the whole survey area (more prominent in area $l>55\degr$). These compact sources are most likely extragalactic sources, and we will discuss them later. \citet{anderson2017} used this dataset and identified 76 new Galactic SNR candidates in the survey area. Although the spectral band of the VGPS continuum ($\sim$1~MHz) is different from our THOR continuum data ($\sim$128~MHz), \citet{anderson2017} showed that the flux retrieved from the combined data is consistent with the literature by comparing the flux density of the known SNRs \citep{green2014}. 

All the reduced continuum data including the THOR+VGPS dataset can be accessed from the THOR survey website located at \url{http://www.mpia.de/thor}.

\begin{table*}
\caption{Synthesized beams of SPWs.}            
\label{tab_beams}      
\centering                          
\begin{tabular}{l c c}       
\hline\hline                 
SPW& Frequency range& Restoring beam size \\  
&$\rm{[MHz]}$ &   \\
\hline 
spw-1060&$989-1117$ & 24.4$\arcsec\times$ 15.1$\arcsec$ to 15.1$\arcsec\times$ 14.7$\arcsec$\\
spw-1310&$1244-1372$ & 19.7$\arcsec\times$ 12.5$\arcsec$ to 12.6$\arcsec\times$ 12.2$\arcsec$\\
spw-1440&$1372-1500$ & 18.1$\arcsec\times$ 11.1$\arcsec$ to 12.0$\arcsec\times$ 11.6$\arcsec$\\
spw-1690&$1628-1756$ & 15.4$\arcsec\times$ 9.1$\arcsec$ to 9.8$\arcsec\times$ 9.5$\arcsec$\\
spw-1820&$1756-1884$ & 14.5$\arcsec\times$ 8.9$\arcsec$ to 9.2$\arcsec\times$ 9.1$\arcsec$\\
spw-1950&$1884-2012$ & 13.1$\arcsec\times$ 8.1$\arcsec$ to 8.6$\arcsec\times$ 8.2$\arcsec$\\
averaged image  & spw-1440 \& spw-1820 & 18.1$\arcsec\times$ 11.1$\arcsec$ to 12.0$\arcsec\times$ 11.6$\arcsec$\\
THOR+VGPS& $1420$ &$25\arcsec\times25\arcsec$\\
\hline 
\end{tabular}
\tablefoot{The averaged images are used for source extraction. The size of the restoring beam varies due to declination and uv-coverage differences among observing blocks.
}
\end{table*}

\section{Source extraction and spectral index determination}
\label{sect_extract}
\begin{figure}
   \centering
  \includegraphics[width=8cm]{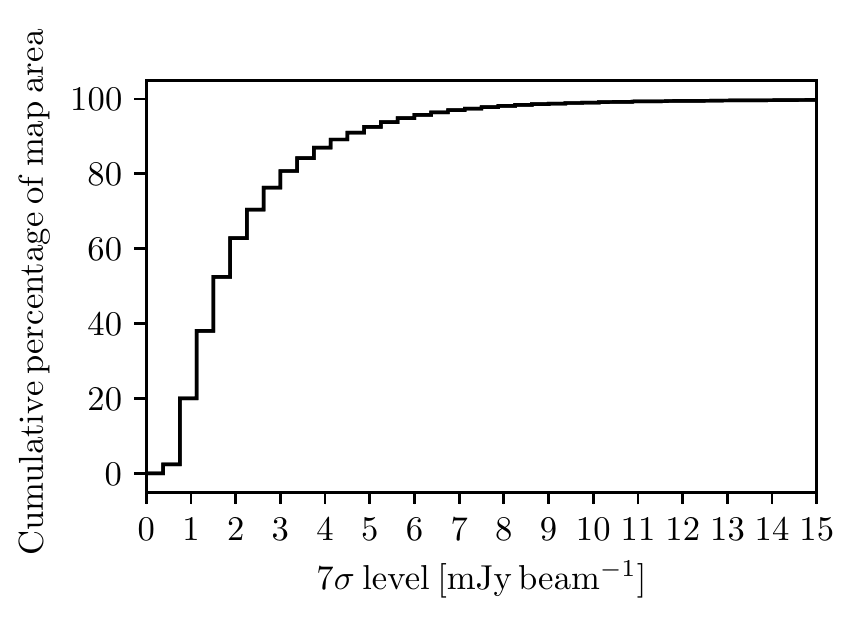}
      \caption{Cumulative fraction noise level diagram. The percentage of the map area as a function of the noise level at a S/N of 7$\sigma$ in mJy~beam$^{-1}$. More than $60\%$ of the survey area has a noise level of 7$\sigma \lesssim$2 mJy beam$^{-1}$.}
         \label{fig_noisestep}
 \end{figure}

While for the 1.4~GHz the combined THOR+VGPS dataset exists, for the other bands only the THIR data are available. Therefore, all spectral index analysis is done on the THOR-only data. To achieve the best signal-to-noise ratio, we chose the two SPWs, spw-1820 and spw-1440, that are least effected by RFI and have the lowest noise, smoothed them into the same resolution, and then averaged them for source extraction. Since the noise of our data is dominated by the sidelobe noise, we followed the method described in \citet{hales2012} and \citet{bihr2016} and constructed noise maps using the averaged residual image from the clean process. The noise maps are shown in Fig. \ref{fig_noise}. We furthermore calculated the cumulative fraction noise level map area at a specific 7$\sigma$ noise level in mJy~beam$^{-1}$ (Fig.~\ref{fig_noisestep}). While the lowest 7$\sigma$ noise level ($\leq 1$~mJy~beam$^{-1}$, dominated by thermal noise) is achieved in $\sim 20\%$ of the survey area, more than 60\% of the survey area has a 7$\sigma$ noise level $\leq2$~mJy~beam$^{-1}$. Comparing to the first half of the survey \citep{bihr2016}, the noise is better for the entire survey as in the large longitude Milky Way regions beyond Galactic longitudes of roughly 51$\degr$ there are fewer strong sources (Fig.~\ref{fig_wholemap} and Fig.~\ref{app_noise}). 

We used the software BLOBCAT \citep{hales2012} to extract the sources from the averaged continuum images. Following the same criteria as in \citet{bihr2016} for the first half of the survey, we set the detection threshold as 5$\sigma$ and the flooding threshold to the standard value of 2.6$\sigma$ \citep{hales2012}, and extracted 10916 sources. We then inspected each source visually, and identified 530 sources as obvious observational sidelobe artifacts and removed them from the catalog. The remaining 10387 sources should be mostly real detections. It was difficult to determine whether some sources with a signal-to-noise ratio between 5 and 7$\sigma$ were real. Therefore, we consider sources with a signal-to-noise ratio higher than 7$\sigma$ with higher confidence. In total, out of the 10916 extracted sources, 7521 sources were detected with a signal-to-noise ratio higher than 7$\sigma$, 2866 have a signal-to-noise ratio between 5 and 7$\sigma$, 530 are observational sidelobe artifacts. 

\begin{figure}
   \centering
  \includegraphics[width=8cm]{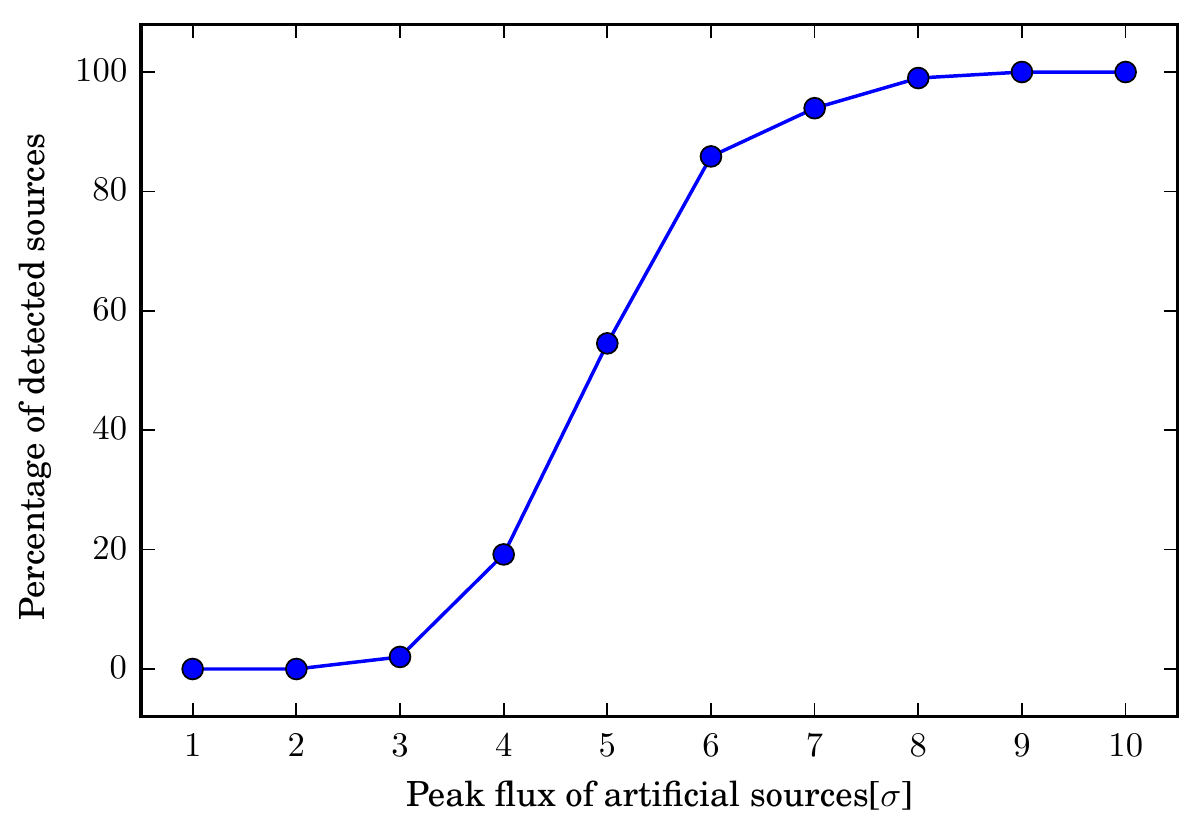}
      \caption{Completeness test plot. Percentage of the added artificial sources detected as a function of the source peak intensity in units of the noise level $\sigma$.}
         \label{fig_complete}
 \end{figure}

Following the same method described in \citet{bihr2016}, we performed a completeness test by extracting artificial compact sources from a $0.5^\circ \times 0.5^\circ$ region with a constant noise level. The result shows that we detected 94$\%$ of all sources with a peak intensity above 7$\sigma$ (Fig.~\ref{fig_complete}). Combining the noise map, we construct completeness maps for different peak intensities that are shown in Appendix~\ref{app_completeness}.

\begin{figure}
   \centering
  \includegraphics[width=8cm]{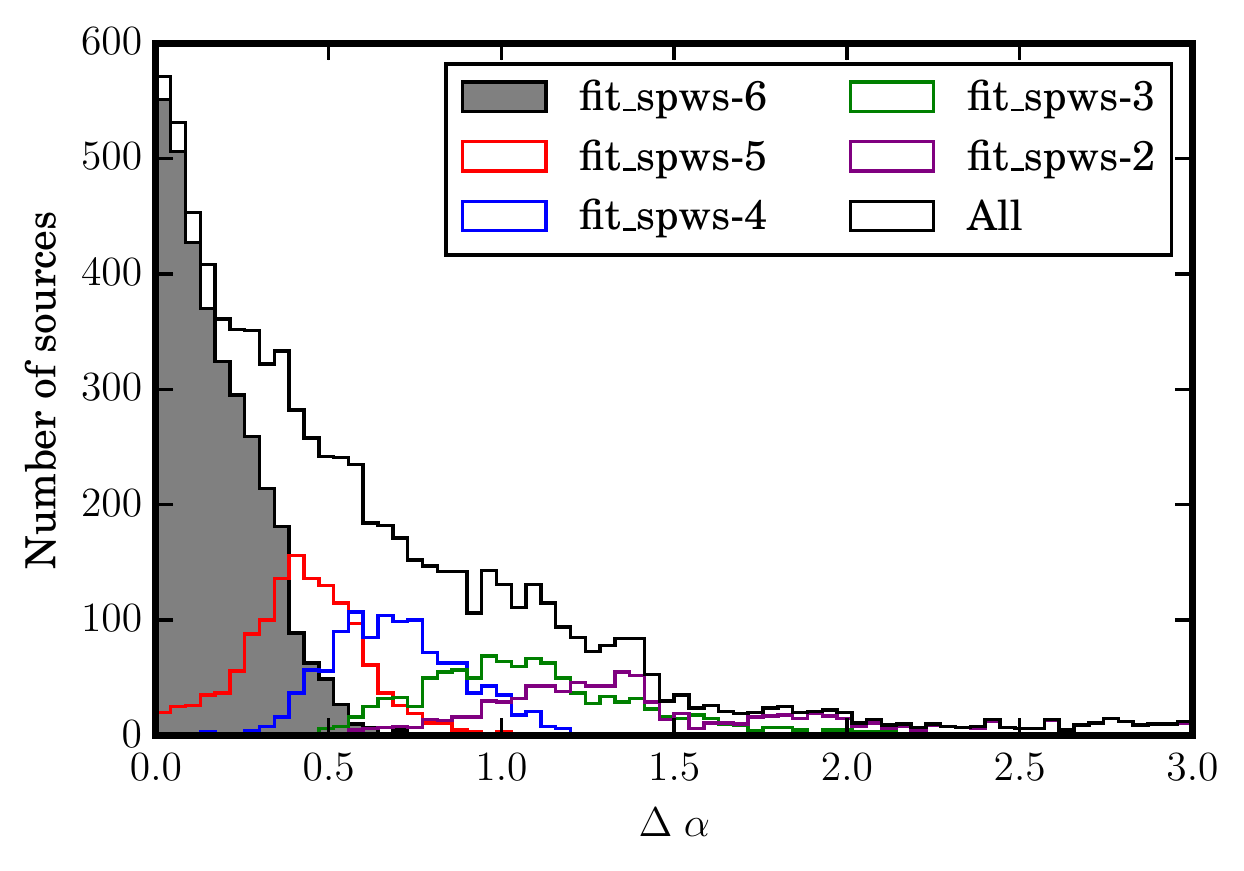}
      \caption{Histogram of the uncertainties of the determined spectral indices. The black line includes all sources for which we are able to determine a spectral index, the grey shaded area represents the sources for which we have an intensity measurement in all six SPWs and therefore the spectral index $\alpha$ is derived from fitting all six SPWs, the red line includes sources fitted with five SPWs, the blue line includes sources fitted with four SPWs, the green line includes sources fitted with three SPWs, and the purple line includes sources fitted with two SPWs.}
         \label{fig_deltaalpha}
 \end{figure}

Since our observations cover a wide range in frequency from 1 to 2~GHz, we can determine the spectral indices for the sources we identified, assuming $I(\nu) \propto \nu^\alpha$. To do this, we first smooth all SPWs to a common resolution of 25$^{\prime\prime}$, then extract the peak intensities. Since the integrated flux density suffers more from filtering effects for extended sources, we use only the peak intensity for the spectral index determination. We use the scipy function ``curve\_fit'' to fit the peak intensity between 1 to 2~GHz of each source we extracted and derive the spectral index $\alpha$ defined as $I(\nu) \propto \nu^\alpha$ together with the uncertainty of $\alpha$. Since a single SPW image has a lower S/N than the averaged image we used for source extraction, we choose a lower threshold (3$\sigma$) for spectral index determination. We fit only the peak intensities that are higher than 3 times of the noise level of the respective SPW, i.e., a reliable intensity \citep[see also][]{bihr2016}. For 8228 sources we can extract a reliable peak intensity from at least two SPWs, and derive a spectral index. We list the number of SPWs that were used for spectral index determination in the continuum source catalog as ``fit\_spws''. For instance, ``fit\_spws''$=$6 means the peak intensity for all 6 SPWs is reliable and used for spectral index determination, ``fit\_spws''$=$4 means the peak intensity for only 4 SPWs is reliable and used for spectral index determination. Figure~\ref{fig_deltaalpha} shows the distribution of the uncertainties of the determined spectral index for all sources and for sources with a different number of SPWs. This clearly demonstrates that   the more SPWs used to fit the spectral index, the smaller the uncertainty is. The mean uncertainty of the spectral indices are $\Delta\alpha = 0.18,\ 0.43,\ 0.70,\ 1.1,\ 2.2$ for sources detected in 6, 5, 4, 3 and 2 SPWs, respectively. We consider all spectral indices fitted with 4 or more SPWs (fit\_spws $\geq4$) reliable. 

\section{Catalog}
\label{sect_cata}
\begin{figure}
   \centering
  \includegraphics[height=0.8\columnwidth]{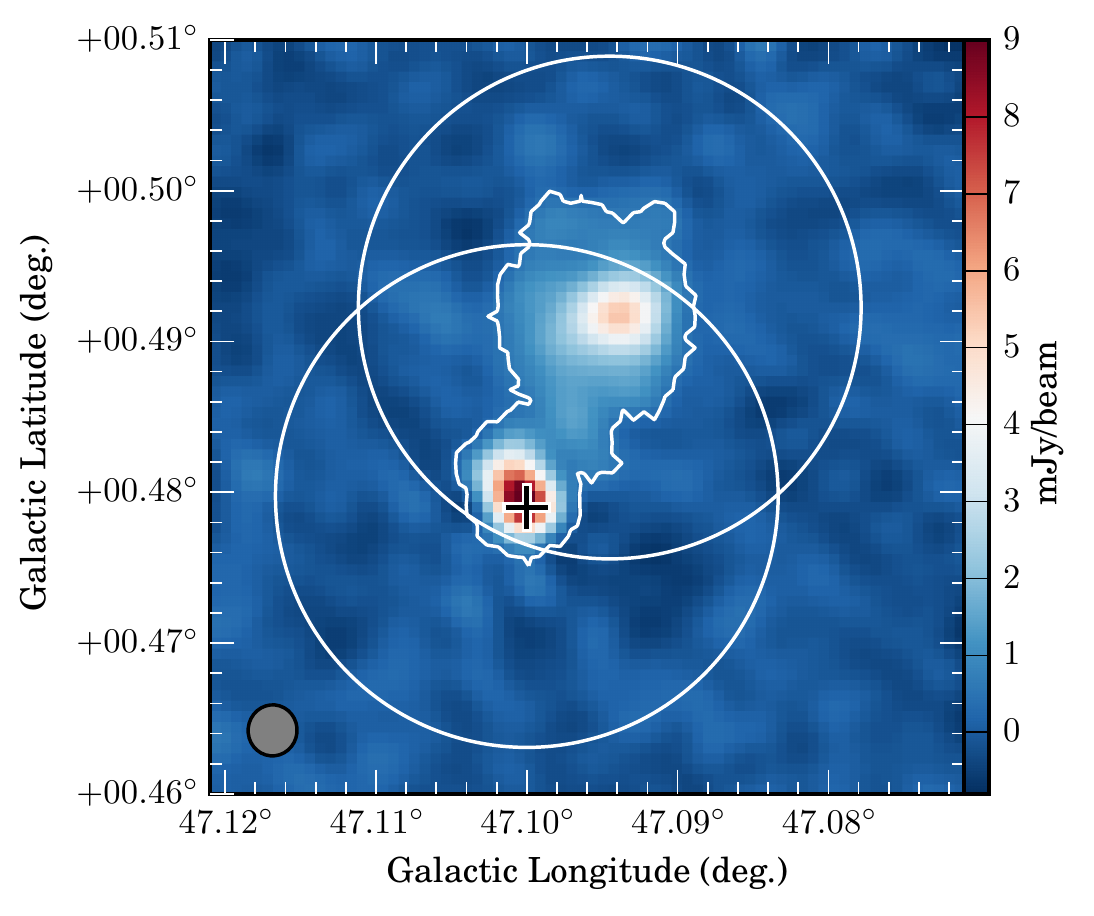}
    \includegraphics[height=0.8\columnwidth]{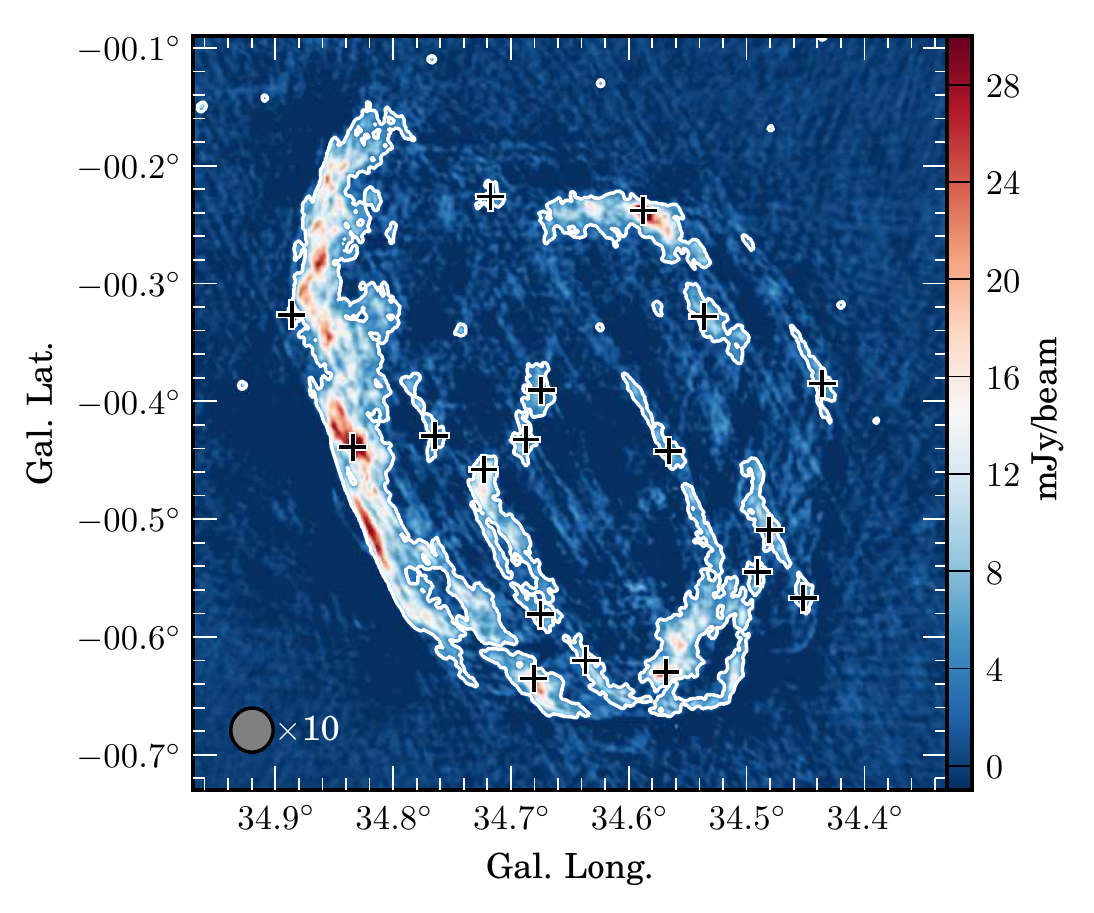}
      \caption{Examples of source identification in the catalog. In both panels, the white contours represent the area of the source extracted by BLOBCAT overlaid on the averaged image of spw-1440 and spw-1820, and the crosses mark the sources in the catalog. {\it Top:} Source G47.100+0.479 consists of two \ion{H}{ii} regions (marked with white circles, \citealt{anderson2014}) close together. {\it Bottom:} The SNR W44 \cite[e.g.,][]{green2014} is resolved into 19 sources in our catalog. The synthesized beam is shown in the bottom left corner of each panel. The beam in the right panel is scaled up to 10 times of its original size.}
         \label{fig_example}
 \end{figure}

Our continuum catalog contains 31 columns for each source as described in detail in Table~\ref{table_catalog_entries}. We summarize the number of extracted sources in Table~\ref{table_num}. Due to the BLOBCAT algorithm, compact sources close to each other or superimposed on large extended emission are identified as one single source. Additionally, large, extended sources, such as SNRs, can be separated into several different sources. As the two examples shown in Fig.~\ref{fig_example},  two \ion{H}{ii} regions are identified into one single object in our catalog and the SNR W44 is resolved into 18 separate objects. Therefore, the exact numbers in Table~\ref{table_num} should be treated cautiously. We cross-matched the THOR continuum catalog with catalogs of different type of Galactic sources (\ion{H}{ii} regions, SNRs, etc., see Sect.~\ref{sect_discuss}) and list all the matched counterparts. For a source matched with multiple sources, we also list the number of counterparts that the continuum source is associated with in column ``Ncounter'' in the catalog (Table~\ref{table_catalog_entries}). We also categorise sources in groups based on counterparts they share, i.e., sources associated with the same \ion{H}{ii} regions or SNRs are in the same group with a ``GroupID''. The size of the groups are also listed as ``GroupSize'' (Table~\ref{table_catalog_entries}). In total, 126 sources have more than 1 counterparts, 153 sources are categorised into 49 groups with GroupSize varies from 2 to 19. We will discuss the association between different Galactic and extragalactic sources in detail in Sect.~\ref{sect_discuss}.


\begin{figure}
   \centering
  \includegraphics[width=\hsize]{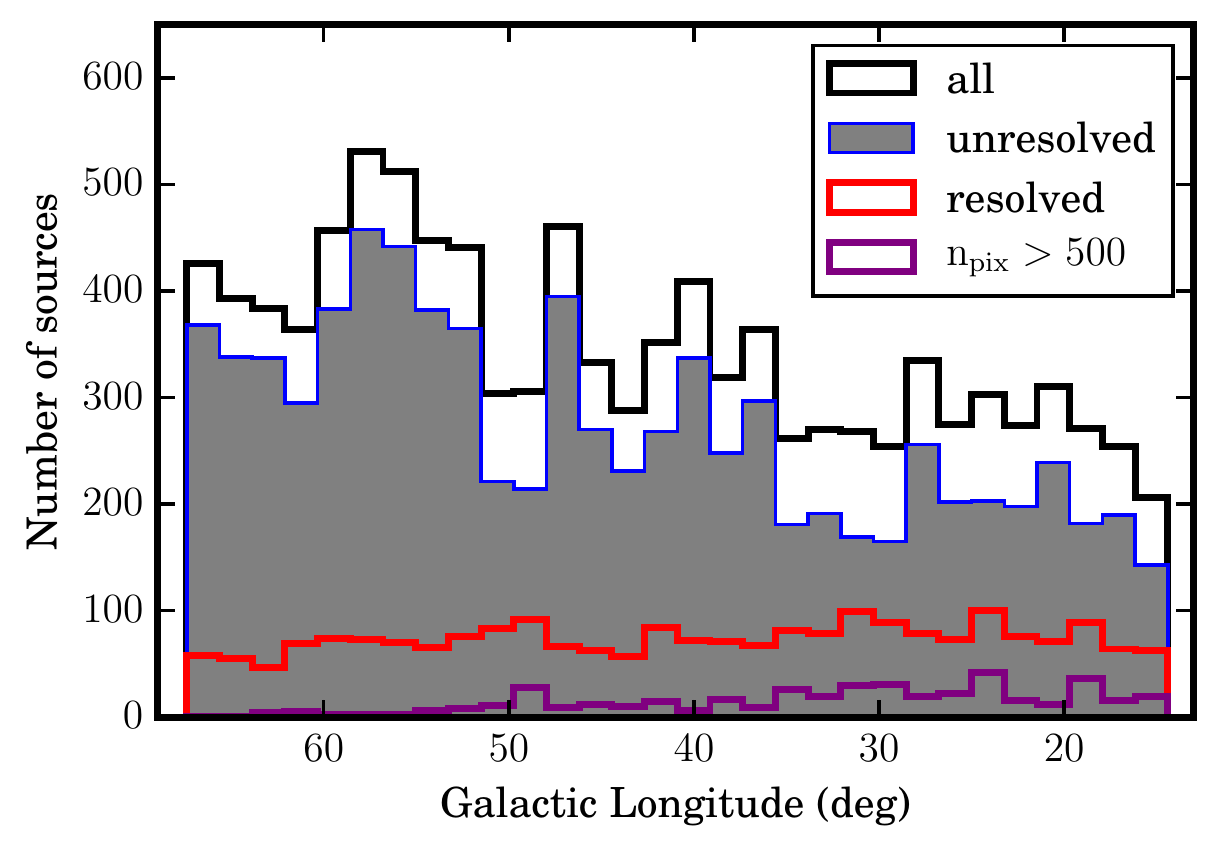}
   \includegraphics[width=\hsize]{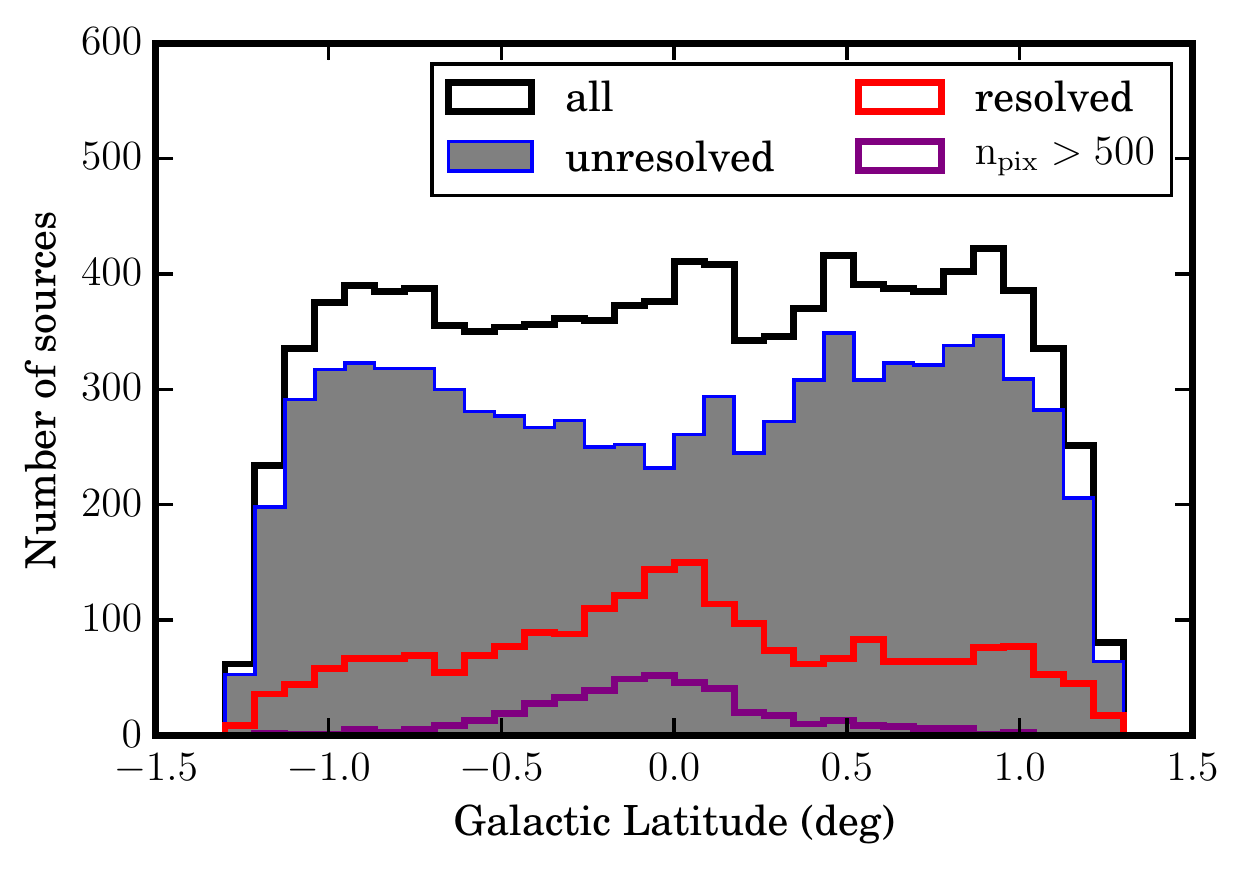}
      \caption{Histograms for the source distribution along Galactic Longitude (top) and latitude (bottom). In both panels, the black histogram shows the distribution of all the sources in the catalog, the other histograms show fractions of the catalog as marked in each panel.}
         \label{fig_distri}
 \end{figure}

The distribution of the extracted sources along Galactic Longitude and latitude is shown in Fig.~\ref{fig_distri}. The majority (79\%, Table~\ref{table_num}) of the extracted sources are unresolved, and a weak gradient for the those sources along Galactic longitude is shown in Fig.~\ref{fig_distri}, where more weak unresolved sources are identified in the larger longitude regions. The resolved sources are evenly distributed along Galactic longitude, but larger sources (n$_{\rm pix} > 500$, effective radius of $\sim32$\arcsec) are more concentrated in the inner Galaxy, since these are mostly Galactic \ion{H}{ii} regions and SNRs. Another possible reason is that there are more mergers of smaller sources in the inner longitudes because of higher source density and longer path-lengths through the Galaxy. The distribution along Galactic latitude reveals that the resolved sources are concentrated close to the Galactic mid-plane, whereas the distribution of the unresolved sources shows a dip at $|b|<0.5^\circ$. The distribution of both resolved and unresolved sources drops at $|b|>1.0^\circ$ due to higher noise at the edges of the survey area. 

We did a Kolmogorov–Smirnov test (KS test) for the dip in the distribution of the unresolved sources at $|b|<0.5^\circ$. We first re-binned the source counts into 100 bins and measured the mean value and the standard deviation of the distribution at $|b|>0.5^\circ$ and $|b|<1.0^\circ$. Then we generated a random artificial distribution with the same mean value and standard deviation. Finally we compute the KS statistic and P-value between the random artificial and the real distribution with the scipy function ``ks\_2samp''. The mean KS statistic and p-value for 100 run is 0.34$\pm0.041$ and 0.001$\pm0.0018$, respectively. So the dip at $|b|<0.5^\circ$ is statistically significant. A similar distribution along Galactic latitude is also found in the first half of our survey \citep{bihr2016} and in the MAGPIS survey \citep{helfand2006}. Less unresolved sources are identified at $|b|<0.5^\circ$ and in the inner Galaxy which could be due to that the strong emissions from the extended Galactic sources such as \ion{H}{ii} regions and SNRs lower our detection completeness to weak sources towards the Galactic mdi-plane and inner Galaxy (see Appendix~\ref{app_completeness}).

\begin{table*}
\caption{Description of the continuum source catalog entries.}             
\label{table_catalog_entries}      
\centering          
\begin{tabular}{c l l l  }
\hline\hline       
Col. Num. & Name & Unit  & Description\\
\hline
\small
1 & Gal. ID & & Name of the source the form G``Gal. longitude''$\pm$``Gal. latitude''\tablefootmark{a}.\\
2 & RA & deg & Right ascension in J2000 of the peak position.\\
3 & Dec & deg & Declination in J2000 of the peak position.\\
4 & S\_p \tablefootmark{b}  & Jy~beam$^{-1}$ & Peak intensity of the aver. image used for source extraction (see Sect. \ref{sect_extract}).\\
5 & S\_int &  Jy & Integrated flux density of the averaged image (see Sect. \ref{sect_extract}).\\
6 & S/N & & Signal-to-noise ratio in the averaged image.\\
7 & BMAJ & arcsec & Major axis of the synthesized beam used for source extraction.\\
8 & BMIN & arcsec & Minor axis of the synthesized beam used for source extraction.\\
9 & BPA & deg & Position angle of the synthesized beam used for source extraction.\\
10 & n\_pix & & Number of pixels flooded by BLOBCAT (see Sect. \ref{sect_extract}).\\ 
11 & resolved\_source & & Resolved source label (see Sec \ref{sect_extract}). 1 = Resolved, 0 = Point.\\
12 & S\_p(spw-1060) \tablefootmark{c}  &  Jy~beam$^{-1}$   & Peak intensity around 1.06~GHz used for spectral index (see Sect. \ref{sect_extract}).\\
13 & delta\_S\_p(spw-1060) \tablefootmark{c}  &  Jy~beam$^{-1}$  & Uncertainty of peak intensity around 1.06~GHz.\\ 
14 & S\_p(spw-1310)  \tablefootmark{c} &  Jy~beam$^{-1}$  & Peak intensity around 1.31~GHz used for spectral index (see Sect. \ref{sect_extract}).\\
15 & delta\_S\_p(spw-1310) \tablefootmark{c}  &  Jy~beam$^{-1}$  & Uncertainty of peak intensity around 1.31~GHz.\\
16 & S\_p(spw-1440) \tablefootmark{c}  &  Jy~beam$^{-1}$  & Peak intensity around 1.44~GHz used for spectral index (see Sect. \ref{sect_extract}).\\
17 & delta\_S\_p(spw-1440) \tablefootmark{c} &  Jy~beam$^{-1}$  & Uncertainty of peak intensity around 1.44~GHz.\\
18 & S\_p(spw-1690) \tablefootmark{c}  &  Jy~beam$^{-1}$  & Peak intensity around 1.69~GHz used for spectral index (see Sect. \ref{sect_extract}).\\
19 & delta\_S\_p(spw-1690) \tablefootmark{c}  &  Jy~beam$^{-1}$  & Uncertainty of peak intensity around 1.69~GHz.\\
20 & S\_p(spw-1820) \tablefootmark{c}  &  Jy~beam$^{-1}$  & Peak intensity around 1.82~GHz used for spectral index (see Sect. \ref{sect_extract}).\\
21 & delta\_S\_p(spw-1820) \tablefootmark{c}  &  Jy~beam$^{-1}$  & Uncertainty of peak intensity around 1.82~GHz.\\
22 & S\_p(spw-1950) \tablefootmark{c}  &  Jy~beam$^{-1}$  & Peak intensity around 1.95~GHz used for spectral index (see Sect. \ref{sect_extract}).\\
23 & delta\_S\_p(spw-1950) \tablefootmark{c}  &  Jy~beam$^{-1}$  & Uncertainty of peak intensity around 1.95~GHz.\\
24 & alpha & & Spectral index of source we derived (see Sect.~\ref{sect_extract}).\\
25 & delta\_alpha & & Uncertainty of spectral index.\\
26 & fit\_spws & & Number of SPWs used to fit the spectral index (see Sect. \ref{sect_extract}).\\
27 & Note\tablefootmark{d}  &  & ``HII'', ``SNR\_green'', ``SNR\_anderson'', ``PN'', ``PSR'', ``Xray'', ``jets''(see Sect.~\ref{sect_discuss}).\\
28 & Counterparts & &The counterparts of the \ion{H}{ii} region\tablefootmark{e}, SNR\tablefootmark{f}, the planetary nebula\tablefootmark{g} and the pulsars\tablefootmark{h}. \\
29 & Ncounter &  & The total number of counterparts that the continuum source is associated with. \\
30 & GroupID &   & The Group of continuum sources associated with one or more same counterparts.\\ 
31 & GroupSize &  & The number of the continuum sources in the same group.\\
\hline        
\end{tabular}
\tablefoot{
\tablefoottext{a}{Indicating the peak position.}
\tablefoottext{b}{The synthesized beam is different for different fields and is given in rows 7-9.}
\tablefoottext{c}{The synthesized beam is smoothed to 25\arcsec$\times$25\arcsec.}
\tablefoottext{d}{We classified the continuum source into different categories. ``HII'': sources associated with \ion{H}{ii} regions from \citet{anderson2014}; ``SNR\_green'': sources associated with SNRs from \citet{green2014}; ``SNR\_anderson'': sources associated with SNR candidates from \citet{anderson2017};  ``PN:'' sources classified as planetary nebula; ``PSR'': sources classified as pulsars; ``Xray'': sources associated with X-ray sources;``jets'': sources classified as extragalactic jets candidates.}
\tablefoottext{e}{\citet{anderson2014}.}
\tablefoottext{f}{\citet{green2014, anderson2017}.}
\tablefoottext{g}{\citet{parker2016}.}
\tablefoottext{h}{\citet{manchester2005}.}

}          
\end{table*}

\begin{table}
\caption{Numbers of the catalog.}             
\label{table_num}      
\centering                          
\begin{tabular}{l c c}        
\hline\hline                 
Description  & Numbers & Percentage \\    
\hline                        
   All 				& 10387 &  100\% \\      
   S/N$>7\sigma$   & 7521 &   72\%\\
   Resolved sources  & 2210 &  21\% \\
   Unresolved sources & 8177 & 79\%\\
   n$_{\rm pix}>500$ &   439 & 4\%\\
   fit\_spws$\geq4$     & 5857 &  56\%\\
   \hline                                   
\end{tabular}
\end{table}

\section{Discussion}
\label{sect_discuss}
\subsection{Comparison with other surveys}
\label{sect_comp}
The comparison between the peak positions of the common sources in the THOR survey (first half, $l=14.0-37.9^\circ$ and $l=47.1-51.2^\circ$), the MAGPIS \citep{helfand2006}, and the CORNISH survey \citep{hoare2012, purcell2013} presented in \citet{bihr2016} shows that the THOR survey has positional accuracy better than 2.5\arcsec\ (FWHM of the Gaussian fit to the position offset distribution, see Fig.~15 in \citealt{bihr2016}). Considering that the pixel size of our continuum images is only 2.5\arcsec, the spatial accuracy of the THOR survey is consistent with previous observations. 

\begin{figure}
   \centering
  \includegraphics[width=\hsize]{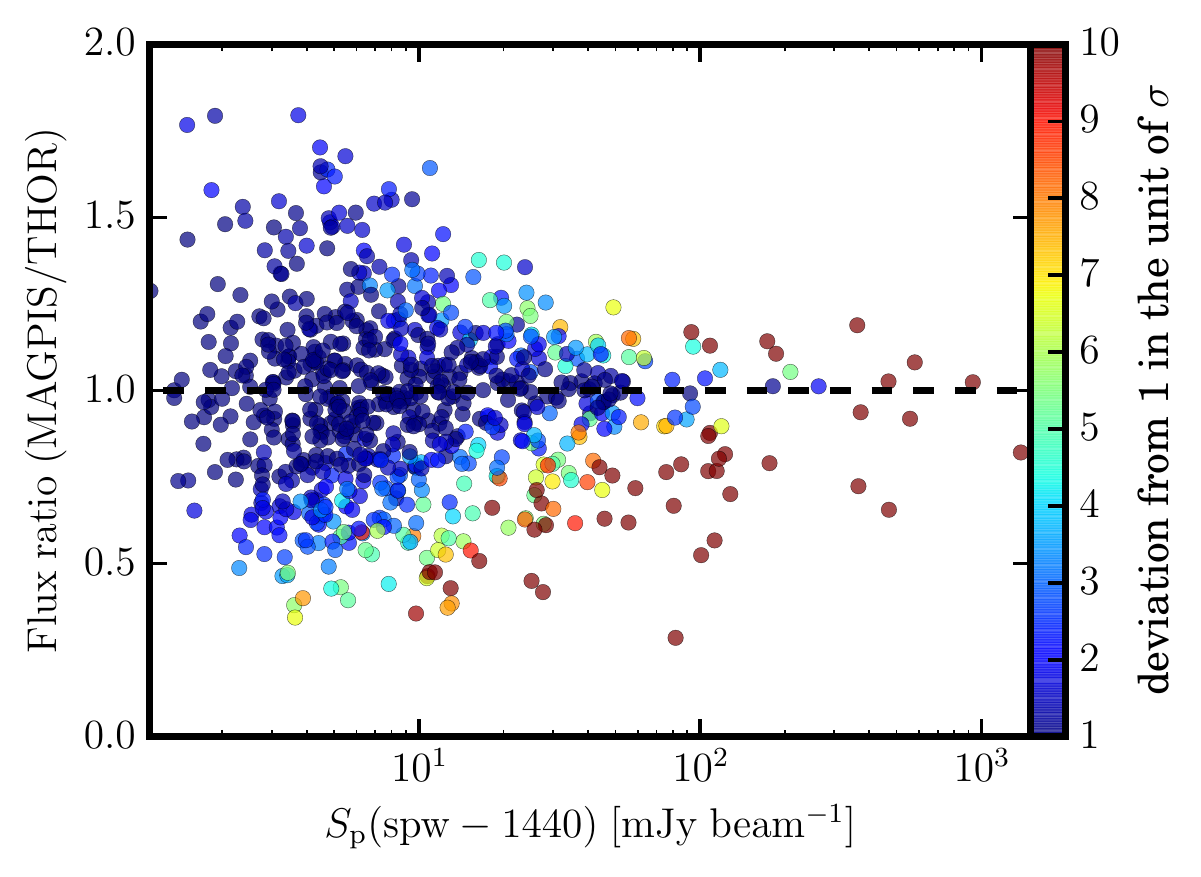}
      \caption{Ratio of the MAGPIS and THOR flux density plotted against the THOR flux density. The points are colored according to their deviation from a flux ratio of unity in units of the uncertainty in the ratio, so the deviation$=(1-\rm{ratio})/\sigma_{\rm ratio}$. The dashed line represents a one-to-one relation.}
         \label{fig_fluxcomp}
 \end{figure}

To check for consistency in the flux density, we compared the THOR with MAGPIS. Within our survey area ($14.2\degr<l<67.4\degr$, $b<|1.25|\degr$), the MAGPIS catalog contains 2256 discrete sources. We cross-matched our catalog with the MAGPIS catalog with a matching radius of 5\arcsec, and we found 1440 matches. We then selected sources with the following criteria: 1) S/N$>$ 3 in spw-1440; 2) unresolved in the THOR catalog; 3) major axis smaller than 10\arcsec in the MAGPIS catalog. We selected 735 sources using the aforementioned criteria. We compared the 1.4~GHz continuum flux density from the spw-1440 in THOR and the flux densities in MAGPIS, as show in Fig.~\ref{fig_fluxcomp}. The MAGPIS/THOR flux ratio shows a tight distribution around 1, with a median value of $\sim$0.97.  Most of the sources are within 5$\sigma$ from the flux ratio equals to one. 67 sources show a deviation larger than 7$\sigma$ from one, among which 56 sources have a ratio smaller than one. If we compare the integrated flux from MAGPIS to the peak flux from THOR for these 56 sources, the mean ratio is 1.08. This indicates that these sources could have been resolved by MAGPIS. For the remaining 11 sources the ratios are smaller than 1.18, therefore we can not rule out the possibility of variability for these sources. Among these 11 sources, 2 are matched with X-ray sources (ratios are 1.02 and 1.08, see also Sect.\ref{sect_xray}). The remaining 9 sources do not match with any identified source and all have negative spectral index, which suggests that they can be extragalactic sources.

With a matching radius of 5\arcsec, we found 1320 unresolved THOR sources matched with NRAO VLA Sky Survey (NVSS, \citealt{condon1998}) sources. We compared the 1.4~GHz flux densities from THOR and NVSS, and the THOR flux at spw-1440 is also tightly correlated with the one from NVSS but slightly higher. The median value of the NVSS/THOR flux ratio is about 0.93. By comparing the THOR 1.4 and 1.8~GHz average flux density of the first half of the survey with the NVSS flux density, \citet{bihr2016} found similar results (see also Figure~17 in \citealt{bihr2016}). This discrepancy is yet undetermined, but this is beyond the scope of this work. The NVSS images show that the NVSS catalog is severely contaminated with obvious false detections, which could be due to sidelobes from strong sources close to the Galactic plane or ghost artifacts \citep{grobler2014}. 

For the second half of the THOR survey, we applied the same data reduction method and source extraction algorithm as the first half \citep{bihr2016}. Since there are overlapping areas between the first and second half of the survey, we further compare the position and flux density of the common sources between the two halves. 94$\%$ of sources have position differences $<$2.5\arcsec (the pixel size of the maps), and 93$\%$ of the sources have flux differences $<$ 1~mJy~beam$^{-1}$. Therefore, the second half of the survey is consistent with the first half and with the previous radio surveys. 

\begin{figure}
   \centering
  \includegraphics[width=\hsize]{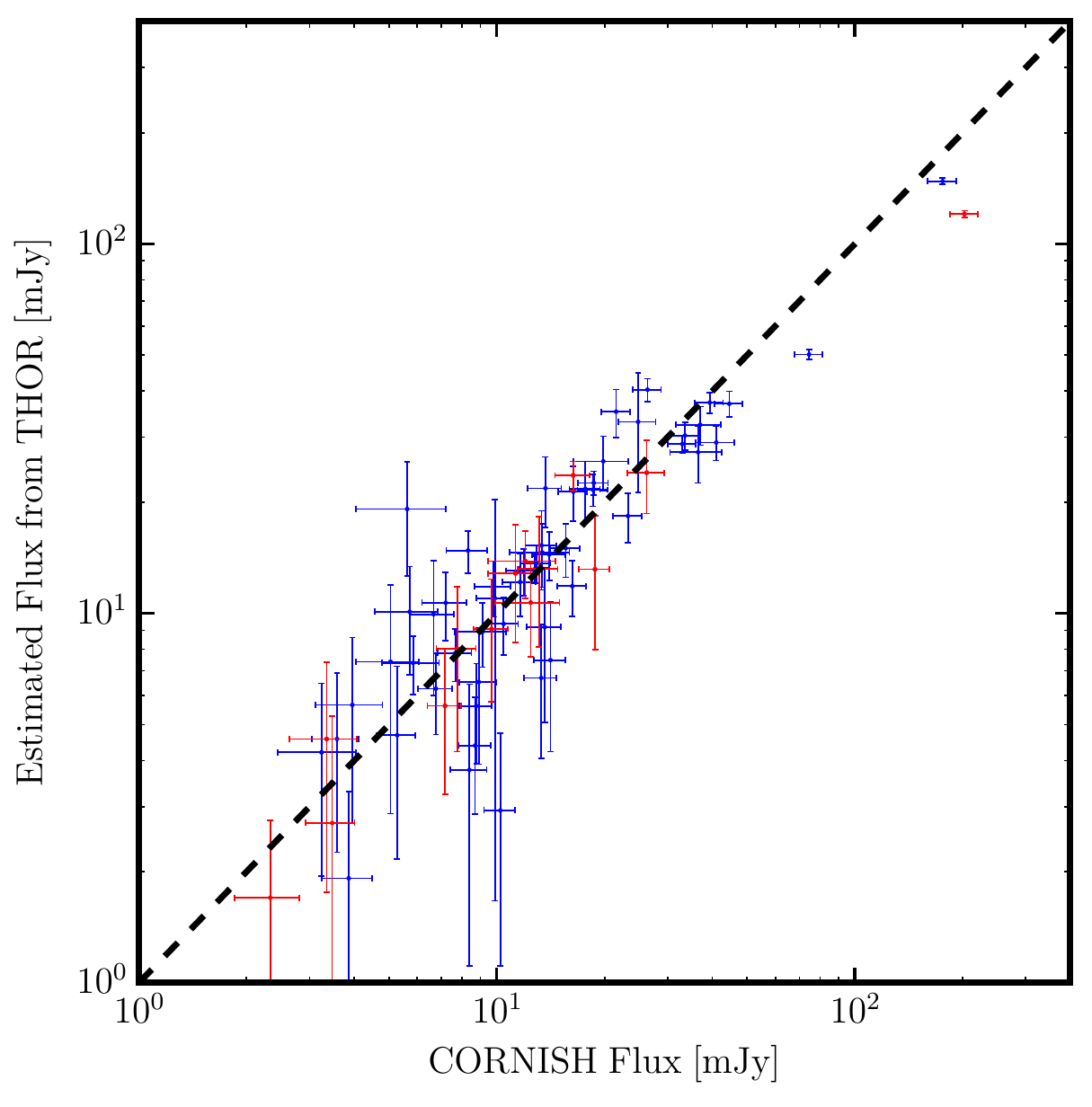}
      \caption{Extrapolated flux densities at 5~GHz for selected sources according to their spectral indices from the THOR catalog plotted against the flux densities in the CORNISH catalog. Sources with a spectral index between 0 and --0.2 are shown in blue, and with a spectral index --0.09 and --0.11 are shown in red. The dashed line represents a one-to-one relation.} 
         \label{fig_extrflux}
 \end{figure}

To verify the reliability of our spectral index determination, we extrapolated the flux density of selected sources to 5~GHz and compared this with the CORNISH catalog. The CORNISH catalog contains 2493 sources within the THOR survey area. With a matching radius of 5\arcsec, we found 1905 matches between THOR and CORNISH. We selected unresolved sources with a THOR spectral index between  0 and --0.2 (optically thin free-free emission has a spectral index --0.1) that are detected in all six spectral windows in THOR (fit\_spws=6). 68 sources were selected this way. We estimated the flux density at 5~GHz of these 68 sources according to their spectral indices from the THOR catalog. The extrapolated flux densities are very close to the flux densities in the CORNISH catalog as shown in Fig.~\ref{fig_extrflux}. Sources with a spectral index between --0.09 and --0.11 (red in Fig.~\ref{fig_extrflux}) are closer to ideal optically thin, and the extrapolated flux for these sources also agree better with the ones from CORNISH as expected.

Furthermore, we also compared the THOR spectral indices with spectral indices \citet{kalcheva2018} derived for the CORNISH UC\ion{H}{ii} regions. \citet{kalcheva2018} used flux density measurements of CORNISH 5~GHz and MAGPIS 20~cm to derive the spectral indices. 11 unresolved THOR sources are associated with the CORNISH UC\ion{H}{ii} regions and have a reliable spectral index measurement in \citet{kalcheva2018}. The comparison again reveals that the spectral indices from these two studies are tightly correlated. 

\subsection{Spectral index}
\label{sect_index}
\begin{figure}
   \centering
  \includegraphics[width=\hsize]{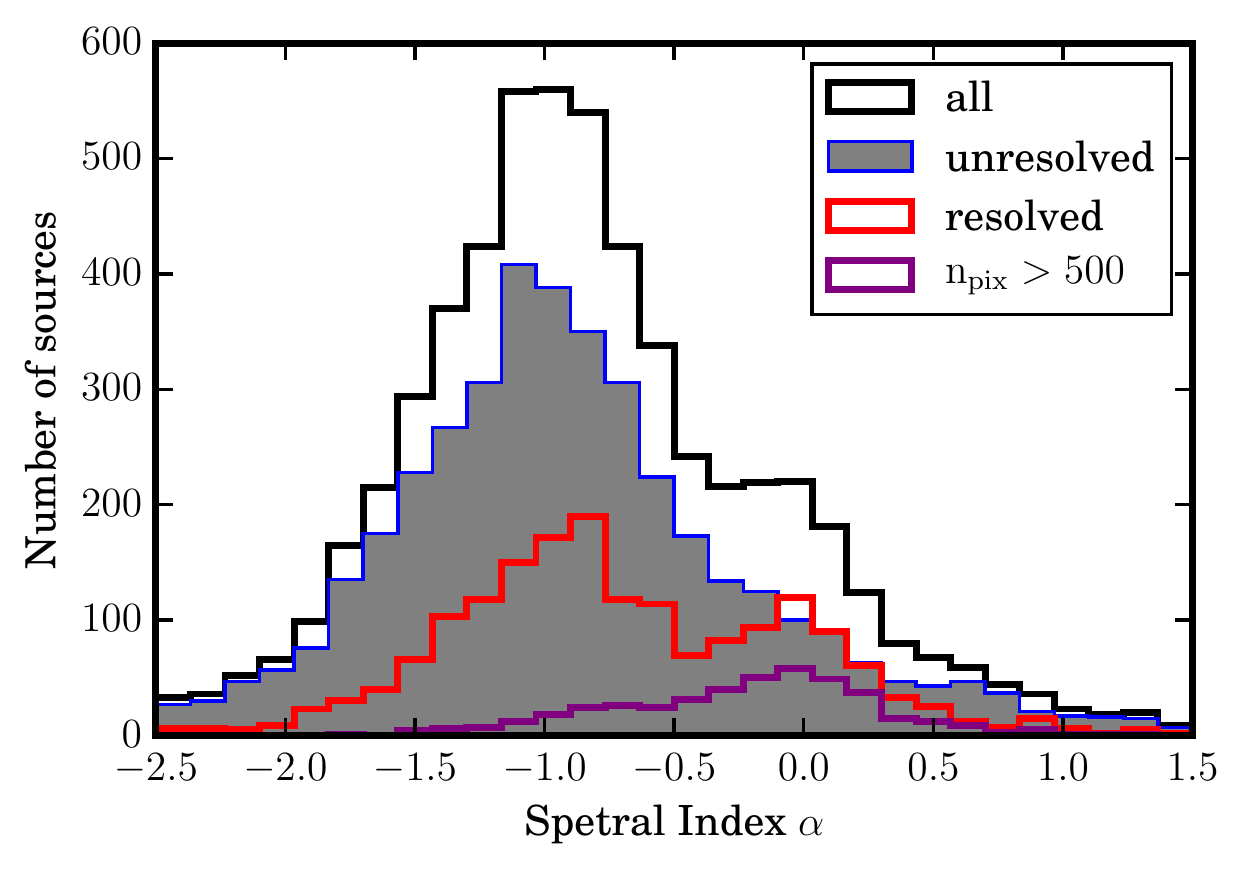}
      \caption{Histogram of the spectral index for all reliably detected sources (fit\_spws$\geq4$) with a reliable spectral index measurement ($\sim 5857$ sources).}
         \label{fig_alpha}
 \end{figure}

As described in Sect.~\ref{sect_extract}, we derived spectral indices for 8228 sources, of which we consider the values for 5857 sources as being reliable (fit\_spws$\geq4$). Using the spectral index information we can distinguish the physical origin of the emission and classify the continuum sources. Figure~\ref{fig_alpha} shows the distribution of the spectral indices for all detected sources with a reliable spectral index measurement. For all sources the spectral index distribution shows a strong peak around $\alpha\sim-1$, and a secondary peak around $\alpha\sim0$. If we consider only the unresolved sources, the spectral index distribution peaks around $\alpha\sim-1$, and the secondary peak around $\alpha\sim0$ diminishes. Most of the unresolved sources show a negative spectral index indicating that they are dominated by non-thermal synchrotron radiation. Considering the unresolved sources are also evenly distributed along Galactic longitude and latitude (Sect.~\ref{sect_cata}), these unresolved sources with a negative spectral index are likely to be mostly extragalactic sources. The spectral distribution of the resolved sources shows two clear peaks around $\alpha\sim-1$ and $\alpha\sim0$. Most of the resolved sources with a flat or positive spectral index are Galactic \ion{H}{ii} regions (see Sect.~\ref{sect_hii}). The ones with negative spectral index are mostly SNRs (see Sect.~\ref{sect_snr}) and radio galaxies (see Sect.~\ref{sect_jet}), or overlapping unresolved sources that were classified as one single source. Considering that only the very large sources that have an area in pixels larger than 500 (${\rm n_{pix}}>500$, effective radius $\gtrsim32$\arcsec), the spectral index distribution peaks around $\alpha\sim0$ and extends to $\alpha \sim -1$. Among these large sources, about 80\% of them are associated with \ion{H}{ii} regions, and the rest of them are classified as SNRs and radio galaxies (see Sect.~\ref{sect_hii}, \ref{sect_snr}, and \ref{sect_jet}). 
  
\subsection{{\it \ion{H}{ii}} regions}
\label{sect_hii}
Using data from the all-sky {\it Wide-Field infrared Survey Explorer (WISE)} satellite, \citet{anderson2014} made the most complete catalog of \ion{H}{ii} regions to date, with a total of more than 8000 sources\footnote{\url{http://astro.phys.wvu.edu/wise/}}. Within the region of the THOR survey, the {\it WISE} \ion{H}{ii} catalog contains $\sim$2400 sources, and $\sim$1500 of them show radio continuum emission. The size of the \ion{H}{ii} regions varies from 10\arcsec\ to 20\arcmin\ from the mid-infrared (MIR) images, so multiple Galactic and extragalactic sources could easily be enclosed within one single large \ion{H}{ii} region. \citet{bihr2016} shows that matching the THOR continuum sources with the {\it WISE} catalog sources with $r<150$\arcsec\ in an automated fashion would produce less than $10\%$ false matches. We take the same radius threshold when doing the automated matching, then we visually inspected the results and removed the false detections. We further visually compared the remaining large {\it WISE} sources ($r>150$\arcsec) with our continuum images, and identified the continuum sources that are associated with \ion{H}{ii} regions. 

In addition to the {\it WISE} catalog, 239 UC\ion{H}{ii} regions are identified in the CORNISH survey \citep{kalcheva2018}, and 205 lie within the THOR coverage. With a matching radius of the effective radius  equal to each THOR continuum source, 202 UC\ion{H}{ii} regions have counter parts in the THOR survey. For sensitivity reason, we do not detect three CORNISH UC\ion{H}{ii} regions \citep[G024.1839+00.1199, G026.1094-00.0937, and G030.0096-00.2734, see][]{kalcheva2018}. All the continuum sources that are associated with UC\ion{H}{ii} regions are also associated with at least one \ion{H}{ii} region in the {\it WISE} catalog except G26.008+0.137 and G37.735-0.113, so we list the CORNISH UC\ion{H}{ii} region counterparts for only these two continuum sources.

In total, we matched 713 continuum sources with \ion{H}{ii} regions. Among the matched \ion{H}{ii} regions, 16 are in the radio quiet group in the {\it WISE} catalog, which means no radio continuum emission is detected in the MAGPIS \citep{helfand2006} and VGPS \citep{stil2006} surveys. \ion{H}{ii} regions close to each other could be identified as one single source in our catalog (left-panel, Fig.~\ref{fig_example}), in total 231 continuum sources are matched with more than one \ion{H}{ii} region. In particular, G43.171+0.007 is matched with 19 \ion{H}{ii} regions in W49, G19.610-0.235 is matched with 7 \ion{H}{ii} regions, G45.122+0.132 encompasses 6 \ion{H}{ii} regions, G48.610+0.027 encompasses 9 \ion{H}{ii} regions, G49.370-0.302 encompasses 8 \ion{H}{ii} regions (W51), and G49.488-0.380 encompasses 11 \ion{H}{ii} regions (W51). The radius of the matched \ion{H}{ii} regions measured at MIR in the {\it WISE} catalog varies from 12\arcsec\ to 20\arcmin\ with a median value of $\sim60$\arcsec.

\begin{figure}
   \centering
  \includegraphics[width=\hsize]{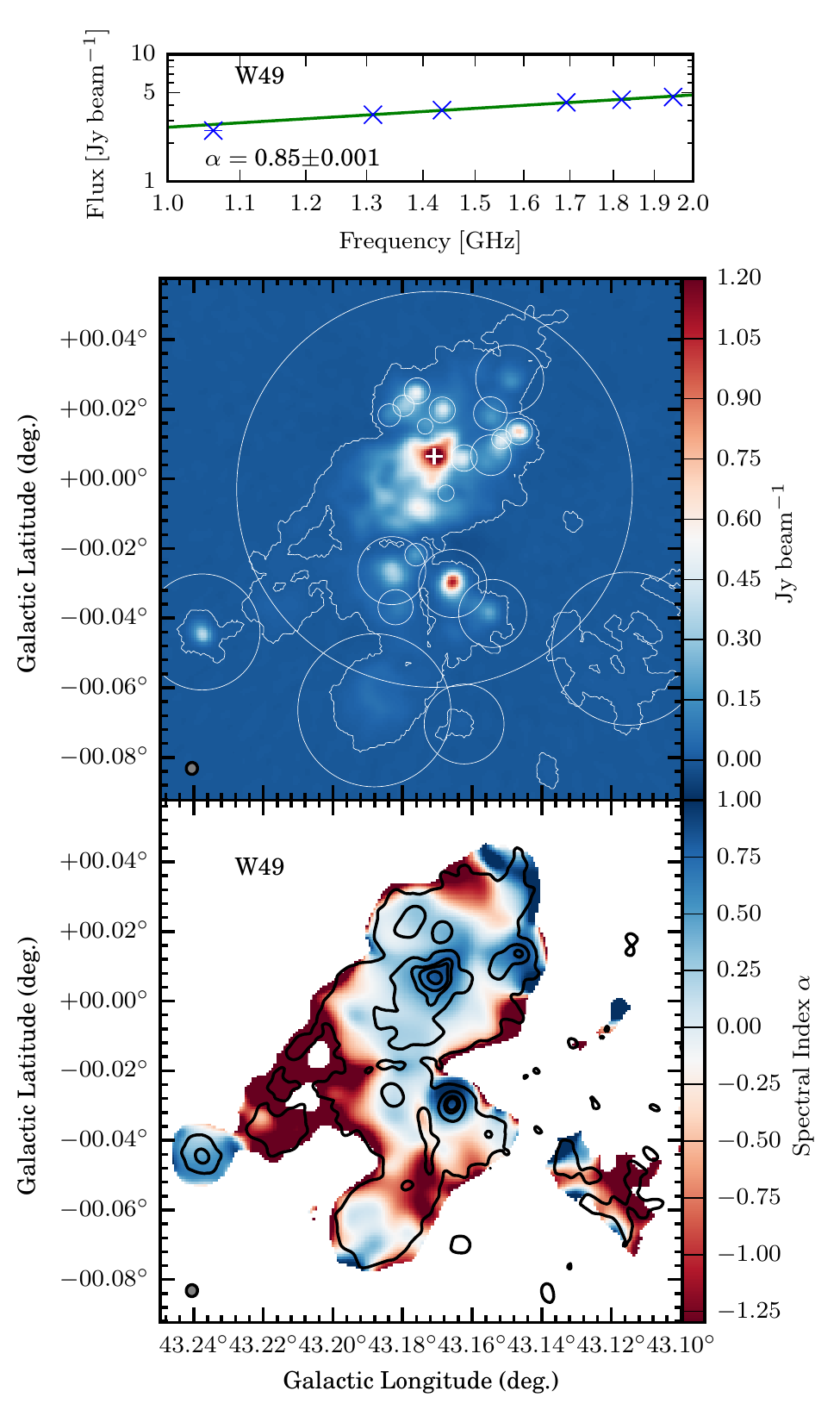}
      \caption{Spectral index map for the \ion{H}{ii} region W49. {\it Top-panel:} Spectral index fitting result of the peak of the continuum source G43.171+0.007 (marked with a cross in the {\it middle-panel}). {\it Middle-panel:} Averaged image of spw-1440 and spw-1820, the white contours represent the area of the source extracted by BLOBCAT, and the cross marks the peak of the source G43.171+0.007 in the catalog. The circles mark the \ion{H}{ii} regions from \citet{anderson2014} matched with the continuum sources in the THOR catalog. {\it Bottom-panel:} The spectral index map produced by fitting the flux of the six SPWs pixel by pixel. The synthesized beams of the averaged image and the spectral index map are shown in the bottom-left corner of each panel.}
         \label{fig_w49}
\end{figure}

We show in Fig.~\ref{fig_w49} one of the extreme cases, G43.171+0.007, which is associated with 19 \ion{H}{ii} regions in W49. By comparing ATLASGAL and CORNISH surveys, \citet{urquhart2013} identified 18 UC\ion{H}{ii} regions associated with this region, and that this region has the highest UC\ion{H}{ii} surface density in the 1st quadrant. The spectral index map of W49 (Fig.~\ref{fig_w49}) reveals a positive or flat spectral index towards most of the area of the source, except the edge regions, where the spectral index fitting is not reliable due to low S/N. This indicates that the \ion{H}{ii} region is dominated by thermal free-free emission. The spectral indices towards some strong continuum peaks are even larger than 0.5, this could indicate optically thick free-free emission.  


To derive the physical properties of the \ion{H}{ii} regions, we fitted the spectral energy distribution (SED) of the \ion{H}{ii} region sources with a simple homogeneous \ion{H}{ii} region model with optical depth to get the emission measure (EM) of the \ion{H}{ii} regions \citep{alvarothesis}. We selected continuum sources with a S/N larger than 7$\sigma$, fit\_spws $=6$ (see Sect.~\ref{sect_index}), and spectral index $\alpha>-0.3$. We further limited our fitting sample to sources with an effective radius smaller than 60\arcsec, since tests done by \citet{bihr2016} show that we are able to recover sources with size up to $\sim120$\arcsec\ reasonably well (80\% flux recovery). Among 713 continuum sources that are matched with \ion{H}{ii} regions, 262 match our selection criteria and are fitted. We fit the peak flux of the sources assuming that the line of sight size of the sources to be the same as the beam size (25\arcsec) to get the EM and the electron density ($n_e$) if the distance information is available in the {\it WISE} catalog \citep{anderson2014}. For 168 sources we got a good fit to the model and list the fitting results in Table~\ref{table_hiifits}. The main uncertainties of the fitting results are from the uncertainties of the peak flux brought in by the flux calibrator ($\sim5\%$ at these wavelengths), uncertainties from the fitting procedure ($<10\%$), and the uncertainties of the sizes of the sources on the line of sight. Two sources have $T_e$ measurements from previous RRLs observations \citep{balser2011}, and we assume a $T_e$ of 8000~K for the remaining sources (the mean $T_e$ estimated by \citealt{balser2011} is $\sim$8000~K for \ion{H}{ii} regions in the THOR survey area). For 151 sources we fitted an $10^4<{\rm EM}<10^5$~cm$^{-6}$~pc. For 124 sources there is distance information in the {\it WISE} catalog so we estimated $n_e$, which is between 0.9$\times10^2$ and 5.4$\times10^2$~cm$^{-3}$. Continuum source G49.477--0.328, which is associated with the \ion{H}{ii} region G049.490--00.381 in W51, has the highest value for EM and $n_e$ from our fitting, which are 4.2$\times10^5$~cm$^{-6}$~pc and 3.1$\times10^4$~cm$^{-3}$, respectively. 

Among the 168 sources where we can get a good \ion{H}{ii} region model fit, 39 are also associated with CORNISH UC\ion{H}{ii} regions. Assuming homogeneous optically thin and the line of sight size is the same as the angular size, \citet{kalcheva2018} calculated EM and $n_e$ for the CORNISH UC\ion{H}{ii} regions. Since CORNISH is at 5~GHz and the angular sizes of the 39 UC\ion{H}{ii} regions they derived are all smaller than 12\arcsec, which is about half beam size of THOR,  we can not compare the EM and $n_e$ from the the two samples directly. However, if we assume all 39 UC\ion{H}{ii} regions have a size of 25\arcsec, and convert the EM and $n_e$ from the CORNISH observations accordingly, they are close to with the values we derived with THOR observations although with large scatter. We can fit the EM$_{\rm CORNISH}$ vs EM$_{\rm THOR}$ and $n_{e\rm CORNISH}$ vs $n_{e\rm THOR}$ plots with a linear function $y=x*0.8+c$. This is a very rough comparison, but shows our \ion{H}{ii} fitting procedure is correct.

\begin{table*}
\caption{Fitting results for \ion{H}{ii} regions}             
\label{table_hiifits}      
\centering                          
\begin{tabular}{l c c c c c c c}        
\hline\hline                 
Gal.ID  & $\alpha$& $\Delta \alpha$&\ion{H}{ii} name  & $T_e$\tablefootmark{a} &d   \tablefootmark{b} & EM&$n_e$\tablefootmark{b}\\    
	&                     &         &     &[K]         & [kpc]& [$\times 10^5$~cm $^{-6} pc]$ & [$\times 10^2$cm$^{-3}$]\\
\hline                        
  G52.753+0.334 & 0.13 &0.01& G052.750+00.334 & 8970 & 9.575 & 2.0 & 5.0\\
  G53.187+0.209 & --0.14 &0.02& G053.188+00.209& --   & 9.96 & 0.7 & 2.9\\
  G56.160+0.077 & --0.21 & 0.04& G056.153+00.076 &--    & 9.9 & 0.2 & 1.7\\
  G56.420+0.423 & --0.12& 0.06&G056.412+00.423 & --   & 9.93 & 0.2 & 1.6\\
  G57.547--0.272 & --0.05 &0.01 &G057.541-00.279 &  --  & 8.805 & 1.3 & 4.2\\
  G58.773+0.646 & --0.10 & 0.04&G058.769+00.648 &   --  & 4.405 & 0.2 & 2.5\\
  G60.883--0.130 & 0.01 & 0.02&G060.881-00.135 & 7463 &--   & 1.4 &--   \\
   \hline       
\end{tabular}
\tablefoot{
\tablefoottext{a}{$T_e$ is assumed to be 8000K if it is not available from the {\it WISE} catalog.}
\tablefoottext{b}{We calculated $n_e$ if distance is available from the {\it WISE} catalog.}
The full table is available at the project website \url{http://www2.mpia-hd.mpg.de/thor/DATA/www/}.
}

\end{table*}

\subsection{Supernova remnants}
\label{sect_snr}
\citet{green2014} provides the most complete catalog of Galactic SNRs with 294 sources, 67 of them lie, or partially lie, in our survey area. By visually comparing the catalog with our continuum images, we identify 92 continuum sources associated with 36 SNRs. As shown in Fig.~\ref{fig_example}, many of the SNRs are resolved into multiple sources by the source-finding algorithm. The rest of the SNRs from \citet{green2014} either have too weak radio continuum emission, which is below our sensitivity, or are too diffuse and are filtered out by the VLA C-configuration \citep[see also][]{bihr2016}. We marked all the matched continuum sources with ``SNR\_green'' in the catalog and list the corresponding ID of the SNRs from \citet{green2014}. 

\begin{figure}
   \centering
  \includegraphics[width=\hsize]{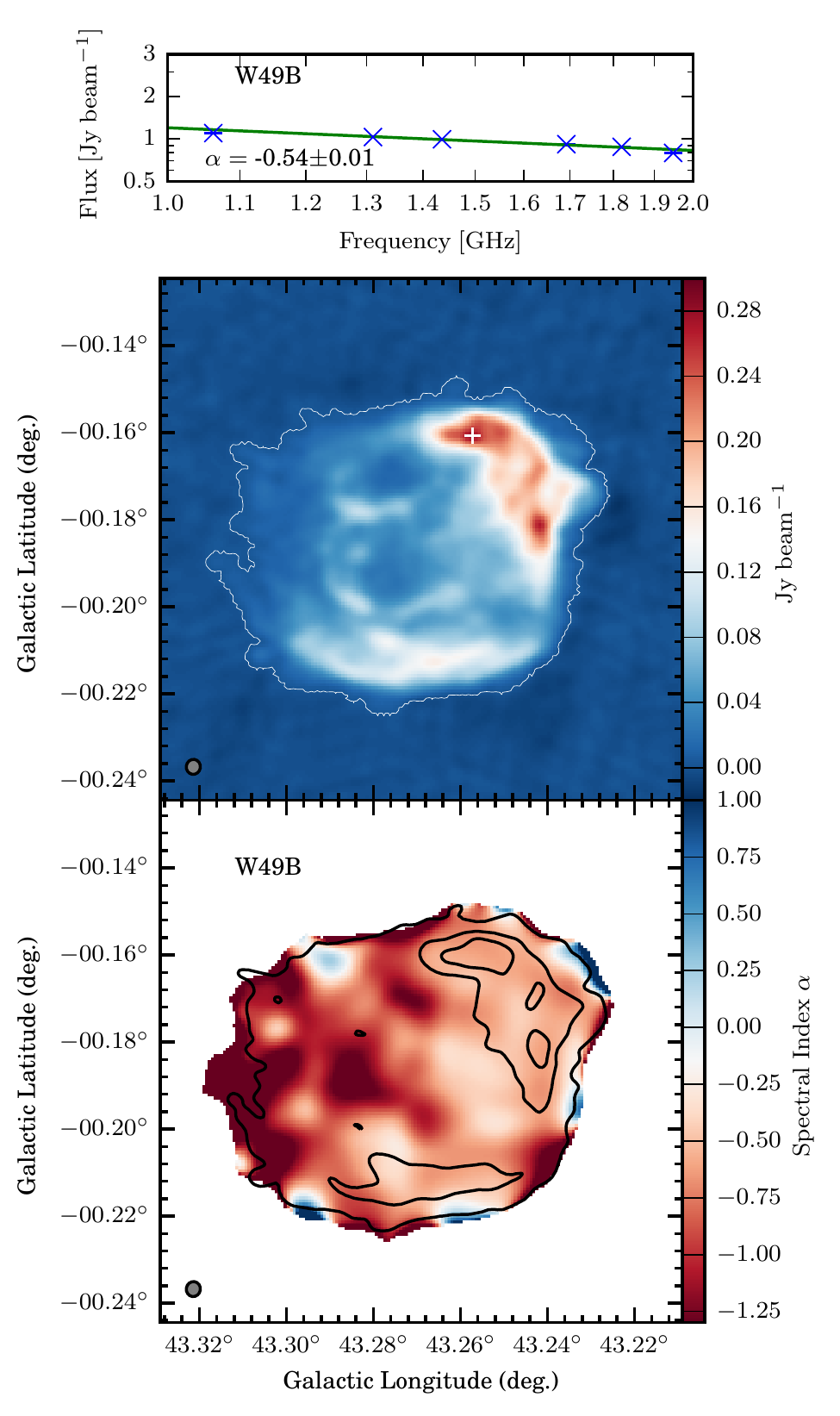}
      \caption{Spectral index map for SNR W49B. {\it Top-panel:} Spectral index fitting result of the peak of the continuum source G43.257--0.161. {\it Middle-panel:} Averaged image of spw-1440 and spw-1820, the white contours represent the area of the source extracted by BLOBCAT, and the cross marks the peak of the source G43.257--0.161 in the catalog. {\it Bottom-panel:} The spectral index map produced by fitting the flux of the six SPWs pixel by pixel. The synthesized beams of the averaged image and the spectral index map are shown in the bottom-left corner of each panel.}
         \label{fig_w49b}
\end{figure}

By studying the compact and extended THOR$+$VGPS 1.4~GHz continuum emission (in the region $l>17.5\degr$), \citet{anderson2017} confirmed the radio emission for 52 SNRs from \citet{green2014}. \citet{anderson2017} found that six of the SNRs from the \citet{green2014} catalog (G20.4+0.1, G21.5 0.1, G23.6+0.3, G54.1+0.3, G59.8+1.2 and G065.8 0.5) are confused with \ion{H}{ii} regions and the radio emission appears to be thermal. Figure~\ref{fig_w49b} shows the spectral index fitting result and the spectral index map of SNR W49B as an example, indicating that the region is dominated by negative spectral index with some variation, and further indicating non-thermal synchrotron emission.

By comparing the large scale diffuse radio continuum emission traced by the combined THOR+VGPS 1.4~GHz continuum data with the mid-infrared {\it Spitzer} GLIMPSE 8.0~$\mu$m \citep{benjamin2003,churchwell2009} and MIPSGAL 24~$\mu$m \citep{carey2009} data, \citet{anderson2017} identified 76 new SNR candidates. Since the radio continuum emission from most of the new SNR candidates is weak and diffuse, we could only detect 13 of them in our continuum catalog that uses only the THOR c-configuration data and hence do not trace the large-scale diffuse emission. Among the new SNR candidates, G17.80-0.02, G26.75+0.73 and G27.06+0.04 are matched with two continuum sources each in our catalog, and G51.21+0.11 is matched with 6 continuum sources. We marked all the matched continuum sources with ``SNR\_anderson'' in the catalog and list the corresponding ID of the SNRs from \citet{anderson2017}.

The MAGPIS survey \citep{helfand2006} identified many SNR candidates, and 33 of them are covered by our THOR survey. \citet{anderson2017} found that 17 MAGPIS SNR candidates (G18.2536--0.3083, G19.4611+0.1444, G19.5800--0.2400, G19.5917+0.0250, G19.6100--0.1200, G19.6600--0.2200, G21.6417+0.0000, G22.7583--0.49171, G22.9917--0.3583, G23.5667--0.0333, G24.1803+0.2167, G25.2222+0.2917, G29.0667--0.6750, G30.8486+0.1333, G31.0583+0.4833, G31.6097+0.3347, and G31.8208--0.1222.) are spatially coincident with a known \ion{H}{ii} region from the  {\it WISE} catalog. G18.2536--0.3083 is also reported as a known \ion{H}{ii} region by \citet{bihr2016}. Furthermore, G16.3583-0.1833 and G17.3361-0.1389 are spatially coincident with known \ion{H}{ii} regions G016.360-00.211 and G017.336-00.146 from the {\it WISE} catalog. G29.0778+0.4542 is a known planetary nebula (PNG029.0+00.4, see also \citealt{anderson2017}). For 10 MAGPIS SNRs \citep{helfand2006} we find matched continuum sources in our catalog, however, these continuum sources are also matched with SNRs from \citet{green2014} or candidates from \citet{anderson2017}, we do not mark them separately in the catalog. For 3 MAGPIS SNRs we do not find matched continuum sources due to THOR sensitivity and the missing flux problem (see also \citealt{bihr2016}).

\subsection{Planetary nebulae}
\label{sect_pn}
The Hong Kong/AAO/Strasbourg H$\alpha$ planetary nebula database (HASH, \citealt{parker2016}) provides up-to date information for all known Galactic planetary nebulae (PNe). In the HASH database, there are 234 PNe with the PN status marked as T (True PN), L (Likely PN), and P (Possible PN) within our survey area. We detected 164 of them in our catalog, among which 75 are True PN, 60 are Likely PN and 29 are Possible PN. We mark them as ``PN'' in our catalog, and list the HASH ID (PNG). In particular PNG029.2+00.0 is associated with the continuum source G29.211--0.069, the extended emission is tracing the known \ion{H}{ii} region G029.165--00.035. Thus we mark the continuum source G29.211--0.069 as ``HII;PN'' and list both the PN and \ion{H}{ii} IDs. Furthermore, 90 of the matched PNe do not have radio emission information at 20~cm in the database; our survey provides this important information. We list all the PNe detected in the THOR survey in Table~\ref{table_pn}. For the remaining 74 sources the peak flux at spw-1440 from THOR observations are in agreement with the radio flux at 20~cm from the database in general, with a median flux difference of 31\% relative to the ones in the HASH database. Since we are comparing flux density in the HASH database with the THOR peak flux, the large difference is expected.

In addition to the 164 detected PNe we include in our catalog, PNG 014.5+00.0, 017.4+00.3 and 018.7+00.0 are also detected in our observations. However, the continuum sources that these PNe are associated with are known \ion{H}{ii} regions (G014.598+00.019, G017.414+00.377 and G018.710+00.000). Since these three PNe are all ``possible PN'', we did not include them in our catalog.

We did not detect the remaining 67 PNe, most likely due to sensitivity reasons. Only 4 of the 67 PNe have 1.4~GHz radio flux information in the HASH catalog. The 1.4~GHz flux density of PNG029.8+00.5 in the HASH catalog is 33~mJy, however, we did not detect any radio counterpart in our THOR-only nor THOR+VGPS data. \citet{bojicic2011} and \citet{luo2005} also reported the radio detection for this PN as suspect. The 1.4~GHz flux of PNG028.7+00.7, 044.9+00.0 and 056.1-00.4 in the HASH catalog is 2.7, 3.5 and 6.6~mJy, respectively.

\subsection{Pulsars}
\label{sect_psr}
The ATNF Pulsar Catalogue\footnote{\url{http://www.atnf.csiro.au/research/pulsar/psrcat}} \citep{manchester2005} provides the most complete catalog of all published rotation-powered pulsars. Within our survey region, the catalog contains 335 pulsars. We cross matched the pulsar catalog with the THOR continuum catalog taking into account the position uncertainty from the pulsar catalog. 38 THOR continuum sources are matched with pulsars, we mark them as ``PSR'' in the continuum catalog, and list the corresponding pulsar names. Except for G38.163-0.151, all matches have a separation smaller than 7.5\arcsec. Considering the synthesized beam size of our observation is $\sim12\arcsec-18\arcsec$, these matches are reliable. G38.163--0.151 is matched with J1901+0435 with a separation of 15\arcsec, however, J1901+0435 has a relatively large position uncertainty of 10\arcsec, which we consider it a good match. We list the matched continuum sources and pulsars in Table~\ref{table_psr}.

Among all the pulsars detected in our survey, J1841-0500 is an extremely intermittent pulsar \citep{camilo2012}, and has no 1.4~GHz radio flux information in the ATNF pulsar catalog. We detect a radio counterpart in our observation with a S/N $\sim$6 (G27.322--0.033).

We did not detect the remaining 297 pulsars in our catalog, which is mainly due to low sensitivity in some sections of the maps. Among the non-detected pulsars, 247 have mean 1.4~GHz radio flux densities in the ATNF pulsar catalog, and their mean flux densities are all below 3~mJy, with a median value of 0.3~mJy. In comparison, the detected ones have a median mean flux density of 1.7~mJy from the ATNF catalog.

In general, the peak flux at spw-1440 from THOR observations are in agreement with the radio flux at 1.4~GHz in the ATNF catalog, with a median flux difference of 30\% relative to the ones in the ATNF catalog.

\subsection{X-ray sources}
\label{sect_xray}
While diffuse radio and X-ray emission can trace \ion{H}{ii} regions (free-free emission and stellar wind shocks, \citealt{silich2005}) and SNRs \citep[e.g., ][]{decourchelle2005}, the compact emission is usually tracing Pulsar Wind Nebulae \citep[e.g.,][]{brisken2005, miller2005}, X-ray binaries \citep[XRB or microquasars; ][]{mirabel1998, mirabel1999}, or active galactic nuclei \citep[AGN, e.g, ][]{bridle1984, harris2006}. 

We cross-matched our continuum catalog with three X-ray source catalogs, 1SXPS {\it SWIFT} X-ray telescope point source catalog \citep{evans2014}, XMM-Newton Serendipitous Source Catalog (3XMM-DR7 Version, \citealt{watson2009,rosen2016}), and Chandra Source Catalog (CSC, v1.1, \citealt{evans2010}). Within our survey area, 1SXPS has $\sim2800$ sources, CSC has $\sim3900$ sources, and 3XMM has $\sim11500$ sources. With a matching radius of 10\arcsec, we found that 79 of the remaining $\sim$9300 THOR sources are associated with X-ray sources (see Table~\ref{table_xray}). Among the 79 X-ray sources we detected, 43 do not have a radio counterpart within a radius of 15\arcsec\ on the SIMBAD Astronomical Database\footnote{\url{http://simbad.u-strasbg.fr/simbad/}} or the NASA/IPAC Extragalactic Database (NED)\footnote{\url{http://ned.ipac.caltech.edu/}}. 65 of the X-ray sources are unresolved, the remaining resolved sources are also relatively compact with an effective radius $<29$\arcsec. In particular, continuum source G45.366--0.219 is associated with the microquasar GRS1915+105 \citep{mirabel1999}. Further \ion{H}{i} absorption studies towards these sources would allow us to identify which ones are galactic. By investigating the X-ray and radio flux ratios, the spectral indices and observations at other (optical/infrared) wavelengths of the Galactic sources in detail, we can further constrain the type of the sources, i.e., low mass XRB, PN, pulsar etc. \citep[e.g.,][]{seaquist1993, maccarone2012, tetarenko2016}.

\subsection{Extragalactic radio sources}
\label{sect_jet}
About 9300 sources in our catalog are not classified. Taking a matching radius of $15\arcsec$, which is the average size of the synthesized beam of the averaged images used for source extraction, more than 7750 sources do not match to anything in SIMBAD. For the matched ones, about $84\%$ of them are matched to only one or more radio sources. With the same matching radius ($15\arcsec$), $\sim$3350 of the sources are matched to one or more infrared sources, and $\sim$2140 of the sources are matched to one or more radio sources, and more than 3760 sources do not have any counterparts within a radius of $15\arcsec$. This means that for $\sim$9000 sources detected in our survey, besides they have radio and/or infrared emission, we do not know exactly what they are.

\begin{figure*}
   \centering
  \includegraphics[width=0.33\hsize]{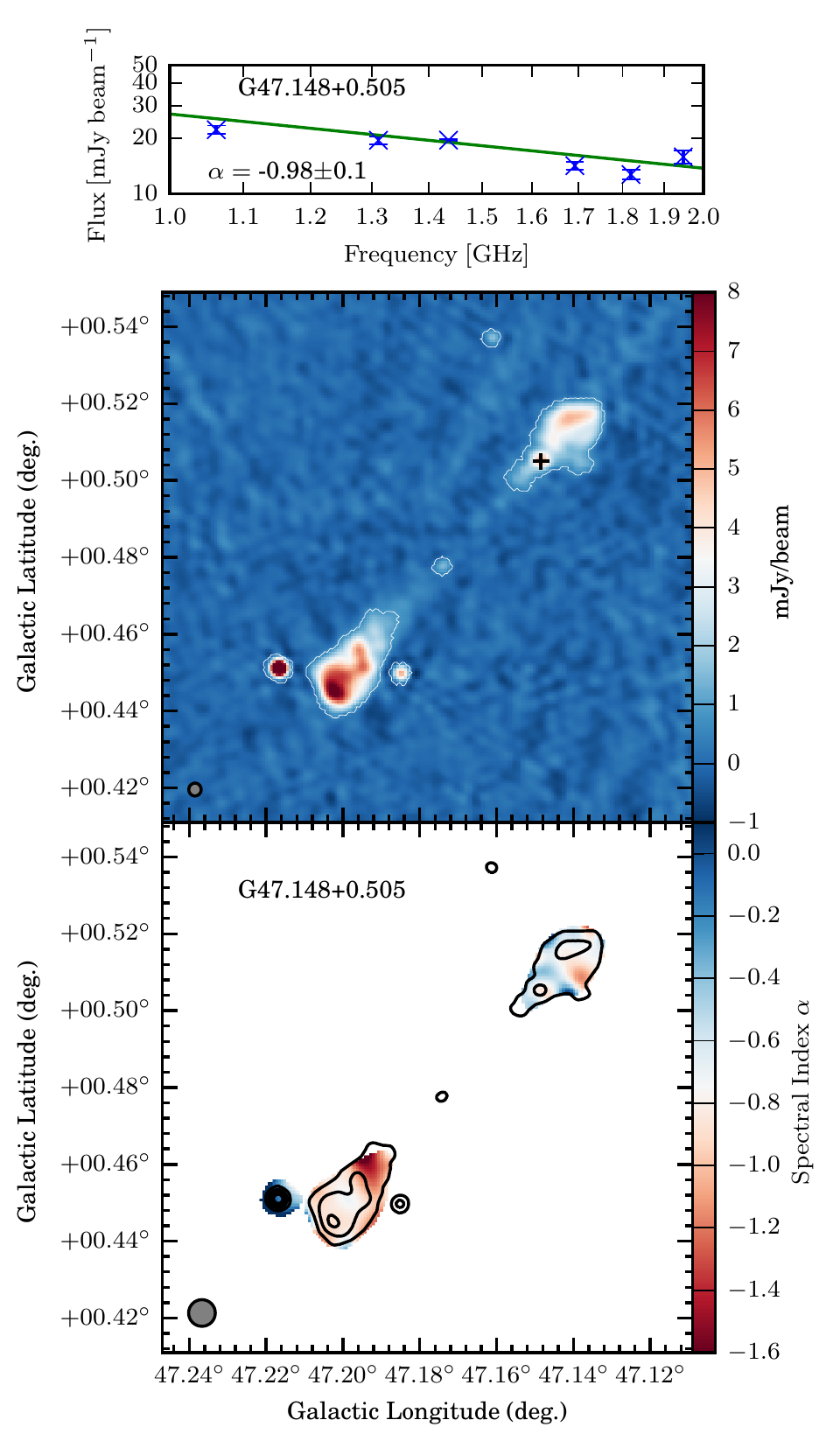}
  \includegraphics[width=0.33\hsize]{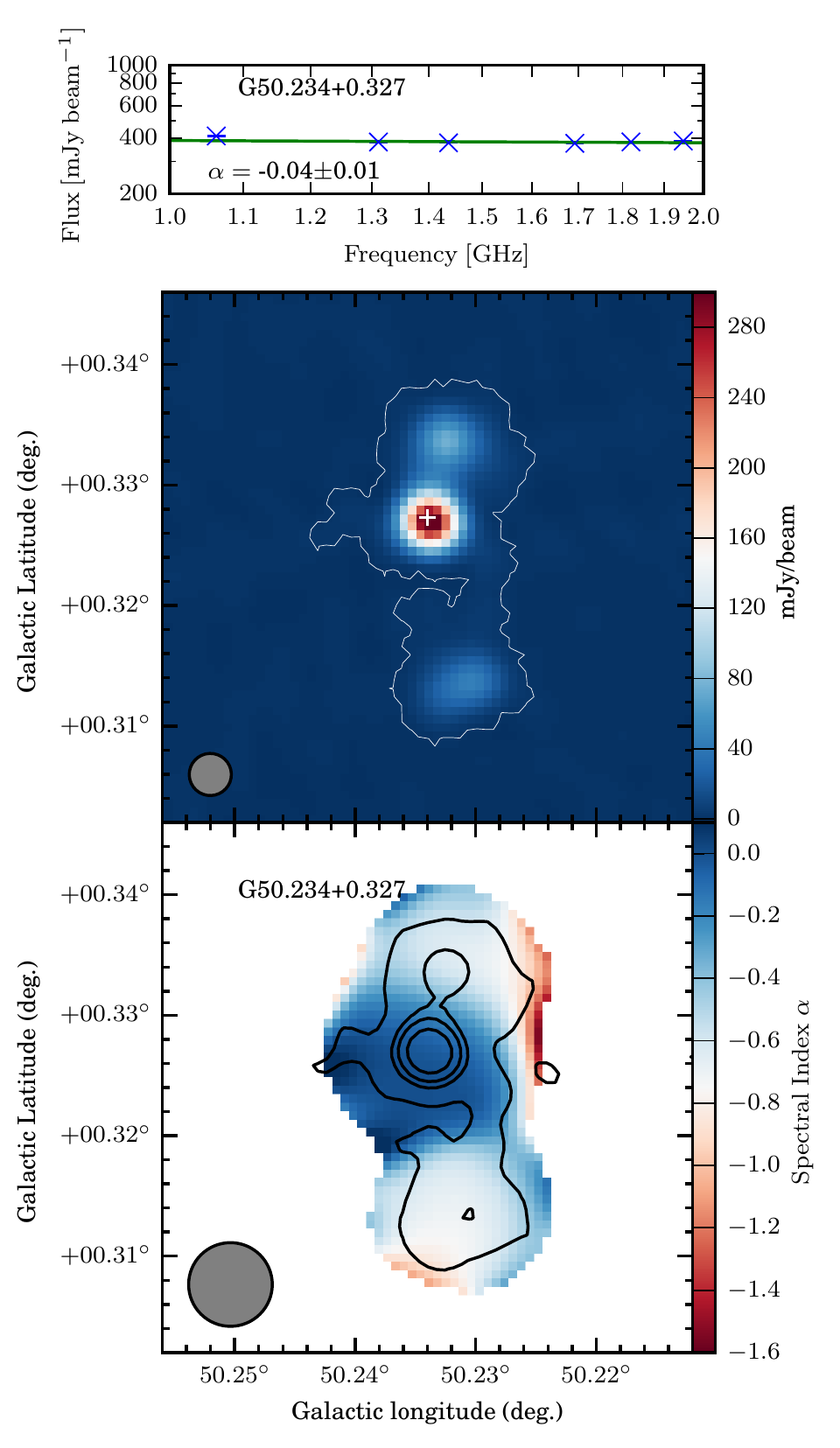}
  \includegraphics[width=0.33\hsize]{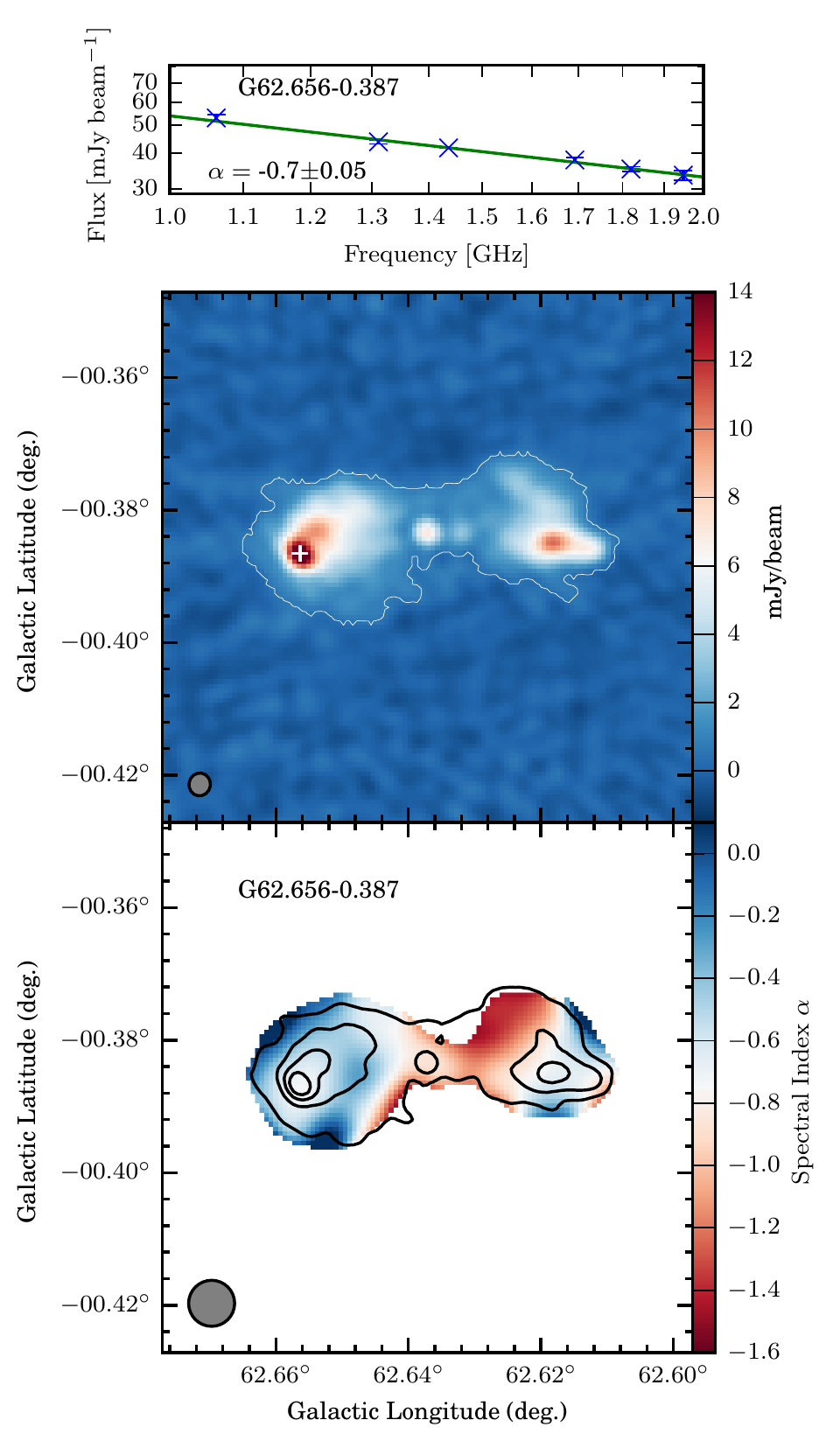}
      \caption{Spectral index maps for three ``jet'' sources. {\it Top-panels:} Spectral indices fitting results of the peak positions. {\it Middle-panels:}  Averaged images of spw-1440 and spw-1820, the white contours represent the area of the sources extracted by BLOBCAT, and the crosses mark the peaks of the sources in the catalog. {\it Bottom-panels:} The spectral index map produced by fitting the flux of the six SPWs pixel by pixel. The synthesized beams of the averaged image and the spectral index map are shown in the bottom-left corner of each panel.}
         \label{fig_jetsmap}
 \end{figure*}

As we mentioned in Sect.~\ref{sect_index}, besides the \ion{H}{ii} regions, SNRs, and PNe, the majority of the sources we detected are actually extragalactic. Many of the those sources are also resolved and show bipolar radio lobe structures. We checked all the resolved sources with a negative or flat spectral index in our catalog by eye, and identified $\sim$300 sources that show clear bipolar jet structure, and mark them as ``jet'' in the catalog. For these ``jet'' sources, we can construct the spectral index maps, which could be used to estimate the source expansion velocity when combined with magnetic field strength information. Since the identification of these ``jet'' sources is done morphologically, two partially overlapped sources could also be categorised into ``jet'' sources. Since we do not have zero-spacing information, we filtered out large scale structures in our observations. \citet{bihr2016} show that the THOR observations are able to recover sources with size up to $\sim$120\arcsec\ reasonably well ($\geq$80\% flux recovery), spectral index maps for sources larger than that should be interpreted with caution. Figure~\ref{fig_jetsmap} shows the spectral index maps of three ``jet'' sources. As the figure shows, the spectral index varies within the source, indicating that the optical depth of the synchrotron emission varies or possible thermal emission exists in different parts of the radio lobes.

\begin{figure}
   \centering
   \includegraphics[width=\hsize]{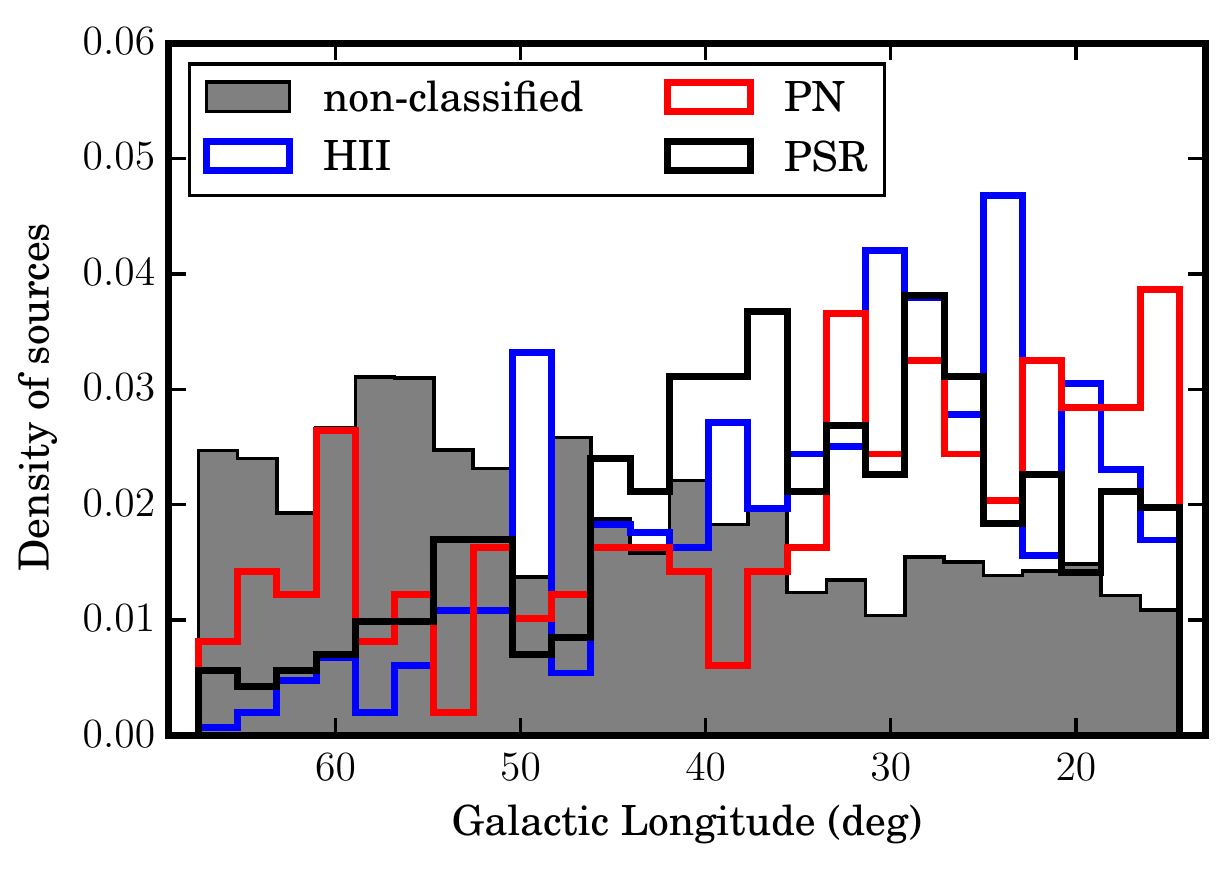}\\
   \includegraphics[width=\hsize]{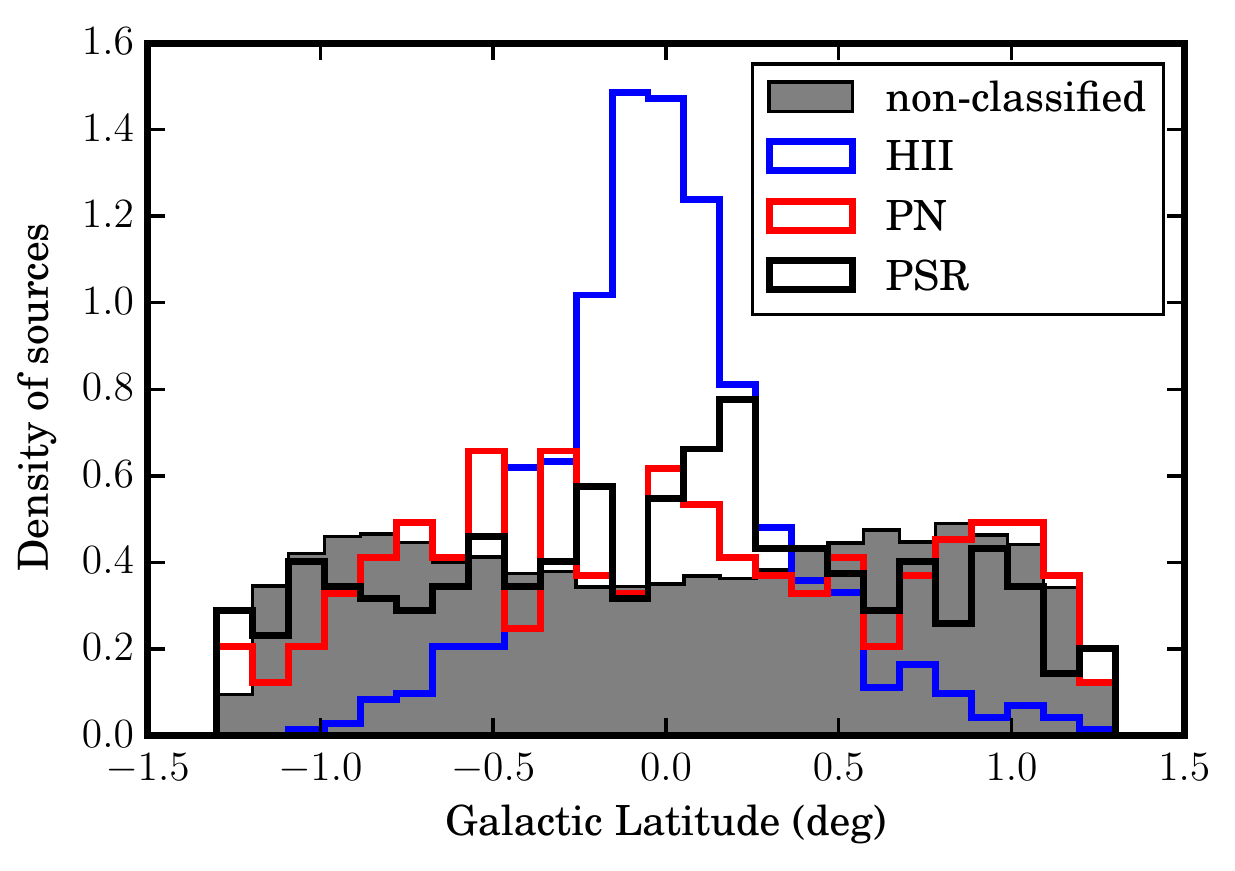}\\
      \caption{Normalized distribution of different types of objects along Galactic Longitude ({\it top-panel}) and Latitude ({\it bottom-panel}) for the non-classified sources, \ion{H}{ii} regions, PNe, and pulsars. For \ion{H}{ii} regions we plot only the ones associated with the THOR continuum sources. For PNe and pulsars, we plot all that lie within our survey area.}
         \label{fig_dist_norm}
 \end{figure}

\begin{figure}
   \centering
   \includegraphics[width=\hsize]{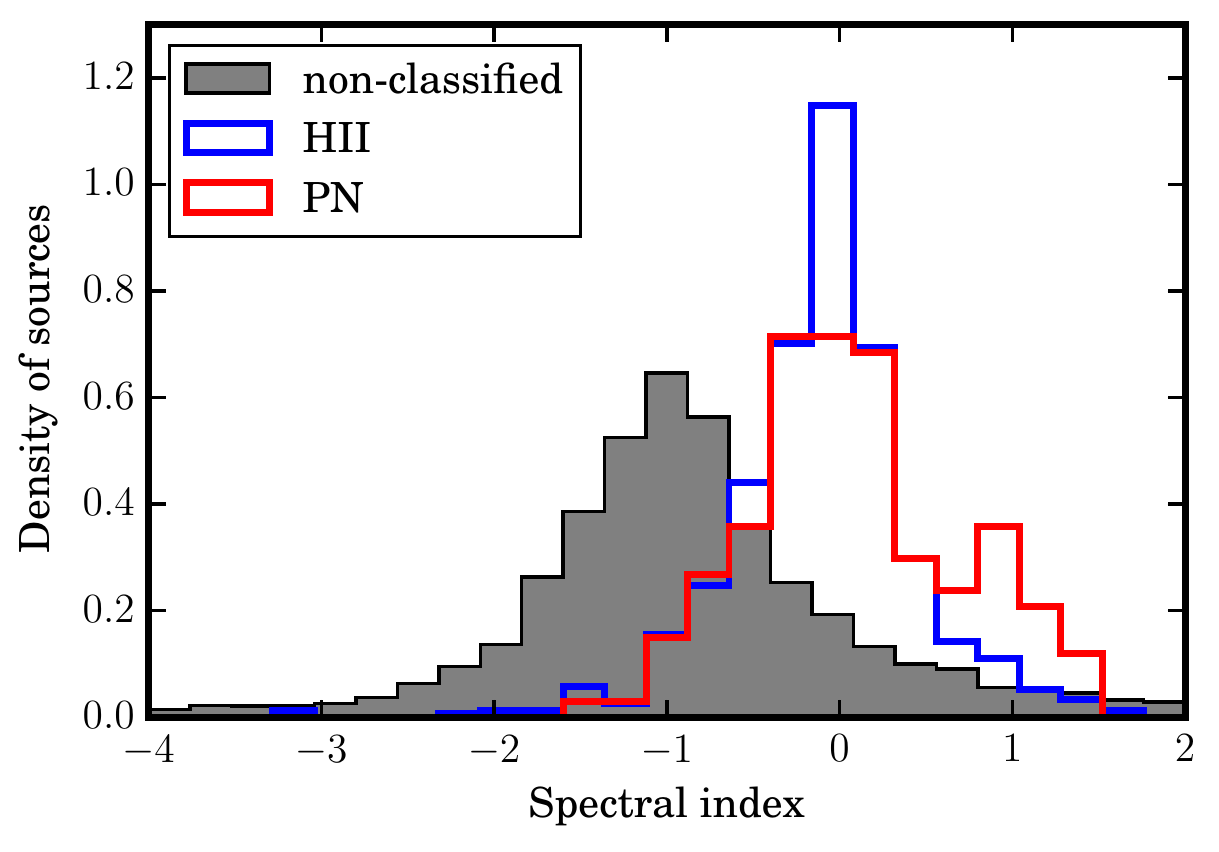}
      \caption{Normalized spectral index distribution of the non-classified sources, \ion{H}{ii} regions, and PNe.}
         \label{fig_spec_norm}
 \end{figure}

In Sect.~\ref{sect_index} we mentioned that the spatial distribution and the spectral index distribution of the unresolved sources indicate the majority of them are extragalactic sources. To further confirm this, we compare the normalized spatial distribution and spectral index distribution of the non-classified sources to the Galactic sources we identified. Figure~\ref{fig_dist_norm} compares the normalized distribution along Galactic longitude and latitude of those non-classified sources with the Galactic objects (\ion{H}{ii} regions, PNe and pulsars). The normalized distribution along Galactic longitude (top-panel, Fig~\ref{fig_dist_norm}) shows that the Galactic objects are more concentrated in the lower longitude region, while more non-classified sources are detected in the higher longitude region. The distribution along Galactic latitude (top-panel, Fig~\ref{fig_dist_norm}) shows that the \ion{H}{ii} regions are concentrated very close to the Galactic mid-plane. Although not as much as the \ion{H}{ii} regions, the PNe and pulsars are also concentrated close to the mid-plane. In contrast, more non-classified sources are found in regions $|b|>0.5\degr$. This distribution difference confirms that the non-classified sources are of mostly extragalactic origin. Furthermore, the spectral index distribution of the non-classified sources also shows clear differences from the Galactic sources and peaks around $\alpha\sim-1$ (Fig.~\ref{fig_spec_norm}), but there are still many of these non-classified sources showing a flat SED ($\alpha \gtrsim0$), which could be Galactic compact \ion{H}{ii} regions that are not listed in the {\it WISE} \ion{H}{ii} catalog. We classify the $\sim5300$ sources with a negative spectral index $\alpha<-0.3$ as extragalactic source candidates, among which $\sim 3970$ sources have a reliable spectral index (fit\_spws$\geq4$).

\begin{figure*}
   \centering
   \includegraphics[width=0.9\hsize]{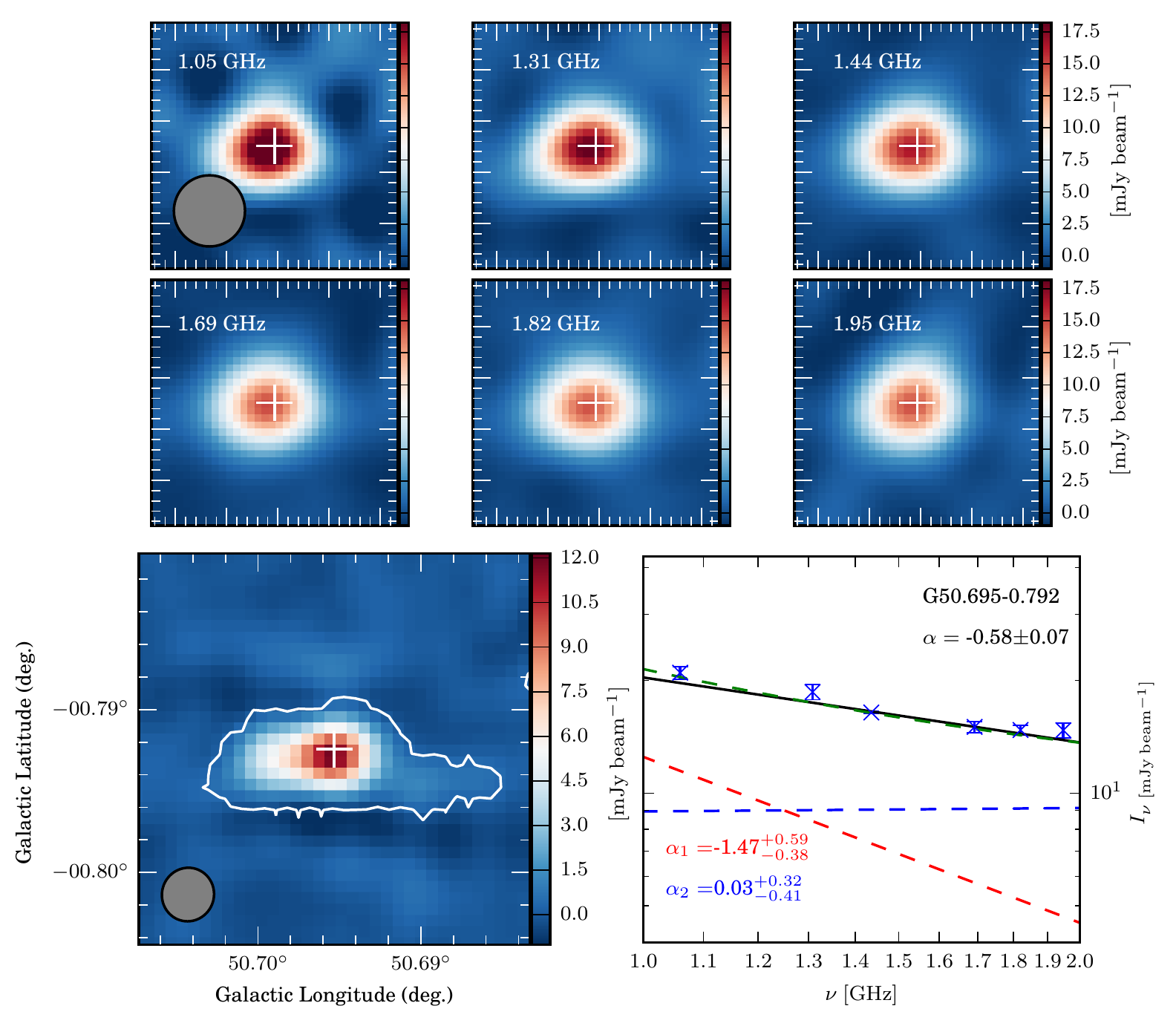}
    \caption{One example source G50.695--0.792 fitted with two indices. The top two panels show each SPW separately. The bottom left large image represents the averaged image of spw-1440 and spw-1820, which we used for the source extraction (see Sect.~\ref{sect_extract}). The white contours show the extent of the source determined by the BLOBCAT algorithm. The cross in each panel marks the peak position, which we used to determine the spectral index. The bottom right panel presents the peak intensity for each SPW and corresponding one single spectral index fitting (black solid line) and two spectral indices fitting (green dashed lines) from two components (red and blue dashed lines).}
      \label{fig_2spec}
\end{figure*}

Extragalactic radio sources often consist of a core and one or more radio lobes, which show different spectral profiles. The core dominates the emission at shorter wavelengths and shows a close to flat or positive spectral index, while the jet dominates at longer wavelengths and has shows negative spectral index \citep[e.g.,][chap.~9]{hey1971}. This transition in the spectral profile happens around $\sim$10 to 20~cm. Assuming the emission of the core and jet at 1 to 2~GHz can both be described with one single power law, then the total flux $I(\nu)$ is proportional to $(c_1\nu^{\alpha_1} + c_2\nu^{\alpha_2})$, where $\alpha_1$ and $\alpha_2$ are the spectral indices of the jet and core, respectively, $c_1$ and $c_2$ are constants. We use the Markov chain Monte Carlo method (MCMC) to fit the extragalactic source candidates which have a S/N ratio larger than 50 with the two spectral indices using the {\it emcee} package \citep{foreman2013}. Then we select the ones that meet the following criteria as good fit: first, the reduced $\chi^2$ of the two-spectral-indices fitting is smaller than the one with single spectral index fitting; second, the spectra of the two components cross with each other within our frequency coverage (1.05 to 1.95~GHz, such as Fig.~\ref{fig_2spec}). In total, we can derive a good fit with two spectral indices for 135 continuum sources. 

We performed an $F$-test \citep[e.g.,][]{philips1982, lomax2013} to compare the two components fitting to the one power low fitting, and calculated the probability values ($p$-value) for all 135 sources. The $p$-values are all above 0.28, so the two components fitting is not significantly better than the simple one power law fitting. However, the two components fitting results still provide alternative insight for those sources. As shown in Fig.~\ref{fig_2spec}, the peak fluxes  of G50.695--0.792 can be fitted with two components with a spectral indices of --1.47 and 0.03, respectively, which could be tracing the radio jet and core respectively. The averaged image in Fig.~\ref{fig_2spec} shows G50.695--0.792 is resolved with an elongated structure, which might be tracing the jet.

Radio sources with Ultra Steep Spectra (USS, $\alpha\leq-1.3$) are efficient tracers of high redshift radio galaxies, since radio sources can be detected uniformly over all redshift ranges, and do not suffer dust extinction at high redshifts \citep[e.g.,]{chambers1996,hughes1997,ivison1998,debreuck2000}. \citet{debreuck2000} found a trend that steeper spectral index sources have higher redshifts, and 50\% of the 4C USS sample \citep{chambers1996} are $z>2$ sources. This $z-\alpha$ correlation is considered to be a combined contribution of a K-correction and an increasing spectral curvature with $z$ of a radio spectrum \citep{krolik1991,carilli1999,vanBreugel1999}. 

In our continuum catalog, we have $\sim$2170 sources that have an spectral index steeper than --1.3, which is our base sample of USS sources. \citet{debreuck2000} further point out that 85\% USS with X-ray detections are in galaxy clusters, so we chose the ones without X-ray emission. Furthermore, we select sources with a reliable spectral index determination (fit\_spws$\geq4$). This selection results 1362 sources that have a good spectral index determination and have a spectral index steeper than --1.3. 663 of these sources have a spectral index steeper than --1.6 and could be Galactic pulsars \citep{debreuck2000}. The remaining 699 USS sources could be high-redshift radio galaxies; further spectroscopic observations are needed to confirm this.

\section{Conclusions}
\label{sect_con}
We observed a large portion of the first Galactic quadrant ($l=14.0-67.4\degr$, $|b|\leq1.25\degr$) using the VLA in C-configuration and achieved a spatial resolution of $\sim10-25\arcsec$ at 1 to 2~GHz with the THOR Galactic plane survey. In this paper, we present the catalog of the continuum sources from the whole survey. We summarise numbers of different types of sources in the catalog in Table~\ref{table_sum}, and the main results below.

\begin{table}
\caption{Summary of different types of sources in the catalog}
\label{table_sum}   
\centering                       
\begin{tabular}{l c}        
\hline\hline           
Source Types  & Number of continuum sources\\    
\hline                       
  \ion{H}{ii} regions &  713 \\      
  SNRs   &  92$(+21)$\\
   PNe &  164 \\
   Pulsars  & 38 (+663)\\
   X-ray sources & 79\\
    extragalactic jets   & 299 \\
    USS   & 699 \\
   \hline                                 
\end{tabular}
\tablefoot{USS stands for sources with Ultra Steep Spectra ($\alpha\leq-1.3$). The numbers of the candidates are in brackets.}
\end{table}

\begin{enumerate}
\item The catalog contains 10387 sources we extracted with the BLOBCAT software after removing the obvious observational sidelobe artifacts. About 72\% (7521 sources) of the extracted sources are detected at a significance higher than 7$\sigma$, and $\sim79\%$ are unresolved. The catalog is complete to at least 94\% above the 7$\sigma$ detection limit. The noise of our data is dominated by the sidelobe noise and spatially varying, although more than 60\% of the observed area has a noise level of $7\sigma\lesssim2$~mJy~beam$^{-1}$. We extracted the peak intensity of the six usable SPWs between 1 to 2~GHz, and we were able to determine a reliable spectral index (spectral index fitted with at least 4 SPWs) for 5657 sources. 

\item We cross-matched the THOR catalog with the {\it WISE} \ion{H}{ii} region catalog and found 713 continuum sources are associated with \ion{H}{ii} regions. Among the matched \ion{H}{ii} regions, 16 are in the radio quiet group in the {\it WISE} catalog which means they did not previously have radio continuum detected. 231 continuum sources are associated with more than one \ion{H}{ii} region. The spectral index distribution shows a single peak around $\alpha=0$, indicating thermal free-free emission. For 168 sources we can fit the SED with a simple homogeneous \ion{H}{ii} region model and derive the emission measure (EM) and the electron density ($n_e$) where the distance information is available in the {\it WISE} catalog.

\item Although the diffuse emission from many of the large scale SNRs is filtered out by our interferometric observations, we identify 92 continuum sources associated with 39 SNRs from the SNR catalog by \citet{green2014}. 13 of the new SNR candidates from \citet{anderson2017} are detected in our continuum catalog.

\item By cross-matching the THOR catalog with the HASH database, we detect 164 PNe in our continuum catalog. As 90 of them do not have radio emission information at 20~cm in the database, our survey provides this important information in Table~\ref{table_pn}. The spectral index distribution is similar to the one of the \ion{H}{ii} regions and shows a single peak around $\alpha=0$, indicating thermal free-free emission.

\item We cross-matched the THOR catalog with the ATNF Pulsar Catalog and found 38 counterparts. One extremely intermittent pulsar J1841-0500 is also detected in our catalog. 663 sources with a spectral index $\alpha < -1.6$ could be Galactic pulsar candidates. 

\item We cross-matched the THOR catalog with X-ray source catalogs 1SXPS, 3XMM- DR7 and CSC, and found 79 overlaps. 12 of the them have a spectral index steeper than $-1.3$ and could be galaxy clusters. 43 of them do not have previous known radio counterparts within a radius of 15\arcsec on SIMBAD or NED.

\item About 300 sources show clear structure of bipolar jets, we mark them as ``jet'' in the catalog and construct spectral index maps. We identified 699 Ultra Steep Spectra (USS) sources and they could be high redshifted radio galaxies ($-1.3>\alpha>-1.6$). Further spectroscopic observations are needed to confirm this. 

\item About 9000 sources in our catalog are not classified specifically. They are likely to be extragalactic background sources. More than 7750 sources do not have counterparts in the SIMBAD Astronomical Database, and more than 3760 sources do not have counterparts in the NED. 
\end{enumerate}

With the THOR continuum catalog, we provide a rich dataset to the community. All the fits images and catalogs can be downloaded from the project website \\footnote{\url{http://www.mpia.de/thor/Overview.html}}. With the follow up observations of particular sources, such as absorption study of the extragalactic sources, or combining with other existing Galactic plane surveys, we can study the the structure of the Milky Way, and the ISM in different phases.

\begin{acknowledgements}
The National Radio Astronomy Observatory is a facility of the National Science Foundation operated under cooperative agreement by Associated Universities, Inc. Y.W., H.B., S.B., and J.D.S. acknowledges support from the European Research Council under the Horizon 2020 Framework Program via the ERC Consolidator Grant CSF-648505, and RSK via the ERC AdvancedGrant 339177 (STARLIGHT). H.B. and M.R. acknowledge support from the Deutsche Forschungsgemeinschaft in the Collaborative Research Center (SFB 881) ``The Milky Way 74 System'' (subproject B1, B2, B8), and from the Priority Program SPP 1573 ``Physics of the Interstellar Medium'' (grant numbers KL 1358/18.1, KL 1358/19.2). F.B. acknowledges funding from the European Union' s Horizon 2020 research and innovation programme (grant agreement No 726384). The authors thank J. Mottram and M. Fouesneau for the productive and fruitful discussions. This research made use of Astropy and affiliated packages, a community-developed core Python package for Astronomy \citep{astropy2018}, Python package {\it SciPy}\footnote{\url{https://www.scipy.org/}}, APLpy, an open-source plotting package for Python \citep{robitaille2012}, and software TOPCAT \citep{taylor2005}.

\end{acknowledgements}

\bibliographystyle{aa}

\bibliography{thor-wang}


\begin{appendix} 

\section{Noise maps}
\label{app_noise}

\begin{figure*}
\centering
   \includegraphics[width= 0.95\hsize]{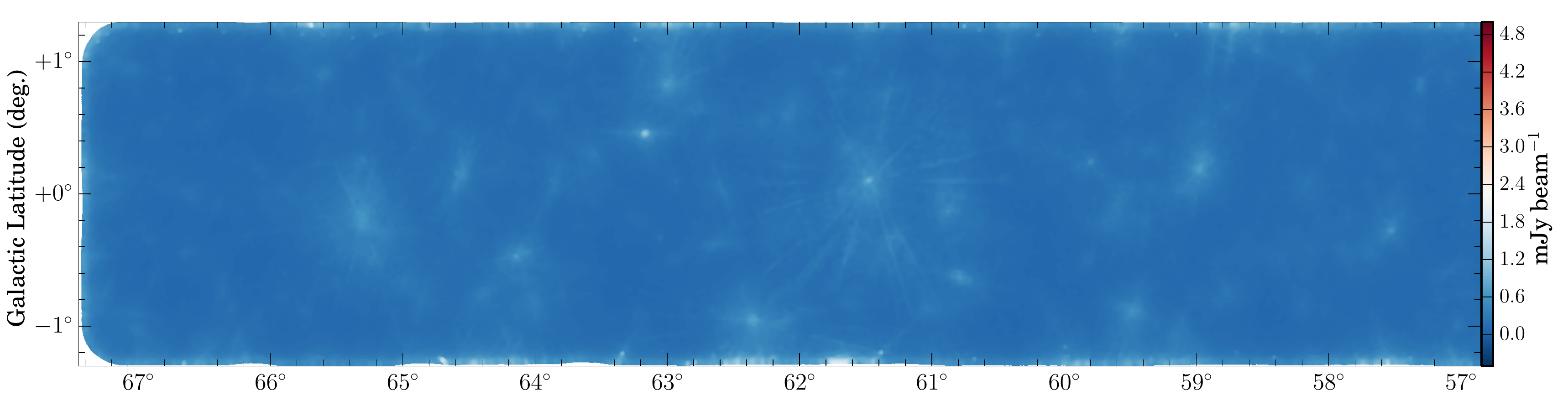}
   \includegraphics[width= 0.95\hsize]{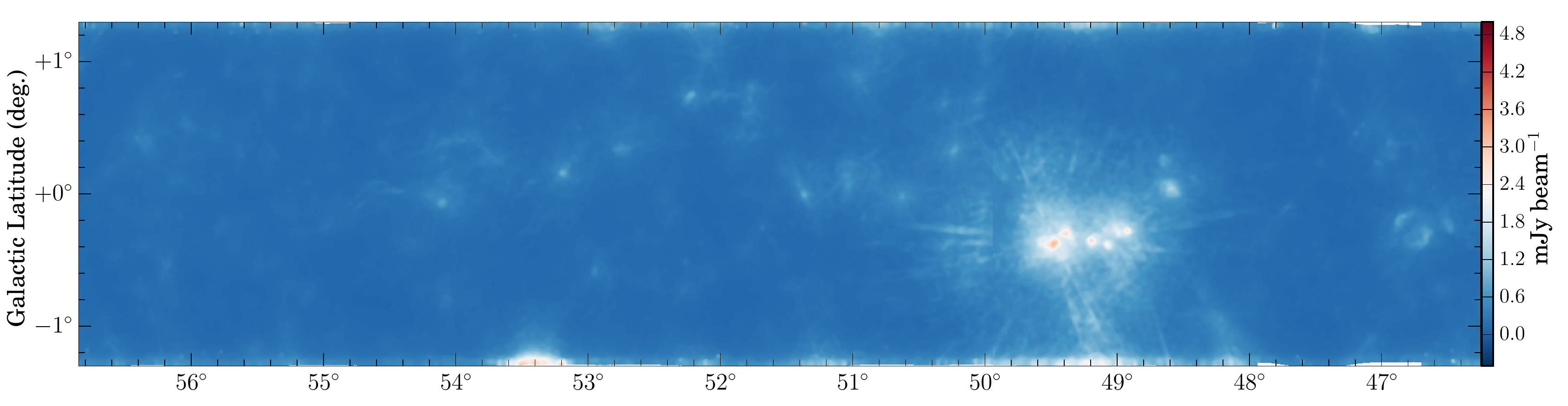}
   \includegraphics[width= 0.95\hsize]{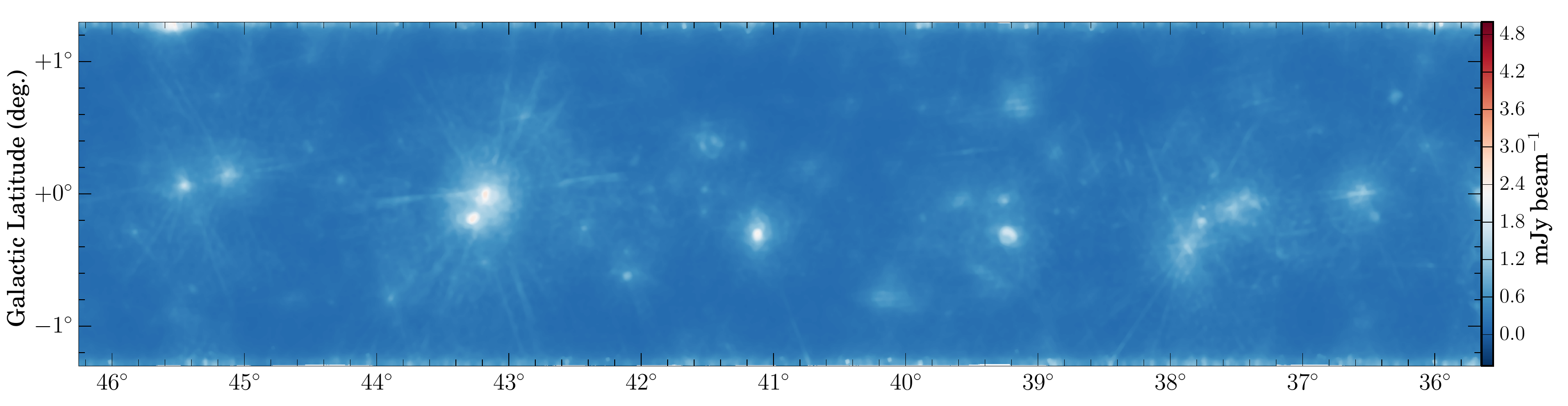}
   \includegraphics[width= 0.95\hsize]{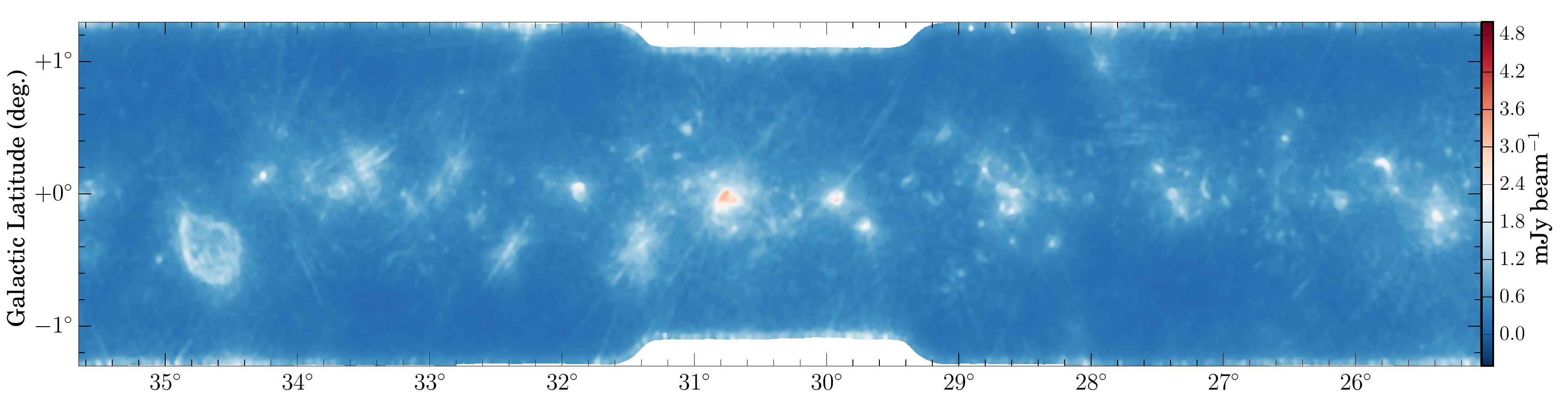}
   \includegraphics[width= 0.95\hsize]{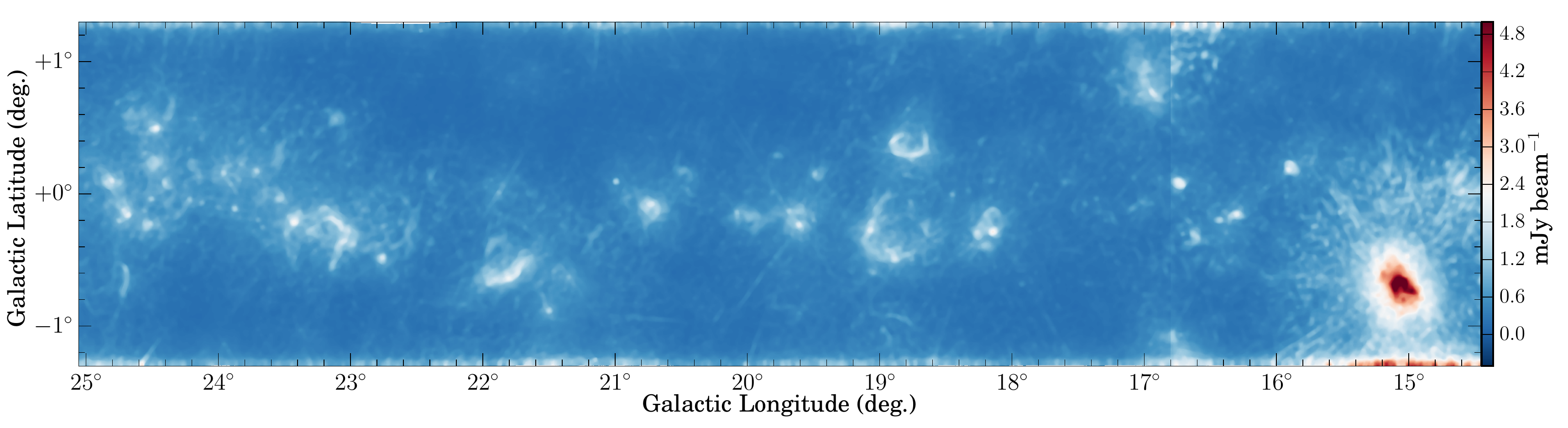}
   \caption{Noise map of the THOR survey using the average of spw-1440 and spw-1820}
    \label{fig_noise}
\end{figure*}

\section{Completeness maps}
\label{app_completeness}

\begin{figure*}
\centering
   \includegraphics[width= 0.95\hsize]{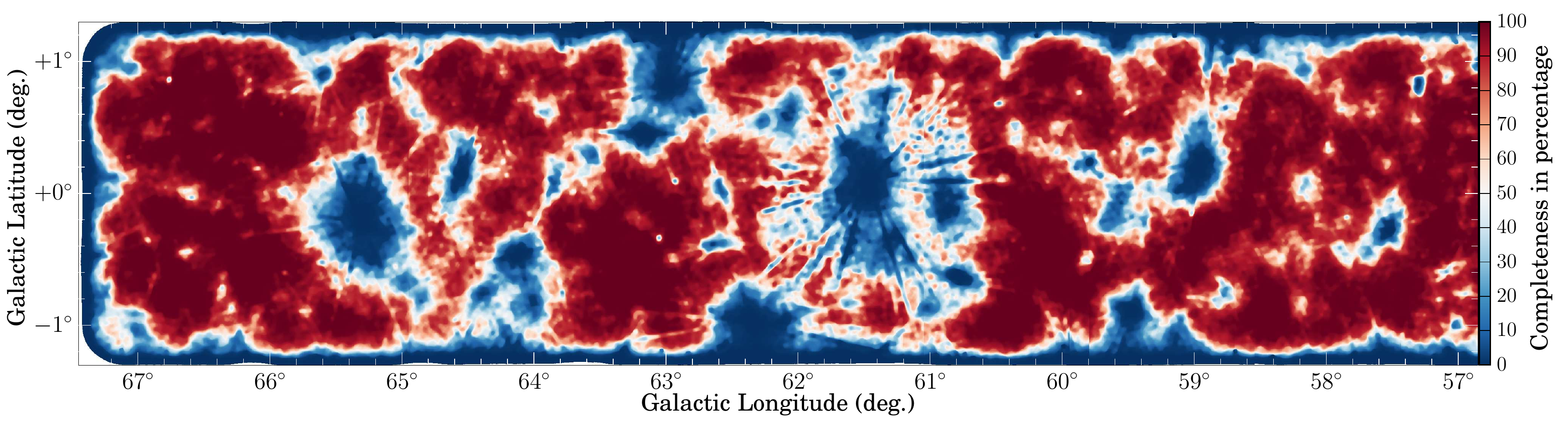}
   \includegraphics[width= 0.95\hsize]{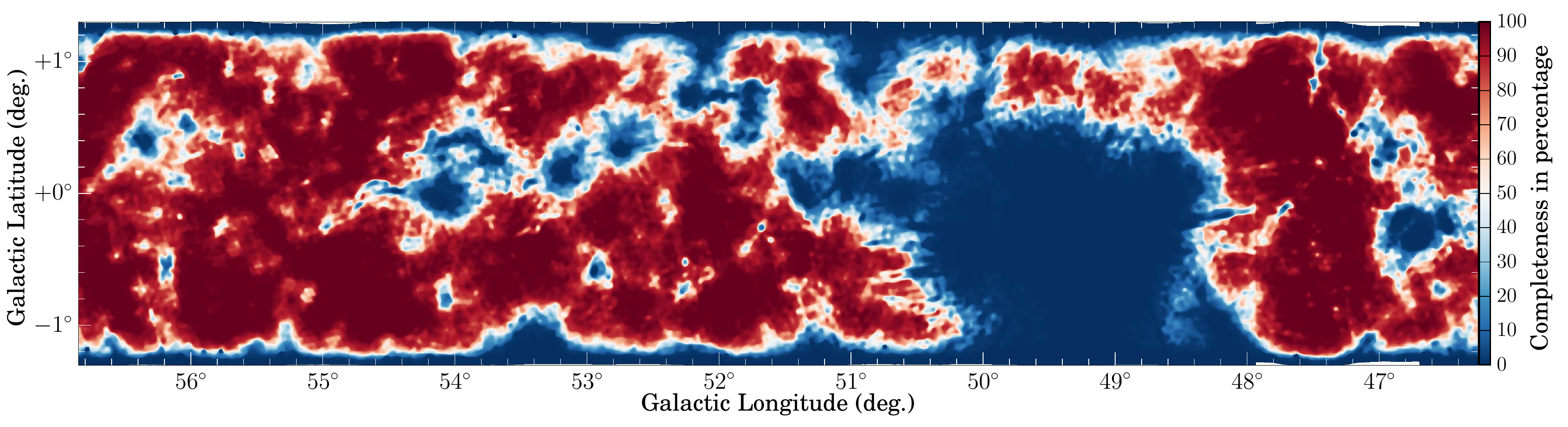}
   \includegraphics[width= 0.95\hsize]{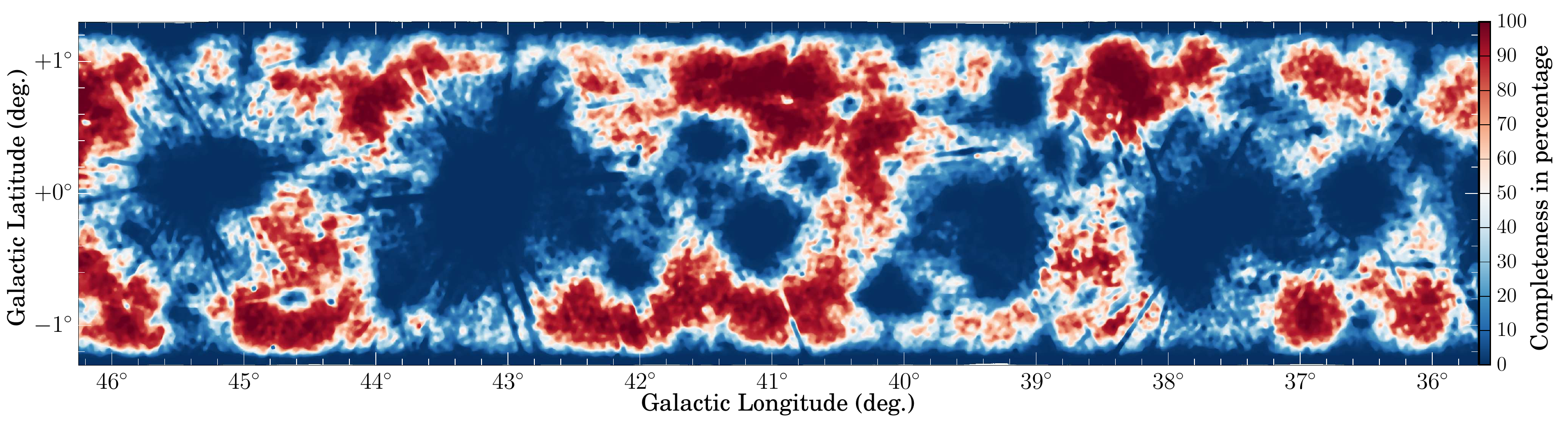}
   \includegraphics[width= 0.95\hsize]{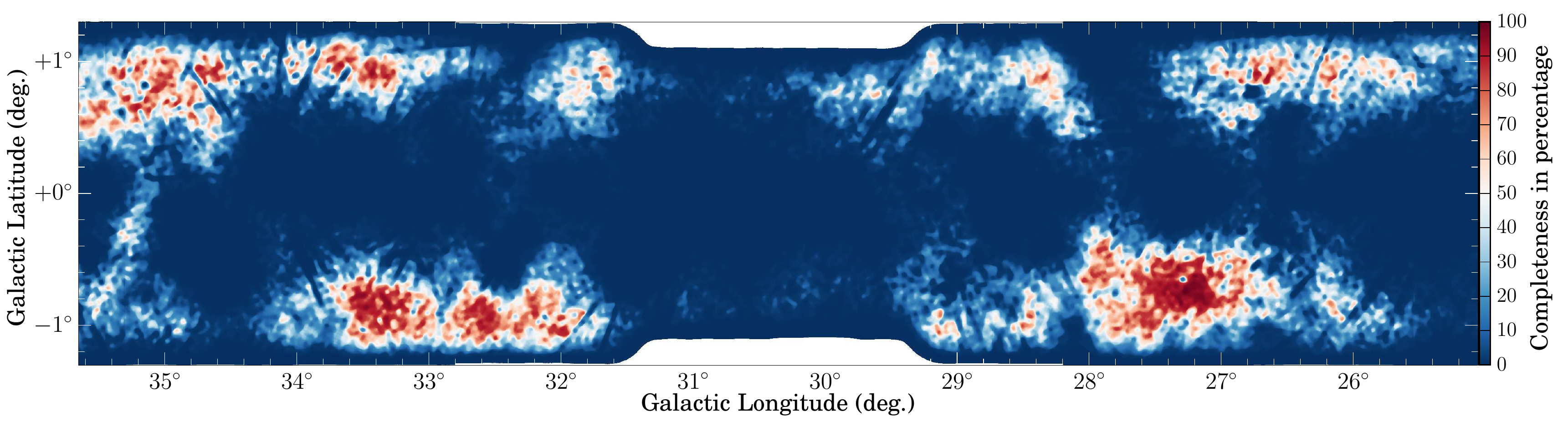}
   \includegraphics[width= 0.95\hsize]{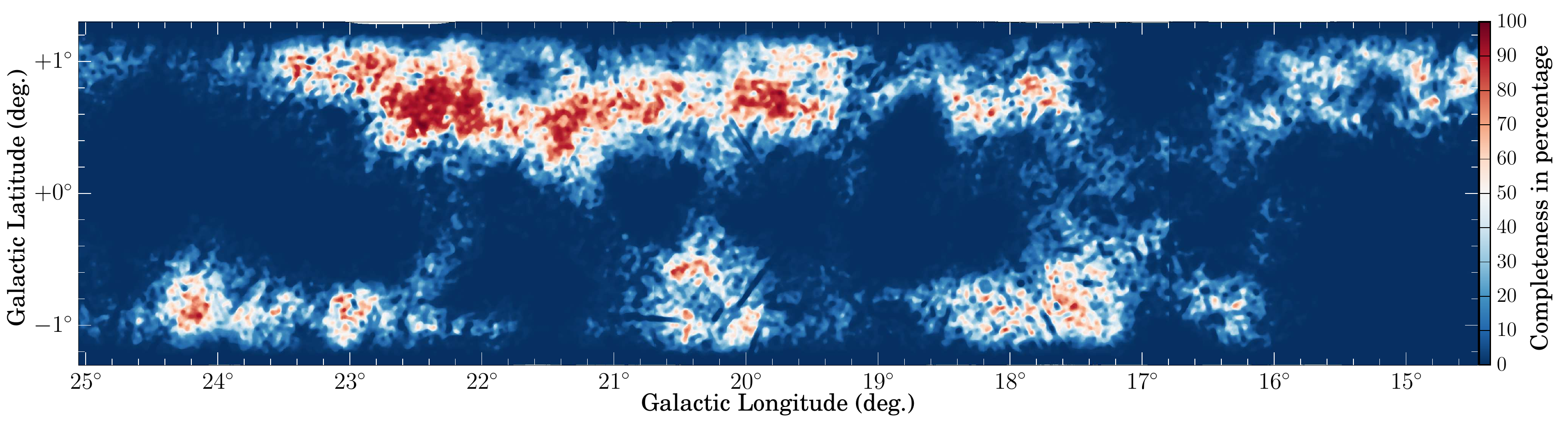}
   \caption{Completeness map in percentage for sources with a peak intensity of 1~mJy~beam$^{-1}$.}
    \label{fig_comp1}
\end{figure*}

\begin{figure*}
\centering
   \includegraphics[width= 0.95\hsize]{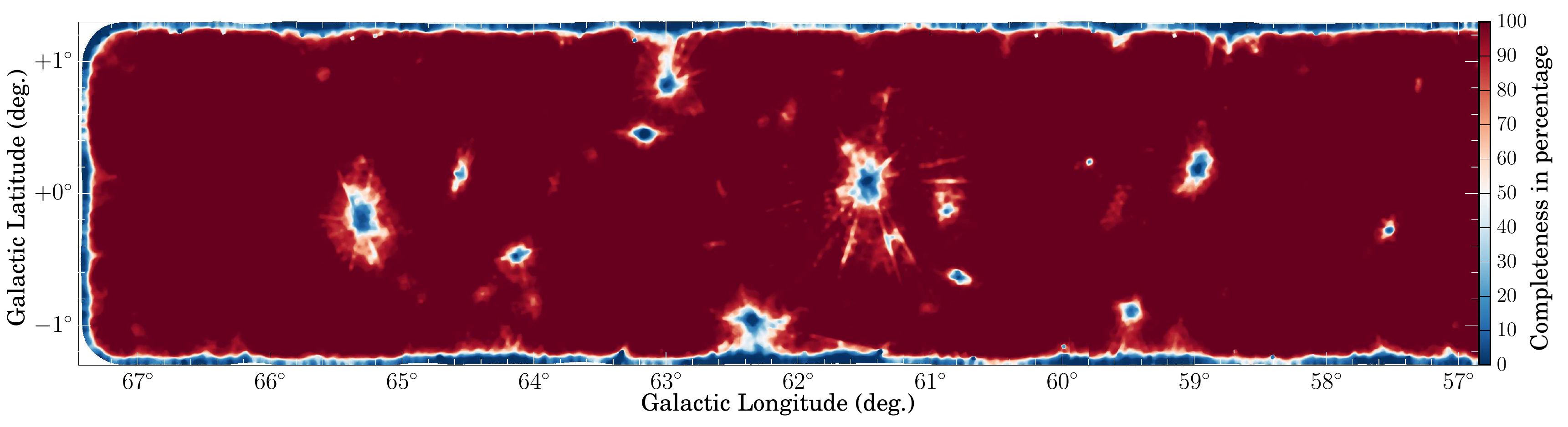}
   \includegraphics[width=0.95 \hsize]{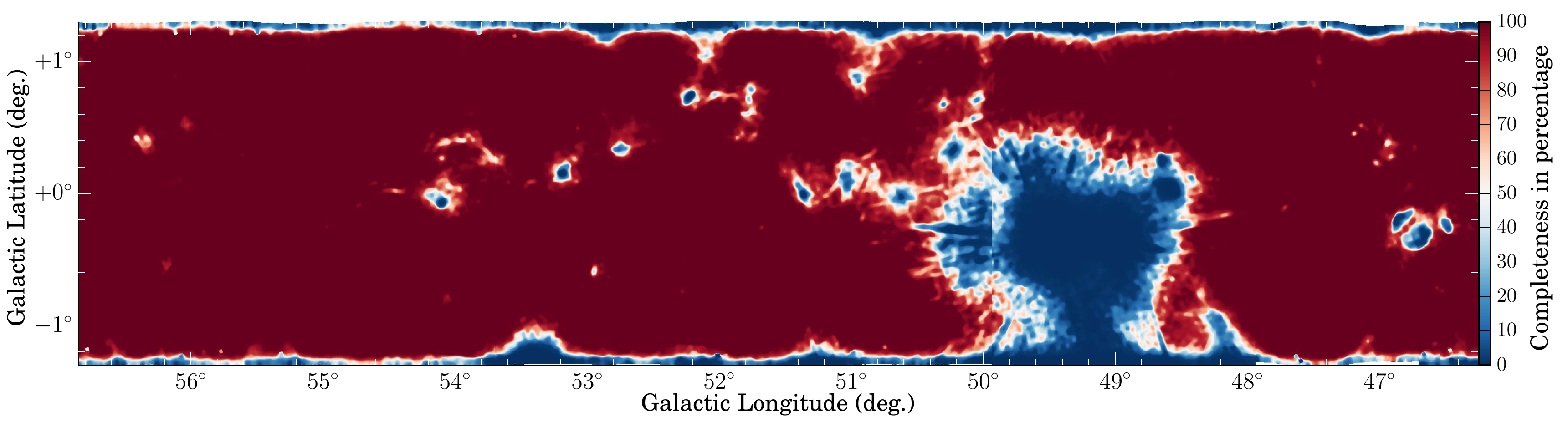}
   \includegraphics[width= 0.95\hsize]{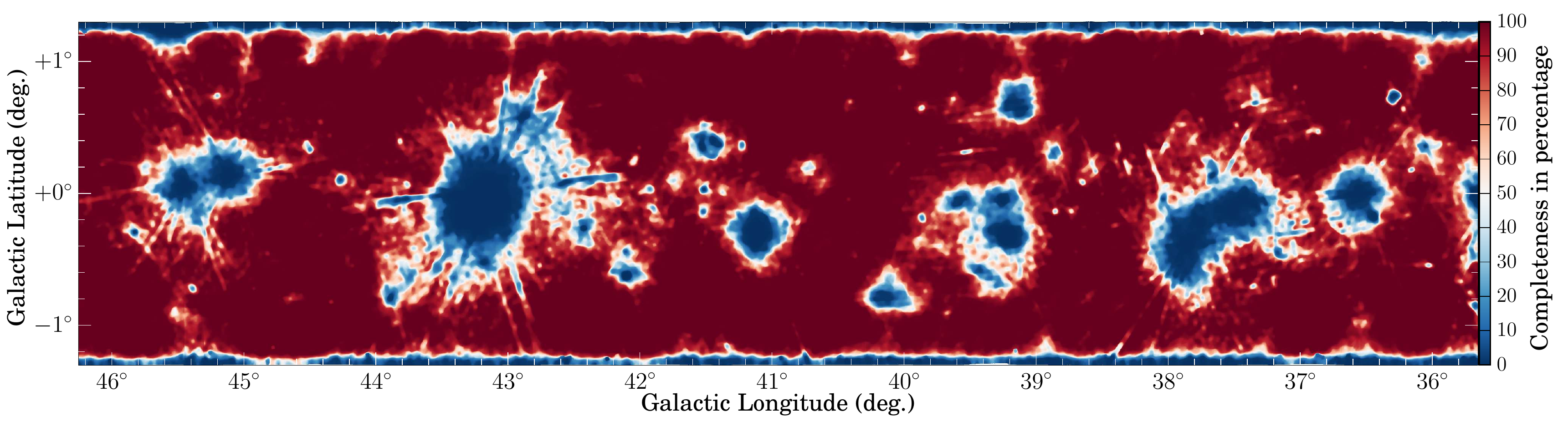}
   \includegraphics[width= 0.95\hsize]{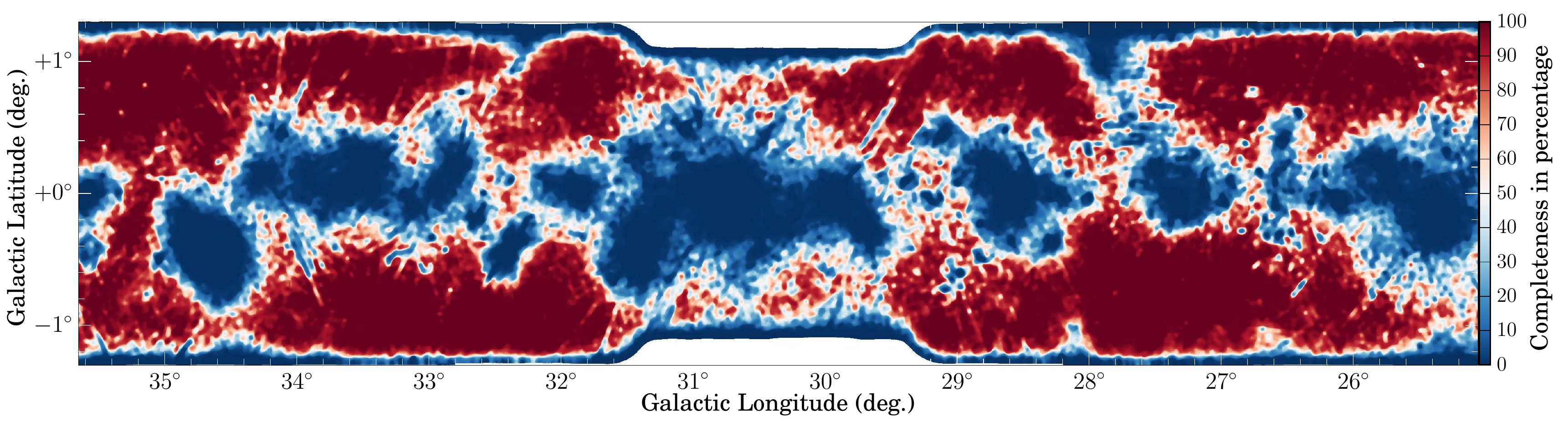}
   \includegraphics[width= 0.95\hsize]{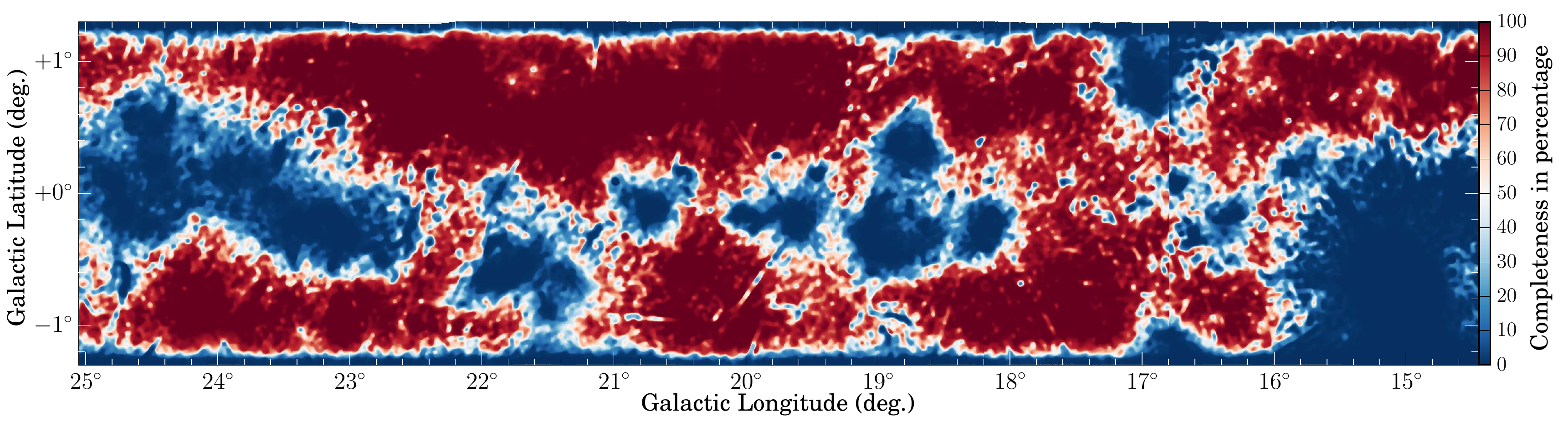}
   \caption{Completeness map in percentage for sources with a peak intensity of 2~mJy~beam$^{-1}$.}
    \label{fig_comp2}
\end{figure*}

\begin{figure*}
\centering
   \includegraphics[width= 0.95\hsize]{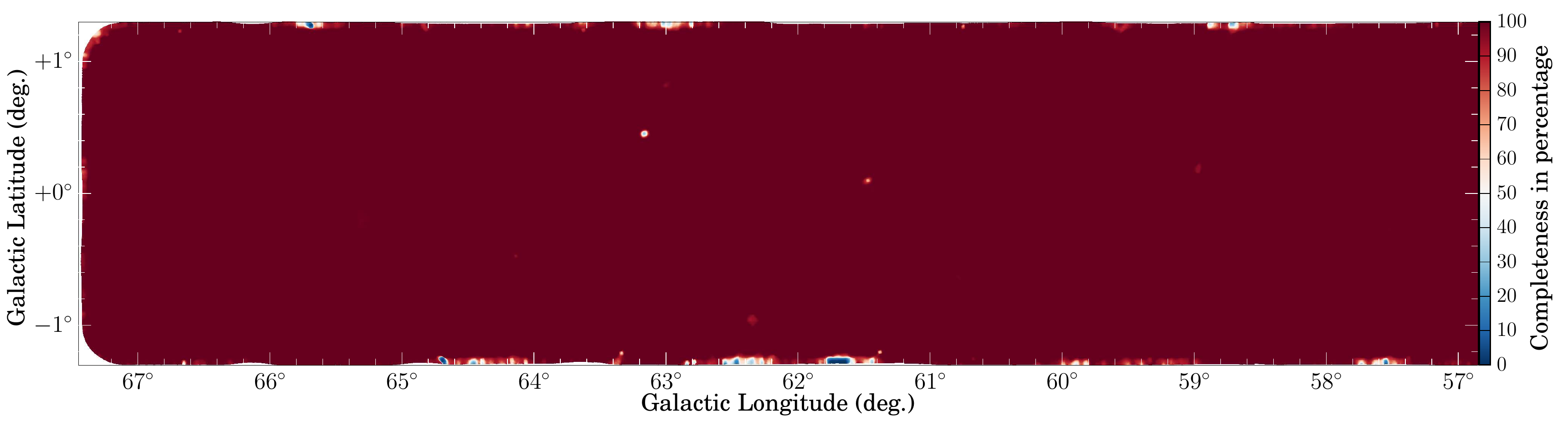}
   \includegraphics[width= 0.95\hsize]{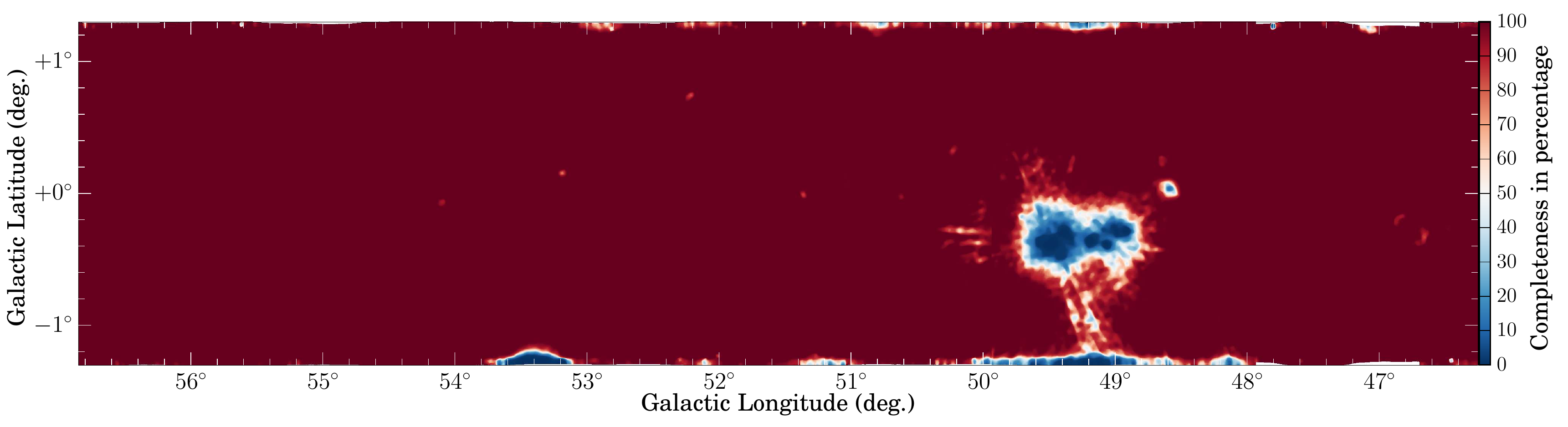}
   \includegraphics[width= 0.95\hsize]{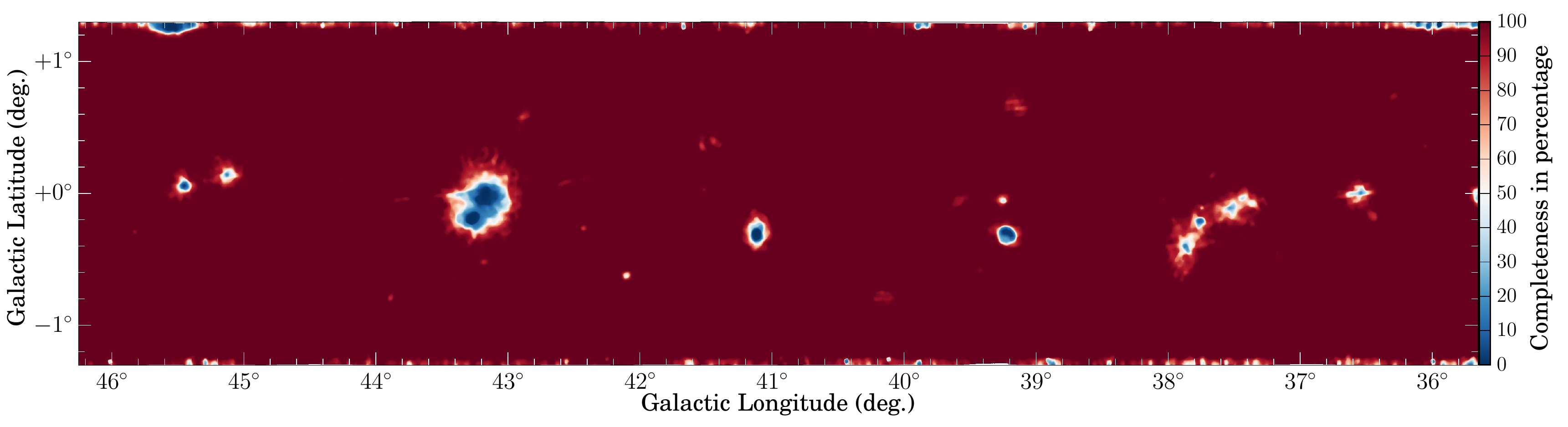}
   \includegraphics[width= 0.95\hsize]{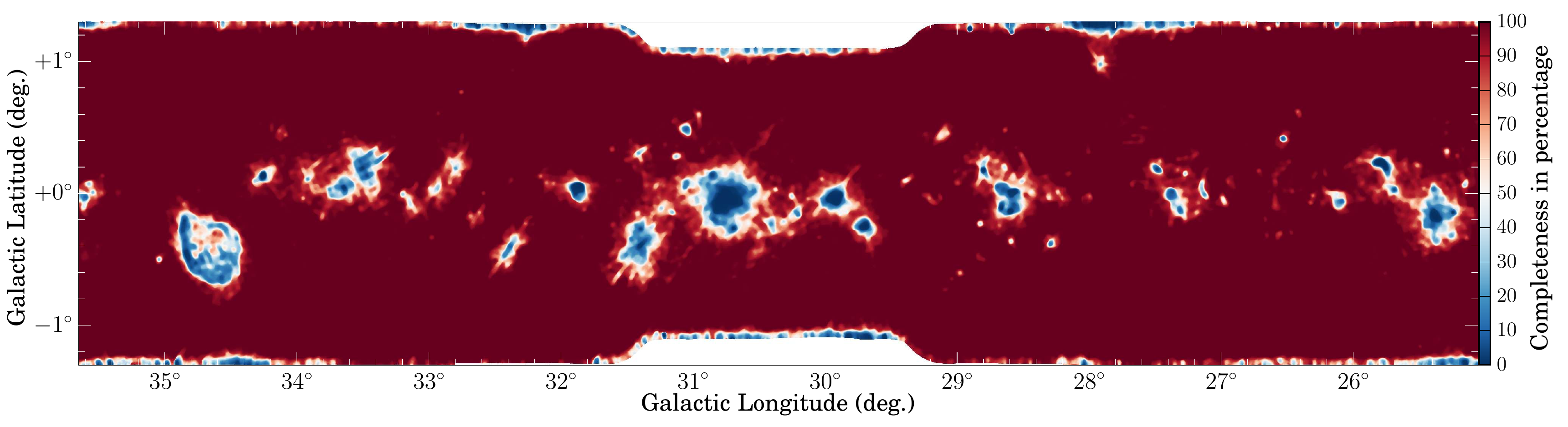}
   \includegraphics[width= 0.95\hsize]{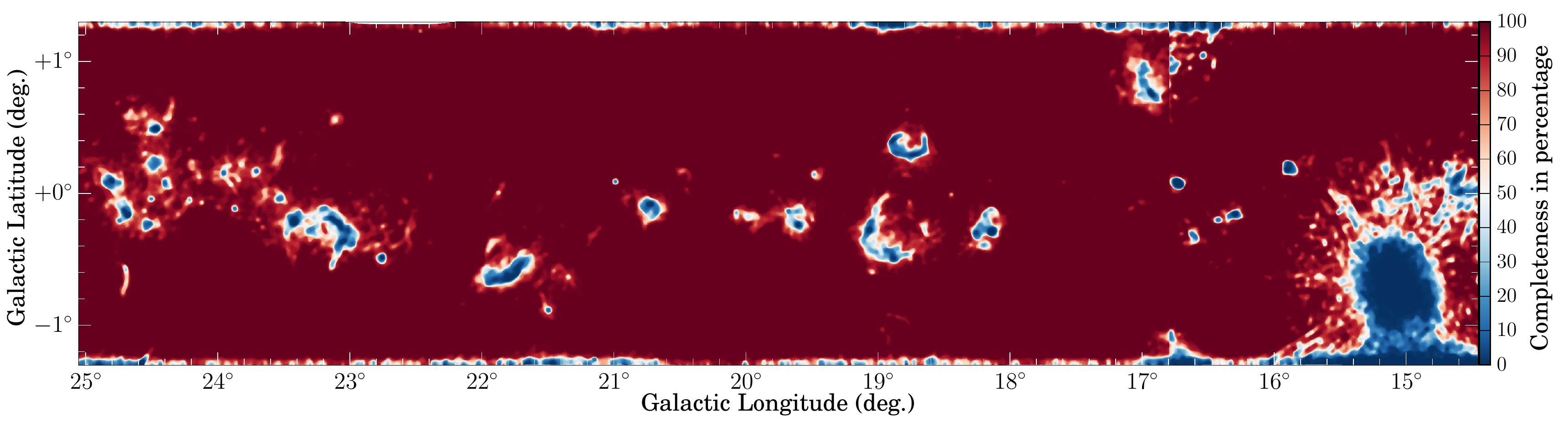}
   \caption{Completeness map in percentage for sources with a peak intensity of 5~mJy~beam$^{-1}$.}
    \label{fig_comp5}
\end{figure*}
 
\begin{figure*}
\centering
   \includegraphics[width= 0.95\hsize]{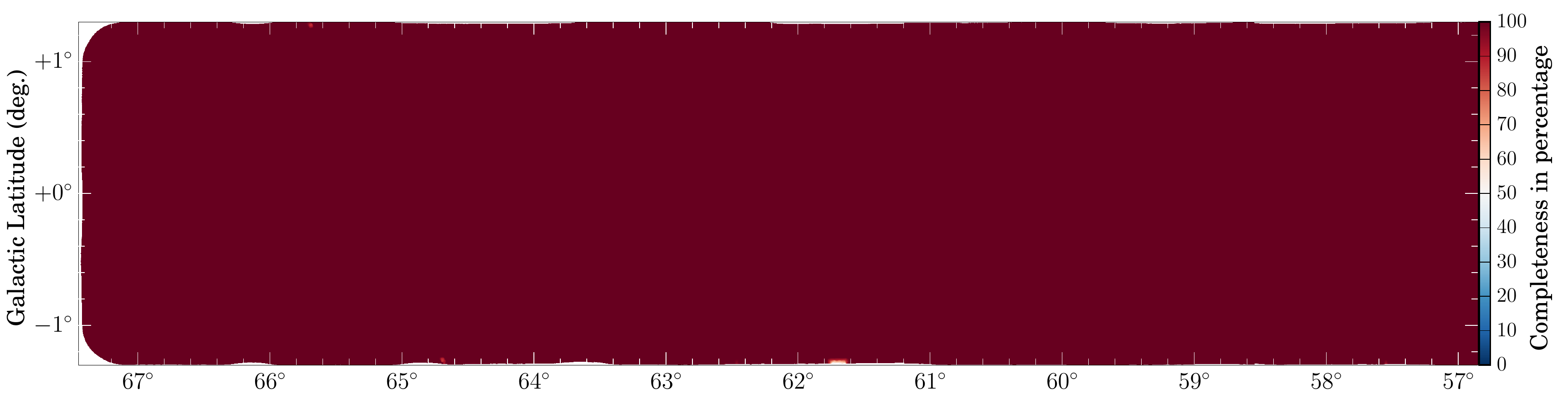}
   \includegraphics[width= 0.95\hsize]{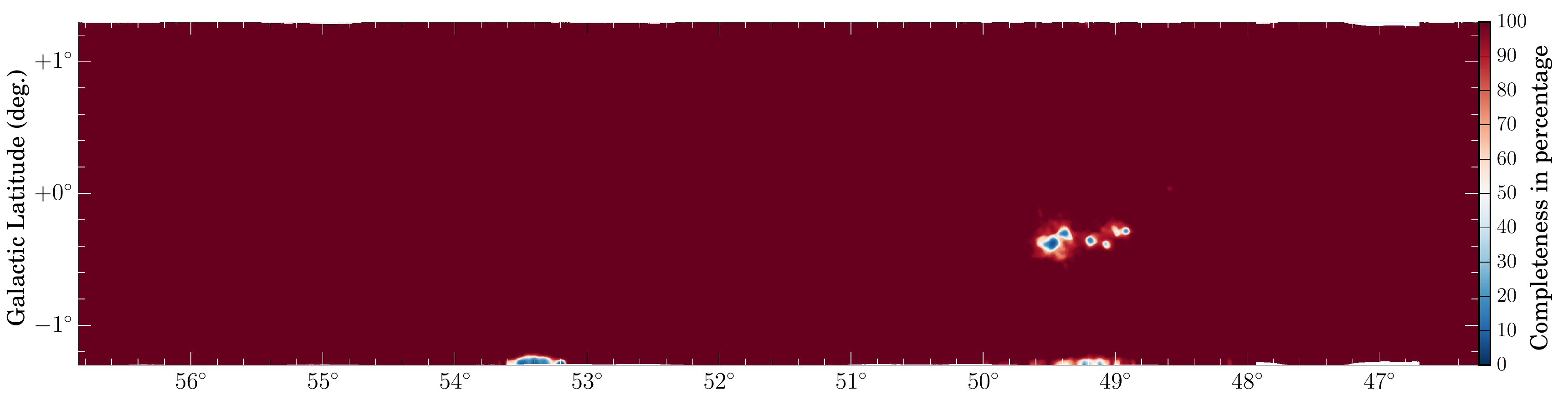}
   \includegraphics[width= 0.95\hsize]{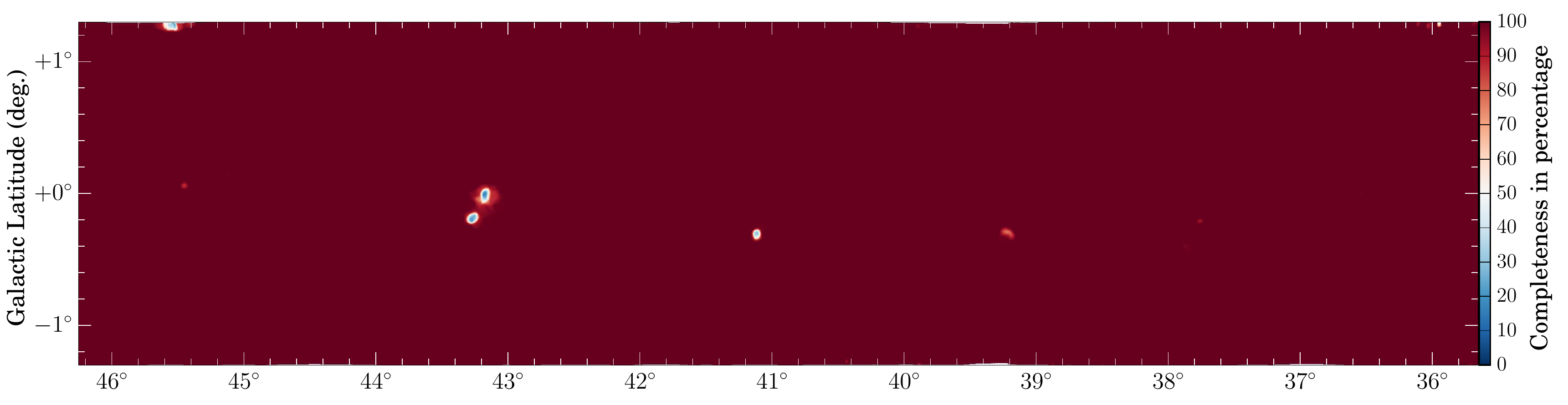}
   \includegraphics[width= 0.95\hsize]{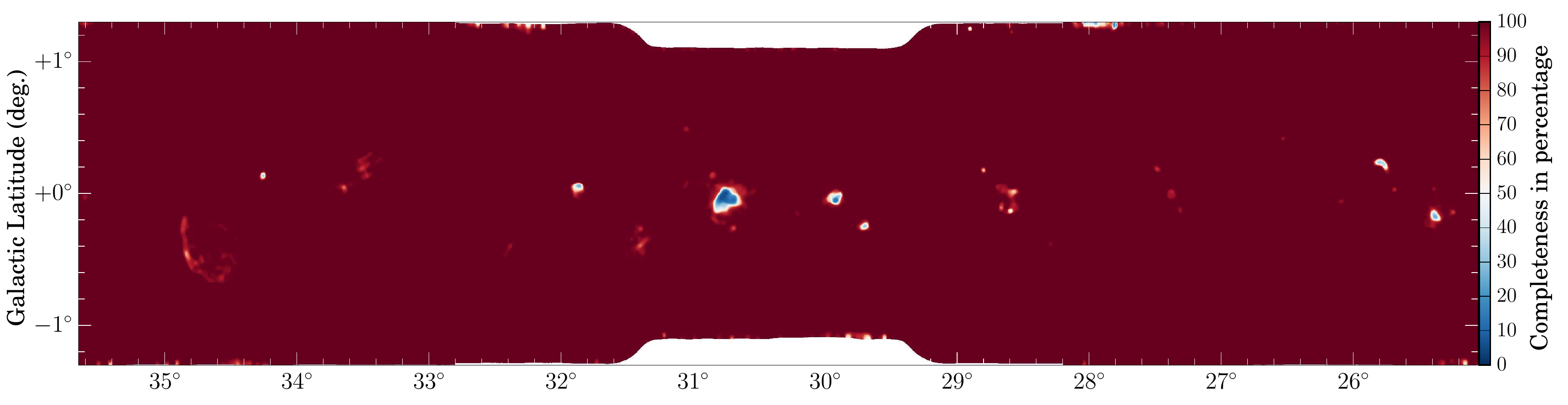}
   \includegraphics[width= 0.95\hsize]{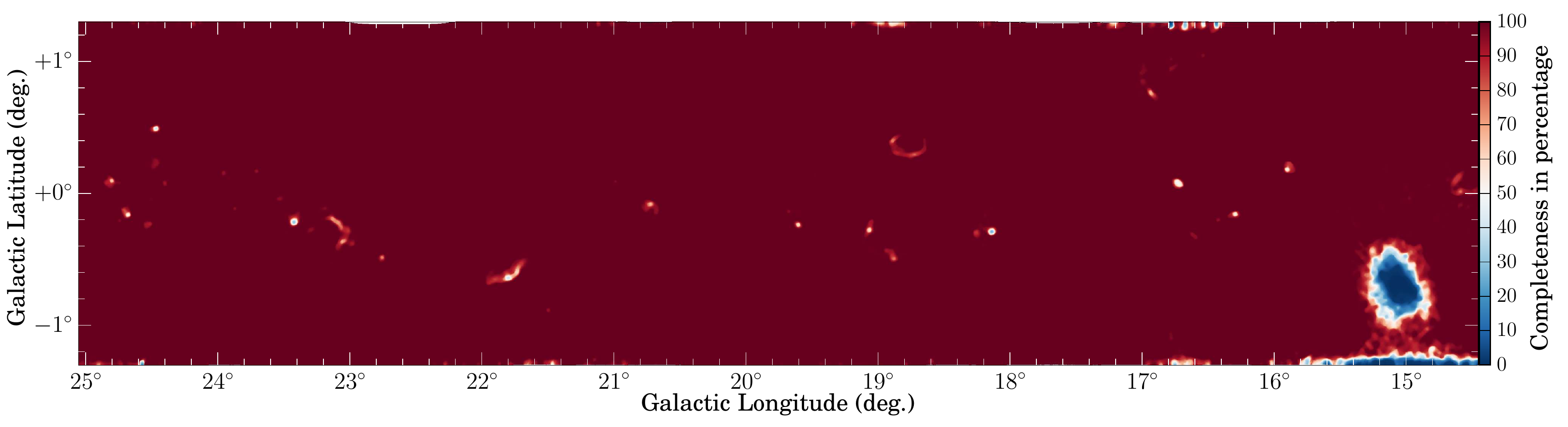}
   \caption{Completeness map in percentage for sources with a peak intensity of 10~mJy~beam$^{-1}$.}
   \label{fig_comp10}
\end{figure*}
 
\section{Compare with CORNISH}
\label{app_cornish}

Sources with a positive spectral index between 1 to 2~GHz are mostly dominated by optically thick free-free emission at these frequencies, however, the emission should be optically thin at higher frequencies and flattening the spectral index there \citep[e.g.,][]{alvarothesis}. If we estimate the flux density at 5~GHz for these sources, we could easily overestimate the flux. Since we do not have flux density measurements between 2~GHz and 5~GHz, we are not sure whether such overestimation would be due to the uncertainty of spectral index or optically depth variation. 

Sources with a spectral index $\lesssim-0.5$ are mostly dominated by synchrotron emission. Depending on the nature of the source, the spectral index could get flatter between 2~GHz and 5~GHz as a possible contribution from free-free emission could increase \citep[e.g.,][]{condon1992}, since free-free emission has a flat or positive spectral index. If we estimate the flux density at 5~GHz for these sources, we would underestimate the flux.

Sources with a flat spectral index ($\sim$--0.1) in the THOR  frequency range are dominated by optically thin free-free emission. The emission at 5~GHz would still be dominated by optically thin free-free emission, since dust continuum will not be relevant until $\sim$100~GHz and synchrotron emission contribution is even smaller than at the lower frequencies. Thus, we can estimate the flux density at 5GHz from the THOR flux density measurements properly and reliably.

\begin{figure}
\centering
   \includegraphics[width= \hsize]{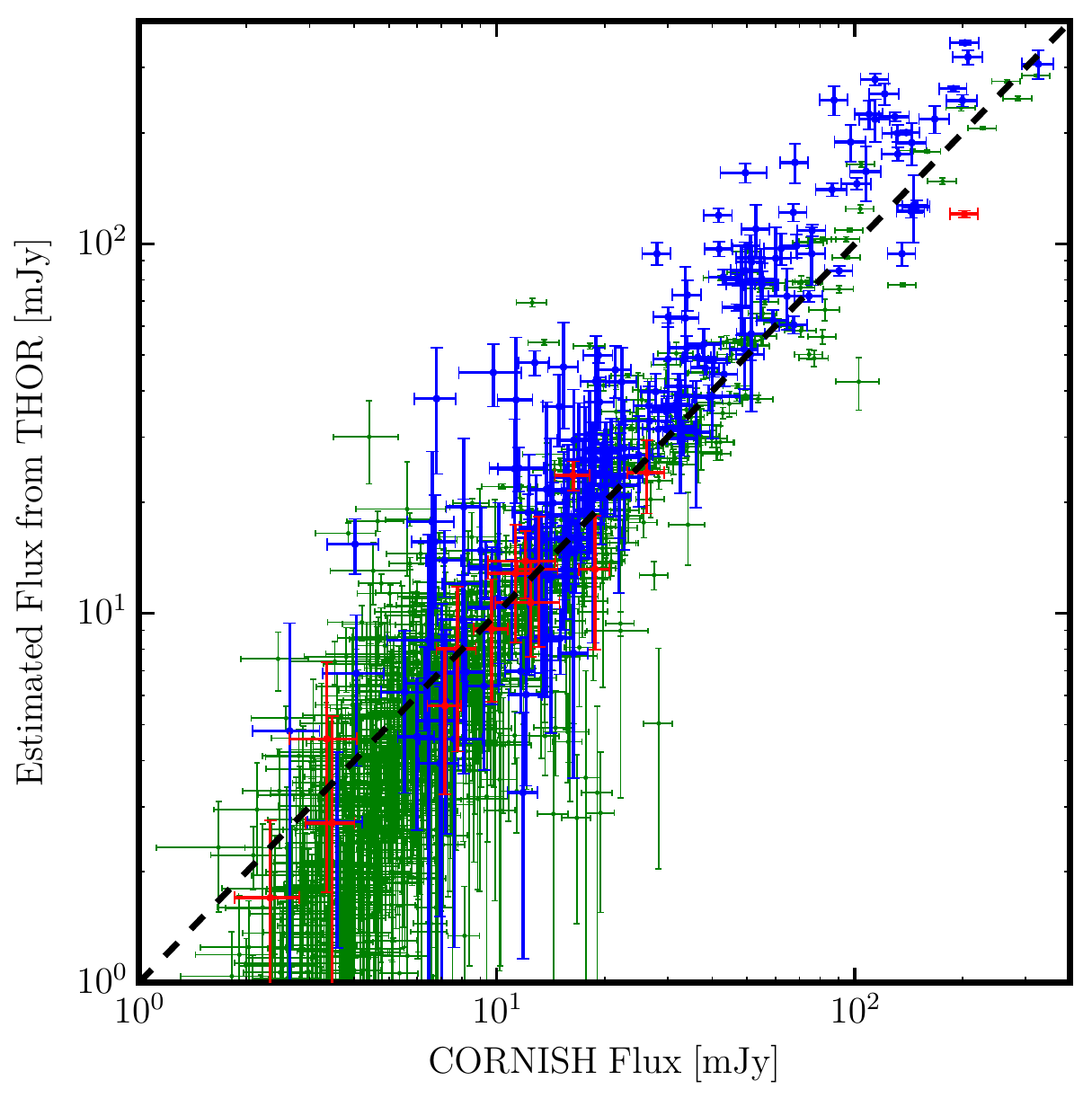}
   \caption{ Extrapolated flux densities at 5~GHz for all 1087 compact sources that are detected in all six spectral windows in THOR (fit\_spws=6) according to their spectral indices from the THOR catalog plotted against the flux densities in the CORNISH catalog. The green ones are sources with a spectral index smaller than 0, the blue ones are the ones with a spectral index larger than 0, the red ones are sources with a spectral index between --0.09 and --0.11 (optically thin free-free emission). The dashed line represents a one-to-one relation.}
   \label{fig_extrflux_all}
\end{figure}

Among the 1905 THOR sources that are matched with a CORNISH source within a radius of 5$\arcsec$, 1087 sources are unresolved and are detected in all six spectral windows in THOR (fit\_spws=6). We extrapolated flux densities at 5~GHz for all these 1087 sources according to their spectral indices from the THOR catalog and plotted against the flux densities in the CORNISH catalog, see Fig.~\ref{fig_extrflux_all}. Figure~\ref{fig_extrflux_all} shows that comparing to the flux densities in the CORNISH catalog, as expected we overestimate the flux density for many sources with a positive spectral index, and similarly we underestimate the flux density for many sources with a negative spectral index. This is expected for such a general extrapolation for the reasons explained above. 

\section{PNe in our continuum catalog}
\label{app_pn}

\longtab[1]{
\begin{landscape}
\begin{longtable}{lccrrrrccr} 
\caption{\label{table_pn} Detected planetary nebulae in THOR continuum catalog.}\\
\hline\hline                 
Gal.ID& R.A. &Dec. &${\rm S_p}$& S/N    & $\alpha$& $\Delta\alpha$&fit\_spws& PNG\footnotemark[1] & Name\footnotemark[1]  \\    
 &      [J2000]&[J2000] & [mJy beam$^{-1}$] &   &  &     &  &    &  \\
\hline    
\endfirsthead
\caption{continued.}\\
\hline\hline
Gal.ID& R.A. &Dec. &${\rm S_p}$& S/N   & $\alpha$& $\Delta\alpha$& fit\_spws & PNG\footnotemark[1] & Name\footnotemark[1] \\    
 &      [J2000]&[J2000] & [mJy beam$^{-1}$] &     &    &  &    & & \\
\hline
\endhead
\hline                        
\endfoot
  G14.585+0.462 & 18:15:21.10 & --16:02:56.3 & 22.84 & 31 & 0.12 & 0.19 & 6 & 014.5+00.4 & MPA J1815--1602\\
  G14.658+1.012 & 18:13:29.12 & --15:43:20.1 & 2.34 & 7 & --0.84 & 0.51 & 4 & 014.6+01.0 & PHR J1813--1543\\
  G14.750--0.250 & 18:18:17.00 & --16:14:31.2 & 23.97 & 28 & --0.37 & 0.32 & 6 & 014.7--00.2$^*$ & \\
  G14.779--0.333 & 18:18:38.70 & --16:15:20.9 & 15.27 & 15 & 2.25 & 0.72 & 4 & 014.7--00.3$^*$ & \\
  G14.896+0.484 & 18:15:53.16 & --15:45:54.9 & 8.01 & 19 & --0.26 & 0.71 & 4 & 014.8+00.4$^*$ & \\
  G15.200--0.086 & 18:18:34.22 & --15:46:06.0 & 11.17 & 9 & --0.49 & 0.64 & 2 & 015.1--00.0$^*$ & \\
  G15.518+1.034 & 18:15:06.47 & --14:57:20.7 & 9.65 & 25 & --0.46 & 0.20 & 6 & 015.5+01.0 & PHR J1815--1457\\
  G15.541+0.336 & 18:17:41.91 & --15:16:04.0 & 13.95 & 30 & 0.55 & 0.29 & 6 & 015.5+00.3$^*$ & \\
  G15.559+0.981 & 18:15:22.87 & --14:56:43.9 & 4.04 & 14 & --0.91 & 0.35 & 5 & 015.5+00.9$^*$ & Pa 99\\
  G15.585+0.400 & 18:17:33.10 & --15:11:56.5 & 11.13 & 25 & 0.81 & 0.38 & 5 & 015.5+00.3$^*$a & \\
  G15.800--0.006 & 18:19:27.38 & --15:12:06.8 & 30.99 & 63 & 0.49 & 0.13 & 6 & 015.7--00.0$^*$ & MSX6C G015.7987--00.0060\\
  G16.055+0.828 & 18:16:54.90 & --14:34:54.6 & 9.92 & 24 & --0.05 & 0.26 & 6 & 016.0+00.8 & MSX6C G016.0545+00.8279\\
  G16.228--0.368 & 18:21:36.82 & --14:59:40.8 & 7.91 & 15 & --0.33 & 0.34 & 6 & 016.2--00.3 & GPSR5 16.228--0.369\\
  G16.428+1.007 & 18:16:59.72 & --14:10:07.4 & 17.03 & 30 & --0.58 & 0.21 & 6 & 016.4+01.0$^*$ & \\
  G16.499+0.115 & 18:20:22.76 & --14:31:39.4 & 14.00 & 29 & --0.78 & 0.17 & 6 & 016.4+00.1$^*$ & \\
  G17.015--0.190 & 18:22:29.76 & --14:12:59.2 & 11.00 & 25 & 0.21 & 0.27 & 6 & 017.0--00.1$^*$ & \\
  G17.221+0.129 & 18:21:43.92 & --13:53:03.9 & 1.73 & 6 & --0.96 & 0.48 & 4 & 017.2+00.1$^*$ & PHR J1821--1353\\
  G17.367+0.523 & 18:20:35.06 & --13:34:15.0 & 14.84 & 49 & 0.24 & 0.15 & 6 & 017.3+00.5 & HRDS G017.364+0.519\\
  G17.449+0.115 & 18:22:13.47 & --13:41:24.1 & 7.37 & 23 & --0.38 & 0.28 & 6 & 017.4+00.1$^*$ & \\
  G17.588+1.068 & 18:19:02.18 & --13:07:03.7 & 7.34 & 25 & --0.23 & 0.26 & 5 & 017.5+01.0 & MPA J1819--1307\\
  G17.616--1.169 & 18:27:13.36 & --14:08:34.0 & 75.50 & 152 & 0.03 & 0.03 & 6 & 017.6--01.1 & VSP 2--18\\
  G17.725--0.243 & 18:24:03.39 & --13:36:50.6 & 50.26 & 141 & 0.11 & 0.04 & 6 & 017.7--00.2$^*$ & \\
  G17.822+0.987 & 18:19:47.13 & --12:57:00.8 & 12.55 & 37 & 0.11 & 0.19 & 6 & 017.8+00.9$^*$ & \\
  G17.865+0.212 & 18:22:40.55 & --13:16:39.1 & 7.87 & 20 & 0.82 & 0.44 & 5 & 017.8+00.2$^*$ & \\
  G18.066+0.854 & 18:20:44.31 & --12:47:52.7 & 26.78 & 97 & 0.11 & 0.06 & 6 & 018.0+00.8 & IRAS 18179--1249\\
  G18.241--0.555 & 18:26:11.05 & --13:18:12.8 & 3.62 & 9 & 1.85 & 4.16 & 2 & 018.2--00.5$^*$ & \\
  G18.241--0.915 & 18:27:29.77 & --13:28:18.2 & 39.56 & 134 & --0.14 & 0.03 & 6 & 018.2--00.9 & MPA J1827--1328\\
  G18.578--0.748 & 18:27:31.93 & --13:05:45.6 & 5.52 & 20 & 0.01 & 0.42 & 6 & 018.5--00.7$^*$ & \\
  G19.003+0.128 & 18:25:10.02 & --12:18:38.5 & 5.73 & 12 & --0.60 & 0.57 & 6 & 019.0+00.1$^*$ & \\
  G19.119+0.818 & 18:22:53.58 & --11:53:09.3 & 10.91 & 46 & --0.41 & 0.15 & 6 & 019.1+00.8$^*$ & MPA J1822--1153\\
  G19.468--0.015 & 18:26:34.37 & --11:58:00.4 & 15.95 & 26 & --0.10 & 0.24 & 6 & 019.4--00.0$^*$ & \\
  G19.533+0.731 & 18:23:59.96 & --11:33:38.9 & 3.56 & 16 & --0.65 & 0.47 & 5 & 019.5+00.7$^*$ & GLIPN1823--1133\\
  G19.610+1.187 & 18:22:30.08 & --11:16:44.2 & 22.79 & 67 & 1.27 & 0.10 & 6 & 019.6+01.1 & MSX6C G019.6095+01.1873\\
  G19.649+0.774 & 18:24:03.96 & --11:26:16.5 & 3.98 & 16 & --0.22 & 0.29 & 6 & 019.6+00.7$^*$ & MPA J1824--1126\\
  G19.930--0.664 & 18:29:47.85 & --11:51:33.1 & 7.33 & 15 & 0.82 & 0.36 & 6 & 019.9--00.6$^*$ & \\
  G19.945+0.913 & 18:24:07.92 & --11:06:41.7 & 26.63 & 109 & 0.24 & 0.05 & 6 & 019.9+00.9 & M 3--53\\
  G20.432+0.357 & 18:27:03.55 & --10:56:25.5 & 9.98 & 37 & 0.16 & 0.20 & 6 & 020.4+00.3$^*$ & MSX6C G020.4320+00.3571\\
  G20.468+0.679 & 18:25:58.08 & --10:45:29.9 & 63.93 & 232 & 0.69 & 0.03 & 6 & 020.4+00.6 & PM 1--231\\
  G20.517+0.478 & 18:26:47.27 & --10:48:30.4 & 17.63 & 66 & 0.05 & 0.09 & 6 & 020.5+00.4$^*$ & \\
  G20.978+0.925 & 18:26:03.07 & --10:11:31.2 & 13.49 & 46 & --0.30 & 0.14 & 6 & 020.9+00.9 & IRAS 18232--1013\\
  G20.983+0.851 & 18:26:19.48 & --10:13:21.8 & 2.70 & 10 & --0.60 & 0.42 & 6 & 020.9+00.8 & PHR1826--1013\\
  G20.999--1.125 & 18:33:28.96 & --11:07:25.5 & 215.77 & 372 & 0.40 & 0.01 & 6 & 020.9--01.1 & M 1--51\\
  G21.165+0.476 & 18:28:01.26 & --10:14:09.8 & 50.44 & 212 & 1.22 & 0.03 & 6 & 021.1+00.4 & GLMP 781\\
  G21.293+0.982 & 18:26:26.62 & --09:53:14.6 & 2.14 & 8 & 0.21 & 0.43 & 5 & 021.2+00.9 & PHR J1826--0953\\
  G21.342--0.842 & 18:33:06.41 & --10:41:21.5 & 4.89 & 10 & 1.91 & 1.40 & 2 & 021.3--00.8$^*$ & IRAS 18303--1043\\
  G21.412+0.074 & 18:29:55.91 & --10:12:13.2 & 1.49 & 5 & --& --&-- & 021.4+00.0$^*$ & 2MASS J18295609--1012119\\
  G21.666+0.811 & 18:27:45.72 & --09:38:14.4 & 21.79 & 71 & 1.43 & 0.10 & 6 & 021.6+00.8 & PM 1--235\\
  G21.685--0.738 & 18:33:22.44 & --10:20:13.4 & 28.96 & 53 & --0.08 & 0.11 & 6 & 021.6--00.7$^*$ & \\
  G21.743--0.672 & 18:33:14.76 & --10:15:18.7 & 6.68 & 8 & --& --& -- & 021.7--00.6$^*$ & M 3--55\\
  G21.820--0.478 & 18:32:41.30 & --10:05:50.0 & 31.58 & 38 & --0.45 & 0.14 & 6 & 021.8--00.4$^*$ & M 3--28\\
  G21.997--0.883 & 18:34:28.96 & --10:07:37.5 & 12.35 & 29 & 0.11 & 0.24 & 6 & 021.9--00.8$^*$ & \\
  G22.221+0.901 & 18:28:28.99 & --09:06:13.1 & 18.02 & 77 & 0.15 & 0.08 & 6 & 022.2+00.9 & IRAS 18257--0908\\
  G22.548--0.106 & 18:32:42.87 & --09:16:49.3 & 27.58 & 49 & 0.29 & 0.13 & 6 & 022.5--00.1$^*$ & \\
  G22.570+1.055 & 18:28:35.28 & --08:43:23.3 & 24.75 & 85 & --0.01 & 0.08 & 6 & 022.5+01.0 & MaC 1--13\\
  G22.658+0.295 & 18:31:28.59 & --08:59:49.4 & 8.63 & 25 & 0.14 & 0.27 & 6 & 022.6+00.2$^*$ & \\
  G23.440+0.745 & 18:31:19.62 & --08:05:42.2 & 9.96 & 26 & 0.55 & 0.28 & 5 & 023.4+00.7$^*$ & PHR J1831--0805\\
  G23.504--0.524 & 18:36:00.05 & --08:37:26.5 & 23.33 & 51 & 0.31 & 0.18 & 5 & 023.5--00.5$^*$ & \\
  G23.890--0.738 & 18:37:29.10 & --08:22:47.0 & 55.10 & 171 & 0.89 & 0.04 & 5 & 023.8--00.7 & GLMP 805\\
  G23.909+1.219 & 18:30:30.31 & --07:27:38.2 & 30.04 & 45 & 0.21 & 0.14 & 5 & 023.9+01.2 & MA 13\\
  G24.179+1.144 & 18:31:16.79 & --07:15:22.1 & 5.54 & 13 & --0.26 & 0.49 & 5 & 024.1+01.1 & PHR J1831--0715\\
  G24.385+0.287 & 18:34:43.75 & --07:28:06.1 & 9.81 & 13 & 0.09 & 0.51 & 6 & 024.3+00.2$^*$ & MSX6C G024.3854+00.2867\\
  G24.430+0.966 & 18:32:22.92 & --07:06:57.5 & 14.20 & 35 & --0.18 & 0.15 & 6 & 024.4+00.9 & MPA J1832--0706\\
  G24.677+0.549 & 18:34:20.00 & --07:05:18.2 & 19.74 & 24 & 1.19 & 0.33 & 5 & 024.6+00.5 & MPA J1834--0706\\
  G24.792--1.004 & 18:40:06.77 & --07:41:59.1 & 10.09 & 22 & 0.09 & 0.27 & 6 & 024.7--01.0$^*$ & \\
  G25.049--0.662 & 18:39:21.55 & --07:18:55.2 & 19.00 & 46 & 0.86 & 0.15 & 6 & 025.0--00.6$^*$ & \\
  G25.847+1.172 & 18:34:16.61 & --05:45:49.6 & 25.83 & 68 & --0.10 & 0.09 & 5 & 025.8+01.1$^*$ & \\
  G25.927--0.984 & 18:42:07.99 & --06:40:54.5 & 8.65 & 32 & --0.33 & 0.19 & 5 & 025.9--00.9 & Pe 1--14\\
  G26.161+0.592 & 18:36:55.59 & --05:45:03.3 & 24.68 & 82 & --0.08 & 0.08 & 5 & 026.1+00.5$^*$ & \\
  G26.715+0.132 & 18:39:35.54 & --05:28:14.0 & 13.01 & 21 & 0.52 & 0.41 & 5 & 026.7+00.1$^*$ & \\
  G26.795--1.050 & 18:43:57.82 & --05:56:21.5 & 5.01 & 17 & --0.27 & 0.42 & 5 & 026.8--01.0 & MPA J1843--0556\\
  G26.832--0.152 & 18:40:49.21 & --05:29:45.9 & 9.05 & 29 & --0.20 & 0.32 & 5 & 026.8--00.1$^*$ & MPA J1840--0529\\
  G27.426--0.250 & 18:42:15.83 & --05:00:46.4 & 6.36 & 11 & --1.37 & 0.64 & 5 & 027.4--00.2$^*$ & \\
  G27.586+1.018 & 18:38:02.06 & --04:17:24.8 & 2.25 & 5 & --& --& -- & 027.5+01.0$^*$ & PHR J1838--0417\\
  G27.659--0.383 & 18:43:10.02 & --04:51:58.5 & 3.46 & 12 & --1.40 & 3.84 & 2 & 027.6--00.3$^*$ & MSX6C G027.6594--00.3838\\
  G27.664--0.827 & 18:44:45.59 & --05:03:54.5 & 6.47 & 28 & 0.13 & 0.22 & 6 & 027.6--00.8 & PHR J1844--0503\\
  G27.702+0.705 & 18:39:21.78 & --04:19:51.0 & 119.82 & 205 & 0.29 & 0.03 & 6 & 027.7+00.7 & M 2--45\\
  G27.886+0.503 & 18:40:25.30 & --04:15:33.9 & 5.00 & 9 & --& --& -- & 027.8+00.5$^*$ & MPA J1840--0415\\
  G28.952+0.256 & 18:43:15.35 & --03:25:27.2 & 22.00 & 33 & --0.15 & 0.14 & 6 & 028.9+00.2$^*$ & PHR J1843--0325\\
  G29.054+0.992 & 18:40:49.32 & --02:59:51.3 & 5.46 & 19 & --0.35 & 0.35 & 6 & 029.0+00.9$^*$ & Pa 112\\
  G29.079+0.458 & 18:42:46.30 & --03:13:09.6 & 21.98 & 21 & --0.19 & 0.08 & 6 & 029.0+00.4 & Abell 48\\
  G29.211--0.069 & 18:44:53.35 & --03:20:32.0 & 117.99 & 156 & 0.52 & 0.04 & 6 & 029.2+00.0 & TDC 1\\
  G29.874--0.819 & 18:48:46.54 & --03:05:41.3 & 76.91 & 149 & 0.92 & 0.05 & 6 & 029.8--00.8 & IRAS 18461--0309\\
  G30.667--0.332 & 18:48:29.28 & --02:10:01.4 & 49.22 & 55 & 0.89 & 0.13 & 6 & 030.6--00.3$^*$ & IRAS 18458--0213\\
  G31.372--0.752 & 18:51:16.16 & --01:43:50.1 & 20.52 & 33 & 0.33 & 0.16 & 6 & 031.3--00.7$^*$ & GPSR 031.374--0.752\\
  G31.907--0.308 & 18:50:39.95 & --01:03:11.1 & 6.82 & 13 & 0.05 & 0.25 & 6 & 031.9--00.3$^*$ & WeSb 4\\
  G32.307+0.154 & 18:49:45.11 & --00:29:08.6 & 16.10 & 42 & 0.54 & 0.19 & 6 & 032.3+00.1$^*$ & 1541--4EAC\\
  G32.498+0.162 & 18:50:04.27 & --00:18:43.6 & 5.610 & 12 & --1.04 & 0.39 & 6 & 032.4+00.1$^*$ & MGE 032.4982+00.1615\\
  G32.548--0.473 & 18:52:25.45 & --00:33:25.9 & 39.54 & 61 & 0.72 & 0.06 & 6 & 032.5--00.4 & MPA J1852--0033\\
  G32.550--0.295 & 18:51:47.44 & --00:28:28.5 & 3.88 & 8 & --& --& -- & 032.5--00.3$^*$ & Te 7\\
  G32.614+0.797 & 18:48:01.27 & +00:04:49.0 & 4.09 & 10 & 1.42 & 1.06 & 3 & 032.6+00.7$^*$ & PM 1--258\\
  G32.859+0.281 & 18:50:18.27 & +00:03:46.4 & 6.44 & 8 & 0.51 & 0.54 & 3 & 032.8+00.2$^*$ & MGE 032.8593+00.2806\\
  G32.939--0.747 & 18:54:06.74 & --00:20:03.7 & 5.43 & 18 & --0.51 & 0.38 & 6 & 032.9--00.7$^*$ & CBSS 3\\
  G33.147+1.044 & 18:48:06.89 & +00:40:03.6 & 26.93 & 105 & --0.36 & 0.06 & 6 & 033.1+01.0$^*$ & \\
  G33.353+0.404 & 18:50:45.90 & +00:33:31.3 & 7.87 & 13 & 1.10 & 0.95 & 4 & 033.3+00.4$^*$ & \\
  G33.454--0.615 & 18:54:34.74 & +00:11:03.8 & 21.32 & 76 & 1.38 & 0.11 & 6 & 033.4--00.6 & GLMP 844\\
  G33.959+1.243 & 18:48:53.01 & +01:28:51.1 & 6.20 & 9 & --& --& -- & 033.9+01.2 & IPHAS J184853.00+012852.2\\
  G33.978--0.985 & 18:56:51.09 & +00:28:53.5 & 3.80 & 13 & --0.81 & 0.45 & 5 & 033.9--00.9$^*$ & PHR J1856+0028\\
  G34.179--0.178 & 18:54:20.74 & +01:01:43.6 & 11.76 & 27 & 0.13 & 0.31 & 6 & 034.1--00.1$^*$ & \\
  G34.420--0.318 & 18:55:17.09 & +01:10:43.6 & 5.97 & 8 & --0.49 & 1.23 & 2 & 034.4--00.3$^*$ & PM 1--265\\
  G34.862--0.063 & 18:55:10.93 & +01:41:21.1 & 45.37 & 60 & 0.40 & 0.12 & 6 & 034.8--00.0$^*$ & IRAS 18526+0137\\
  G35.565--0.491 & 18:57:59.60 & +02:07:07.7 & 71.20 & 115 & --0.01 & 0.05 & 6 & 035.5--00.4 & PHR J1857+0207\\
  G35.770--1.245 & 19:01:03.04 & +01:57:23.8 & 4.21 & 6 & --& --& -- & 035.7--01.2 & UWISH2 PN 3\\
  G35.814--0.253 & 18:57:36.10 & +02:26:57.8 & 1.69 & 5 & --& --& -- & 035.8--00.2$^*$ & UWISH2 PN 2\\
  G36.012--0.256 & 18:57:58.29 & +02:37:25.2 & 11.57 & 49 & 0.92 & 0.19 & 6 & 036.0--00.2$^*$ & \\
  G36.539+0.201 & 18:57:18.44 & +03:18:05.4 & 18.89 & 42 & --0.17 & 0.12 & 6 & 036.5+00.2$^*$ & \\
  G36.954--1.181 & 19:02:59.44 & +03:02:19.3 & 8.91 & 31 & --0.52 & 0.16 & 6 & 036.9--01.1 & HaTr 11\\
  G37.903--0.275 & 19:01:30.36 & +04:17:50.6 & 19.29 & 25 & 0.57 & 0.21 & 6 & 037.9--00.2$^*$ & \\
  G37.960+0.454 & 18:59:00.53 & +04:40:52.2 & 10.58 & 31 & 0.91 & 0.26 & 6 & 037.9+00.4 & IRAS 18564+0436\\
  G40.261--0.276 & 19:05:51.00 & +06:23:32.4 & 3.73 & 17 & 0.59 & 0.58 & 6 & 040.2--00.2$^*$ & \\
  G40.336--1.010 & 19:08:36.72 & +06:07:15.9 & 11.37 & 49 & --0.14 & 0.12 & 6 & 040.3--01.0$^*$ & \\
  G40.370--0.474 & 19:06:45.51 & +06:23:53.9 & 13.47 & 46 & --0.05 & 0.04 & 6 & 040.3--00.4 & Abell 53\\
  G40.554--0.091 & 19:05:43.83 & +06:44:13.9 & 4.84 & 24 & --0.13 & 0.29 & 5 & 040.5--00.0$^*$ & IPHASX J190543.8+064413\\
  G41.271--0.697 & 19:09:13.60 & +07:05:44.0 & 3.73 & 18 & --0.88 & 0.29 & 6 & 041.2--00.6$^*$ & HaTr 14\\
  G41.354+0.539 & 19:04:57.38 & +07:44:16.2 & 36.43 & 131 & 0.32 & 0.05 & 6 & 041.3+00.5$^*$ & \\
  G42.242+1.180 & 19:04:18.01 & +08:49:14.9 & 1.89 & 6 & --& --& -- & 042.2+01.1$^*$ & IPHASX J190417.9+084916\\
  G42.663--0.865 & 19:12:25.41 & +08:15:07.3 & 16.75 & 76 & --0.24 & 0.10 & 6 & 042.6--00.8 & PM 1--288\\
  G42.767+0.822 & 19:06:33.70 & +09:07:21.1 & 9.82 & 24 & --0.32 & 0.18 & 6 & 042.7+00.8 & [GKF2010] MN102\\
  G43.029+0.140 & 19:09:30.19 & +09:02:26.6 & 66.52 & 77 & 1.13 & 0.06 & 6 & 043.0+00.1 & GLMP 879\\
  G43.294--0.646 & 19:12:49.29 & +08:54:48.0 & 38.97 & 89 & 1.28 & 0.08 & 6 & 043.2--00.6 & IRAS 19104+0849\\
  G43.580+0.026 & 19:10:56.62 & +09:28:37.0 & 9.82 & 23 & 0.93 & 0.29 & 5 & 043.5+00.0$^*$ & 1631--598D\\
  G44.638+0.483 & 19:11:17.20 & +10:37:35.0 & 14.63 & 54 & 0.45 & 0.14 & 6 & 044.6+00.4 & IRAS 19089+1032\\
  G44.735+0.260 & 19:12:16.21 & +10:36:33.4 & 8.02 & 23 & 0.32 & 0.34 & 6 & 044.7+00.2 & AGP 1\\
  G44.949+0.898 & 19:10:22.18 & +11:05:38.6 & 7.78 & 26 & --0.01 & 0.20 & 6 & 044.9+00.8 & IPHASX J191022.1+110538\\
  G45.283--0.627 & 19:16:30.44 & +10:40:56.8 & 20.12 & 72 & 0.80 & 0.10 & 6 & 045.2--00.6 & 1648--2717\\
  G45.661+1.040 & 19:11:11.93 & +11:47:27.1 & 4.11 & 13 & --0.23 & 0.46 & 6 & 045.6+01.0$^*$ & Pa 128\\
  G46.975+0.270 & 19:16:29.09 & +12:35:51.5 & 17.15 & 60 & 0.06 & 0.12 & 6 & 046.9+00.2$^*$ & \\
  G47.611+1.082 & 19:14:45.09 & +13:32:18.4 & 1.37 & 9 & --0.54 & 0.66 & 4 & 047.6+01.0$^*$ & IPHASX J191445.1+133219\\
  G47.636--1.232 & 19:23:11.47 & +12:28:37.3 & 17.02 & 36 & --0.29 & 0.11 & 6 & 047.6--01.2$^*$ & Pa 132\\
  G47.688--0.302 & 19:19:55.74 & +12:57:37.7 & 3.73 & 28 & 1.27 & 0.33 & 4 & 047.6--00.3$^*$ & PM 1--296\\
  G47.987+0.202 & 19:18:40.40 & +13:27:39.9 & 8.07 & 46 & --0.22 & 0.17 & 6 & 047.9+00.2$^*$ & \\
  G48.162+1.163 & 19:15:30.59 & +14:03:49.0 & 17.81 & 67 & 1.32 & 0.12 & 6 & 048.1+01.1 & K 3--29\\
  G48.732+0.930 & 19:17:27.34 & +14:27:35.8 & 17.12 & 76 & 0.40 & 0.08 & 6 & 048.7+00.9 & IPHASX J191727.2+142735\\
  G50.040+1.096 & 19:19:22.93 & +15:41:36.7 & 15.18 & 44 & 0.17 & 0.14 & 6 & 050.0+01.0 & IRAS 19171+1536\\
  G50.556+0.045 & 19:24:14.60 & +15:39:12.4 & 93.33 & 215 & 0.72 & 0.02 & 6 & 050.5+00.0 & NVSS J192414+153909\\
  G50.895+0.057 & 19:24:52.01 & +15:57:28.7 & 7.97 & 23 & --0.43 & 0.36 & 6 & 050.8+00.0$^*$ & \\
  G51.022--0.489 & 19:27:06.85 & +15:48:37.5 & 1.92 & 14 & 0.07 & 0.48 & 5 & 051.0--00.4$^*$ & MGE 051.0214--00.4885\\
  G51.510+0.167 & 19:25:40.57 & +16:33:04.8 & 107.19 & 314 & --0.04 & 0.01 & 6 & 051.5+00.2 & KLW 1\\
  G51.606+0.914 & 19:23:07.27 & +16:59:21.5 & 11.96 & 72 & 0.80 & 0.10 & 6 & 051.6+00.9$^*$ & \\
  G51.834+0.284 & 19:25:53.53 & +16:53:31.9 & 54.21 & 252 & 0.61 & 0.02 & 6 & 051.8+00.2 & IPHAS J192553.53+165331.4\\
  G52.150--0.376 & 19:28:56.77 & +16:51:17.4 & 40.02 & 251 & --0.65 & 0.03 & 6 & 052.1--00.3$^*$ & \\
  G54.712+0.420 & 19:31:10.80 & +19:29:04.4 & 1.06 & 10 & --1.20 & 0.77 & 4 & 054.7+00.4$^*$ & UWISH2 PN 4\\
  G55.507--0.558 & 19:36:26.87 & +19:42:23.6 & 96.49 & 560 & 1.12 & 0.01 & 6 & 055.5--00.5 & M 1--71\\
  G56.402--0.903 & 19:39:35.73 & +20:19:03.3 & 9.60 & 60 & 0.80 & 0.11 & 6 & 056.4--00.9 & K 3--42\\
  G56.422--0.373 & 19:37:40.14 & +20:35:43.1 & 3.60 & 21 & --0.15 & 0.17 & 6 & 056.4--00.3 & IPHASX J193740.4+203547\\
  G57.980--0.768 & 19:42:26.08 & +21:45:21.4 & 2.66 & 21 & --0.72 & 0.27 & 5 & 057.9--00.7 & Kn 7\\
  G58.179--0.811 & 19:43:01.32 & +21:54:25.9 & 1.52 & 8 & --0.68 & 0.56 & 4 & 058.1--00.8 & IPHASX J194301.3+215424\\
  G58.641+0.919 & 19:37:29.40 & +23:09:47.2 & 3.54 & 22 & --0.12 & 0.47 & 5 & 058.6+00.9 & PM 1--309\\
  G59.399--0.788 & 19:45:34.20 & +22:58:33.6 & 27.54 & 110 & 0.26 & 0.05 & 6 & 059.4--00.7 & PM 1--313\\
  G59.778--0.829 & 19:46:33.11 & +23:16:59.3 & 3.31 & 22 & --1.05 & 0.35 & 5 & 059.7--00.8 & IPHASX J194633.0+231659\\
  G59.824--0.536 & 19:45:32.94 & +23:28:11.2 & 17.95 & 98 & 0.69 & 0.05 & 6 & 059.8--00.5 & 2MASS J19453289+2328105\\
  G59.991+0.686 & 19:41:16.63 & +24:13:25.5 & 0.64 & 5 & --& --& -- & 059.9+00.6$^*$ & PM 1--311\\
  G60.249+0.822 & 19:41:19.09 & +24:30:53.2 & 7.78 & 57 & --0.58 & 0.10 & 6 & 060.2+00.8 & Kn 11\\
  G60.524--0.318 & 19:46:15.60 & +24:11:04.6 & 3.32 & 21 & --0.61 & 0.50 & 5 & 060.5--00.3 & K 3--45\\
  G60.987--0.570 & 19:48:14.27 & +24:27:27.8 & 6.65 & 27 & 0.60 & 0.40 & 4 & 060.9--00.5 & IRAS 19461+2419\\
  G61.206--0.088 & 19:46:53.58 & +24:53:23.0 & 4.70 & 20 & --1.02 & 0.35 & 6 & 061.2--00.0$^*$ & PM 1--314\\
  G62.494--0.270 & 19:50:28.47 & +25:54:30.5 & 19.84 & 104 & --0.24 & 0.06 & 6 & 062.4--00.2 & M 2--48\\
  G62.702+0.060 & 19:49:40.93 & +26:15:20.2 & 4.15 & 21 & --0.67 & 0.38 & 5 & 062.7+00.0 & IPHASX J194940.9+261521\\
  G62.755--0.727 & 19:52:48.91 & +25:53:58.1 & 15.39 & 105 & 0.09 & 0.07 & 6 & 062.7--00.7 & IPHASX J195248.8+255359\\
  G63.045+0.598 & 19:48:23.27 & +26:49:26.7 & 5.73 & 22 & --0.38 & 0.32 & 6 & 063.0+00.5$^*$ & \\
  G63.523+0.089 & 19:51:26.54 & +26:58:37.7 & 3.59 & 20 & --0.10 & 0.47 & 5 & 063.5+00.0 & IPHASX J195126.5+265839\\
  G63.889+0.123 & 19:52:09.18 & +27:18:31.8 & 29.92 & 116 & 0.03 & 0.04 & 6 & 063.8+00.1 & K 3--48\\
  G64.730--0.518 & 19:56:35.12 & +27:41:53.5 & 7.17 & 37 & --0.18 & 0.18 & 6 & 064.7--00.5$^*$ & Pa 139\\
  G65.893--0.866 & 20:00:41.34 & +28:30:26.4 & 0.68 & 5 & 0.76 & 0.71 & 3 & 065.8--00.8 & IPHASX J200041.5+283023\\
  G65.914+0.599 & 19:55:01.93 & +29:17:22.8 & 2.26 & 15 & --0.57 & 0.13 & 6 & 065.9+00.5 & NGC 6842\\
\end{longtable}
\footnotetext[1]{These information are from the HASH planetary nebula database \citep{parker2016}. PNe that have no 20~cm radio flux information in the HASH catalog are marked with $^*$.} 
\end{landscape}
}

\section{Detected pulsars in our continuum catalog}
\label{app_psr}

\longtab[2]{
\begin{longtable}{lccrrrrcr} 
\caption{\label{table_psr} Detected pulsars in our continuum catalog}\\
\hline\hline                 
Gal.ID& R.A. &Dec. &${\rm S_p}$& S/N & $\alpha$& $\Delta\alpha$ & fit\_spws&  PSR Name\\    
      &  [J2000] &[J2000] & [mJy beam$^{-1}$] &    &  & & &   \\
\hline    
\endfirsthead
\caption{continued.}\\
\hline\hline
Gal.ID& R.A. &Dec. &${\rm S_p}$& S/N & $\alpha$& $\Delta\alpha$ & fit\_spws&  PSR Name\\    
        &  [J2000] &[J2000] & [mJy beam$^{-1}$] &    &  & & &   \\
\hline
\endhead
\hline                     
\endfoot
  G16.405+0.610 & 18:18:23.69 & --14:22:38.1 & 5.60 & 14 & --2.95 & 0.30 & 6 & B1815--14\\
  G17.160+0.483 & 18:20:19.69 & --13:46:17.0 & 2.40 & 7 & -- & -- & -- & B1817--13\\
  G18.001--0.691 & 18:26:13.18 & --13:34:47.9 & 3.55 & 16 & --1.29 & 0.48 & 5 & B1823--13\\
  G19.767+0.946 & 18:23:40.30 & --11:15:10.5 & 3.48 & 17 & --1.55 & 0.32 & 6 & B1820--11\\
  G19.810+0.741 & 18:24:29.54 & --11:18:40.0 & 1.01 & 6 & --1.05 & 1.73 & 2 & B1821--11\\
  G21.286+0.798 & 18:27:05.49 & --09:58:43.4 & 1.74 & 8 & --2.72 & 4.08 & 2 & B1824--10\\
  G23.385+0.063 & 18:33:40.15 & --08:27:32.1 & 4.89 & 8 & --1.35 & 1.10 & 3 & B1830--08\\
  G25.172+0.762 & 18:34:29.38 & --06:33:03.3 & 1.77 & 5 & 0.06 & 1.36 & 2 & J1834--0633\\
  G27.073--0.941 & 18:44:05.09 & --05:38:32.7 & 1.93 & 9 & --2.03 & 0.59 & 4 & B1841--05\\
  G27.322--0.033 & 18:41:17.95 & --05:00:21.6 & 6.27 & 6 & -- & -- & -- & J1841--0500\\
  G27.818+0.279 & 18:41:05.74 & --04:25:21.6 & 2.11 & 6 & 0.97 & 1.57 & 2 & B1838--04\\
  G28.193--0.785 & 18:45:34.76 & --04:34:29.2 & 2.19 & 8 & 0.4 & 1.93 & 2 & B1842--04\\
  G28.347+0.174 & 18:42:26.55 & --03:59:59.6 & 6.83 & 12 & --1.88 & 0.89 & 3 & B1839--04\\
  G28.876--0.939 & 18:47:22.78 & --04:02:15.3 & 3.01 & 10 & --1.90 & 0.60 & 3 & B1844--04\\
  G31.339+0.039 & 18:48:23.56 & --01:23:58.9 & 10.64 & 14 & --1.15 & 0.77 & 3 & B1845--01\\
  G32.763+0.091 & 18:50:48.27 & --00:06:31.1 & 12.99 & 17 & --0.66 & 0.35 & 6 & J1850--0006\\
  G36.007+0.057 & 18:56:50.89 & +02:45:45.4 & 1.50 & 6 & --2.84 & 5.76 & 2 & J1856+0245\\
  G37.213--0.637 & 19:01:31.72 & +03:31:05.4 & 3.27 & 9 & --3.21 & 1.17 & 2 & B1859+03\\
  G38.163--0.151 & 19:01:32.37 & +04:35:08.1 & 5.45 & 17 & --2.86 & 0.31 & 5 & J1901+0435\\
  G39.814+0.335 & 19:02:50.39 & +06:16:33.2 & 1.52 & 7 & --4.68 & 0.64 & 2 & B1900+06\\
  G40.569+1.056 & 19:01:39.01 & +07:16:34.4 & 1.78 & 10 & --1.00 & 0.82 & 4 & B1859+07\\
  G40.604--0.304 & 19:06:35.18 & +06:41:03.9 & 2.27 & 10 & -- & -- & -- & B1904+06\\
  G40.944+0.065 & 19:05:53.73 & +07:09:19.0 & 1.25 & 6 & -- & -- & -- & B1903+07\\
  G41.520--0.871 & 19:10:18.84 & +07:14:09.5 & 1.07 & 5 & --0.89 & 1.23 & 2 & J1910+0714\\
  G41.740--0.772 & 19:10:22.07 & +07:28:35.8 & 1.19 & 8 & --1.08 & 1.31 & 2 & J1910+0728\\
  G43.501--0.684 & 19:13:20.74 & +09:04:42.2 & 2.61 & 7 & -- & -- & -- & J1913+0904\\
  G44.557--1.019 & 19:16:32.43 & +09:51:26.8 & 1.53 & 11 & 0.01 & 2.34 & 2 & B1914+09\\
  G44.707--0.650 & 19:15:29.98 & +10:09:43.9 & 1.15 & 7 & --2.09 & 3.31 & 2 & B1913+10\\
  G44.832+0.992 & 19:09:48.76 & +11:02:02.5 & 1.75 & 9 & 0.51 & 1.12 & 3 & B1907+10\\
  G47.576+0.451 & 19:16:58.69 & +13:12:49.4 & 1.84 & 18 & --0.73 & 0.48 & 5 & B1914+13\\
  G48.260+0.624 & 19:17:39.80 & +13:53:56.6 & 4.68 & 33 & --1.74 & 0.22 & 6 & B1915+13\\
  G49.096+0.866 & 19:18:23.54 & +14:45:04.0 & 2.10 & 13 & --0.45 & 1.08 & 3 & B1916+14\\
  G51.859+0.063 & 19:26:45.32 & +16:48:32.0 & 1.42 & 10 & 0.76 & 1.37 & 3 & B1924+16\\
  G55.575+0.639 & 19:32:07.99 & +20:20:47.6 & 0.88 & 7 & --3.67 & 3.30 & 2 & B1929+20\\
  G57.509--0.290 & 19:39:38.59 & +21:34:58.7 & 11.05 & 21 & --3.78 & 0.13 & 6 & B1937+21\\
  G57.903+0.308 & 19:38:14.12 & +22:13:11.6 & 0.74 & 5 & 2.73 & 1.29 & 2 & J1938+2213\\
  G65.839+0.443 & 19:55:27.97 & +29:08:43.4 & 1.13 & 10 & --0.69 & 1.62 & 3 & B1953+29\\
  G65.924+0.772 & 19:54:22.47 & +29:23:16.3 & 8.10 & 43 & --3.45 & 0.13 & 6 & B1952+29\\
\end{longtable}
}

\section{X-ray sources detected in our continuum catalog}
\label{app_xray}

\longtab[3]{
 \small
\begin{landscape}
\begin{longtable}{lccrrrrcr} 
\caption{\label{table_xray} X--ray sources detected in our continuum catalog. Sources do not have radio counterparts in SIMBAD or NED are marked with $^*$.}\\
\hline\hline                 
Gal.ID&R. A. &Dec. & ${\rm S_p}$& S/N  & $\alpha$& $\Delta\alpha$ & fit\_spws& X-ray ID\\    
           & [J2000]&[J2000] & [mJy beam$^{-1}$] &       &       \\
\hline    
\endfirsthead
\caption{continued.}\\
\hline\hline
Gal.ID&R. A. &Dec. & ${\rm S_p}$& S/N  & $\alpha$& $\Delta\alpha$ & fit\_spws& X-ray ID\\    
           & [J2000]&[J2000] & [mJy beam$^{-1}$] &       &       \\
\hline
\endhead
\hline                        
\endfoot
  G14.926+0.040 & 18:17:34.12 & --15:56:56.5 & 11.58 & 13 & --0.41 & 0.39 & 6 & 3XMM J181734.1--155655\\
  G15.179--0.844$^*$ & 18:21:18.47 & --16:08:39.4 & 12.95 & 5 & --1.47 & 0.46 & 4 & CXO J182118.0--160838\\
  G15.487+0.340 & 18:17:34.65 & --15:18:46.5 & 82.29 & 132 & --1.08 & 0.03 & 6 & 3XMM J181734.5--151847\\
  G16.471+0.150 & 18:20:11.91 & --14:32:12.0 & 17.02 & 35 & 1.27 & 0.23 & 6 & 3XMM J182011.7--143210\\
  G16.667+0.693$^*$ & 18:18:36.18 & --14:06:25.9 & 5.45 & 9 & 0.84 & 2.73 & 2 & 3XMM J181836.3--140626\\
  G16.733--1.185 & 18:25:35.01 & --14:55:54.9 & 354.21 & 329 & --0.99 & 0.01 & 6 & CXO J182534.8--145556;3XMM J182534.7--145554\\
  G17.071+0.607$^*$ & 18:19:42.09 & --13:47:30.6 & 3.97 & 8 & -- & -- & -- & CXO J181941.9--134735\\
  G17.227+0.880 & 18:19:00.97 & --13:31:31.2 & 107.64 & 168 & --0.69 & 0.03 & 6 & CXO J181900.9--133131\\
  G17.772+0.409$^*$ & 18:21:46.95 & --13:16:01.5 & 8.18 & 21 & --0.72 & 0.24 & 6 & 3XMM J182146.9--131602\\
  G17.777--0.737 & 18:25:57.42 & --13:47:56.2 & 18.36 & 62 & --0.54 & 0.09 & 6 & 1SXPS J182557.4--134753\\
  G18.156--0.823$^*$ & 18:26:59.79 & --13:30:11.7 & 1.76 & 7 & --2.61 & 4.66 & 2 & 3XMM J182659.9--133012\\
  G19.235--0.127$^*$ & 18:26:32.04 & --12:13:29.9 & 3.31 & 6 & 0.29 & 0.71 & 3 & 3XMM J182631.7--121336\\
  G19.236+0.495 & 18:24:17.02 & --11:56:00.9 & 98.64 & 274 & --0.75 & 0.02 & 6 & 3XMM J182416.7--115558\\
  G19.384--0.461$^*$ & 18:28:01.47 & --12:14:55.4 & 5.29 & 13 & --1.66 & 0.39 & 5 & 3XMM J182801.4--121458\\
  G20.367+0.449 & 18:26:36.22 & --10:57:17.9 & 102.55 & 293 & --0.36 & 0.02 & 6 & 3XMM J182636.4--105719\\
  G20.409+0.100$^*$ & 18:27:56.70 & --11:04:47.9 & 3.63 & 6 & -- & -- & -- & 3XMM J182756.7--110447\\
  G20.923+0.213 & 18:28:30.65 & --10:34:20.3 & 124.64 & 268 & --0.89 & 0.02 & 6 & 1SXPS J182830.6--103415\\
  G21.104--0.976$^*$ & 18:33:08.47 & --10:57:46.4 & 2.03 & 6 & --0.34 & 0.59 & 2 & 3XMM J183308.8--105752\\
  G21.347--0.629 & 18:32:20.73 & --10:35:10.7 & 930.3 & 917 & --0.00 & 0.00 & 6 & 1SXPS J183220.9--103509;CXO J183220.8--103510;3XMM J183220.8--103509\\
  G21.622--0.742$^*$ & 18:33:16.21 & --10:23:41.2 & 6.36 & 13 & --0.82 & 0.72 & 5 & 1SXPS J183316.2--102339;CXO J183316.2--102341;3XMM J183316.2--102341\\
  G21.655--0.361 & 18:31:57.49 & --10:11:23.9 & 29.25 & 53 & 0.94 & 0.14 & 6 & 3XMM J183157.5--101123\\
  G22.155--0.156 & 18:32:09.44 & --09:39:04.5 & 34.01 & 76 & 0.18 & 0.05 & 6 & 1SXPS J183208.9--093905;CXO J183208.9--093905;3XMM J183208.9--093905\\
  G22.933--0.076 & 18:33:19.61 & --08:55:28.9 & 120.22 & 155 & --0.67 & 0.02 & 6 & 3XMM J183319.5--085528\\
  G25.266--0.161 & 18:37:57.96 & --06:53:31.6 & 534.1 & 393 & --0.69 & 0.01 & 6 & 1SXPS J183757.9--065331;CXO J183758.0--065331;3XMM J183758.0--065331\\
  G25.320--0.098 & 18:37:50.40 & --06:48:56.7 & 34.56 & 28 & --1.02 & 0.14 & 6 & CXO J183750.5--064856;3XMM J183750.5--064854\\
  G26.239--0.080 & 18:39:28.20 & --05:59:26.6 & 89.3 & 123 & --0.89 & 0.04 & 5 & CXO J183928.1--055925\\
  G26.398--0.104$^*$ & 18:39:51.15 & --05:51:36.5 & 2.28 & 5 & -- & -- & -- & 3XMM J183951.1--055135\\
  G26.560--0.175$^*$ & 18:40:24.13 & --05:44:53.7 & 10.77 & 20 & --0.75 & 0.38 & 5 & 1SXPS J184024.4--054446;3XMM J184023.9--054445\\
  G26.757--0.413$^*$ & 18:41:36.98 & --05:40:55.2 & 1.49 & 5 & --2.27 & 1.76 & 2 & 3XMM J184136.8--054045\\
  G26.760--0.403$^*$ & 18:41:35.31 & --05:40:30.3 & 1.46 & 5 & -- & -- & -- & 1SXPS J184135.2--054030;3XMM J184135.2--054029\\
  G28.098--0.781 & 18:45:23.42 & --04:39:28.8 & 61.15 & 196 & --1.02 & 0.02 & 6 & 3XMM J184523.4--043927\\
  G28.402+0.479 & 18:41:27.26 & --03:48:44.2 & 124.6 & 228 & 0.70 & 0.02 & 6 & CXO J184127.3--034844\\
  G29.159--0.722$^*$ & 18:47:07.32 & --03:41:13.1 & 11.46 & 29 & --0.49 & 0.24 & 6 & 1SXPS J184707.3--034110\\
  G29.719--0.031 & 18:45:40.95 & --02:52:24.1 & 11.35 & 15 & --1.04 & 0.41 & 6 & CXO J184541.1--025227;3XMM J184541.1--025225\\
  G30.236--0.572$^*$ & 18:48:33.32 & --02:39:36.4 & 5.94 & 14 & --0.29 & 0.96 & 3 & 3XMM J184833.2--023937\\
  G30.246--0.910 & 18:49:46.75 & --02:48:20.9 & 5.95 & 16 & --1.24 & 0.39 & 5 & 1SXPS J184946.9--024815\\
  G30.437--0.206 & 18:47:37.24 & --02:18:51.6 & 12.75 & 12 & --0.99 & 0.52 & 6 & 1SXPS J184737.2--021848\\
  G31.150--0.189 & 18:48:51.75 & --01:40:18.8 & 20.52 & 30 & --0.82 & 0.22 & 6 & CXO J184852.0--014027;3XMM J184852.1--014026\\
  G31.897+0.316 & 18:48:25.43 & --00:46:36.0 & 26.08 & 62 & --1.23 & 0.10 & 6 & CXO J184825.4--004635\\
  G32.482--0.123 & 18:51:03.44 & --00:27:23.6 & 21.69 & 45 & 0.07 & 0.13 & 6 & 3XMM J185103.7--002726\\
  G32.491--1.066$^*$ & 18:54:25.64 & --00:52:43.1 & 1.23 & 7 & -- & -- & -- & 3XMM J185425.7--005241\\
  G33.275+1.069$^*$ & 18:48:15.38 & +00:47:33.6 & 4.01 & 14 & --0.76 & 0.36 & 6 & 3XMM J184815.3+004733\\
  G33.498+0.194 & 18:51:46.74 & +00:35:31.3 & 822.8 & 496 & --0.35 & 0.01 & 6 & 3XMM J185146.7+003533\\
  G33.595+0.184$^*$ & 18:51:59.40 & +00:40:26.3 & 5.07 & 5 & -- & -- & -- & 3XMM J185159.6+004013\\
  G34.576--0.320$^*$ & 18:55:34.59 & +01:19:01.0 & 4.67 & 5 & -- & -- & -- & 3XMM J185535.1+011855\\
  G35.786+0.064$^*$ & 18:56:25.03 & +02:34:07.8 & 2.64 & 6 & 0.42 & 2.67 & 2 & 1SXPS J185625.1+023410;3XMM J185625.1+023406\\
  G36.551+0.002 & 18:58:02.33 & +03:13:15.2 & 832.33 & 651 & --0.23 & 0.01 & 6 & 3XMM J185802.3+031316\\
  G39.907--0.972$^*$ & 19:07:40.91 & +05:45:27.7 & 1.34 & 6 & 4.82 & 3.31 & 2 & CXO J190740.8+054528\\
  G40.437--0.797$^*$ & 19:08:02.35 & +06:18:32.9 & 1.56 & 9 & -- & -- & -- & 3XMM J190802.5+061833\\
  G42.895+0.573 & 19:07:41.80 & +09:07:19.2 & 352.36 & 415 & --0.51 & 0.01 & 6 & 1SXPS J190741.6+090715;3XMM J190742.0+090713\\
  G43.407--0.021 & 19:10:47.32 & +09:18:09.5 & 40.93 & 40 & --1.04 & 0.07 & 6 & 3XMM J191047.7+091806\\
  G45.366--0.219 & 19:15:11.52 & +10:56:44.4 & 108.14 & 188 & 0.10 & 0.03 & 6 & 3XMM J191511.4+105645\\
  G49.093--0.462$^*$ & 19:23:13.84 & +14:07:27.7 & 11.9 & 10 & --0.28 & 0.42 & 5 & 3XMM J192313.4+140727\\
  G49.113+0.935$^*$ & 19:18:10.57 & +14:47:55.8 & 1.22 & 6 & -- & -- & -- & 3XMM J191810.7+144754\\
  G53.769+0.161$^*$ & 19:30:13.64 & +18:32:01.4 & 1.32 & 5 & -- & -- & -- & 1SXPS J193013.6+183203\\
  G54.116--1.014$^*$ & 19:35:16.00 & +18:16:11.7 & 0.74 & 5 & -- & -- & -- & 1SXPS J193516.3+181613\\
  G55.237+0.387$^*$ & 19:32:22.82 & +19:55:45.4 & 1.31 & 12 & --1.93 & 0.82 & 4 & 1SXPS J193222.6+195543\\
  G56.082+0.105 & 19:35:10.44 & +20:31:54.7 & 319.89 & 1206 & 0.44 & 0.00 & 6 & 1SXPS J193510.2+203156\\
  G57.157+0.796$^*$ & 19:34:49.41 & +21:48:24.9 & 1.53 & 11 & --0.36 & 1.00 & 4 & 3XMM J193449.4+214824\\
  G57.161+0.746$^*$ & 19:35:01.48 & +21:47:11.1 & 1.08 & 7 & --1.44 & 3.13 & 2 & 3XMM J193501.3+214710\\
  G57.414+0.855$^*$ & 19:35:08.35 & +22:03:36.7 & 0.69 & 5 & -- & -- & -- & 3XMM J193508.2+220336\\
  G59.031+0.850$^*$ & 19:38:35.28 & +23:28:08.3 & 2.07 & 13 & --0.55 & 0.55 & 4 & 1SXPS J193835.5+232804\\
  G59.774+0.047$^*$ & 19:43:14.14 & +23:43:05.1 & 0.96 & 6 & --2.29 & 3.21 & 2 & CXO J194314.5+234307\\
  G59.908+0.199$^*$ & 19:42:57.05 & +23:54:34.4 & 4.01 & 14 & 1.00 & 0.68 & 4 & CXO J194257.3+235435\\
  G61.444+1.227 & 19:42:22.78 & +25:45:13.4 & 3.53 & 7 & --1.82 & 1.35 & 2 & 1SXPS J194222.7+254518\\
  G63.271--1.158 & 19:55:38.63 & +26:07:10.1 & 3.23 & 13 & --1.02 & 0.52 & 5 & 1SXPS J195538.5+260708;3XMM J195538.4+260708\\
  G63.356--0.976$^*$ & 19:55:08.74 & +26:17:11.9 & 0.95 & 7 & -- & -- & -- & 1SXPS J195508.3+261712\\
  G63.658+1.188 & 19:47:29.19 & +27:39:01.6 & 34.15 & 70 & --1.16 & 0.05 & 6 & 3XMM J194729.1+273903\\
  G63.750+1.071$^*$ & 19:48:09.22 & +27:40:14.9 & 1.58 & 8 & 0.99 & 1.12 & 3 & 3XMM J194809.5+274023\\
  G63.791--0.176$^*$ & 19:53:04.71 & +27:04:15.1 & 0.65 & 6 & 2.21 & 1.38 & 2 & 3XMM J195304.4+270413\\
  G63.905+0.998$^*$ & 19:48:47.56 & +27:46:04.7 & 1.59 & 10 & --1.02 & 0.84 & 3 & CXO J194847.4+274605;3XMM J194847.4+274603\\
  G64.075+0.077$^*$ & 19:52:45.54 & +27:26:39.9 & 1.19 & 8 & --3.31 & 1.15 & 3 & 3XMM J195245.4+272637\\
  G65.665+1.042$^*$ & 19:52:42.21 & +29:18:16.8 & 1.74 & 8 & --1.11 & 0.73 & 3 & CXO J195241.8+291820;3XMM J195241.9+291821\\
  G65.678+1.205$^*$ & 19:52:05.48 & +29:23:58.1 & 3.82 & 9 & --1.31 & 0.81 & 3 & CXO J195206.1+292353\\
  G65.731+1.063 & 19:52:46.28 & +29:22:19.7 & 6.19 & 24 & --1.64 & 0.23 & 6 & CXO J195246.4+292213\\
  G65.779--0.337$^*$ & 19:58:21.98 & +28:41:17.0 & 1.45 & 9 & --7.52 & 2.33 & 2 & 1SXPS J195821.6+284118\\
  G65.950--0.184 & 19:58:10.66 & +28:54:50.2 & 14.12 & 85 & --1.03 & 0.07 & 6 & 1SXPS J195810.6+285449\\
  G66.766+0.458 & 19:57:37.30 & +29:56:39.4 & 7.20 & 50 & --0.13 & 0.14 & 6 & 1SXPS J195737.5+295640\\
  G66.891+0.083$^*$ & 19:59:23.87 & +29:51:19.9 & 3.24 & 22 & --0.47 & 0.35 & 5 & 1SXPS J195923.8+295115\\

\end{longtable}
\end{landscape}
}

\end{appendix}
\end{document}